\documentclass[11pt]{article}
\usepackage[T1]{fontenc}
\usepackage{jheppub}
\usepackage{mathtools}
\usepackage{amssymb}
\usepackage{amsmath,amsthm}
\usepackage{cancel}
\usepackage{empheq}
\usepackage{mathrsfs}
\usepackage{comment}
\usepackage{physics}
\usepackage{xcolor}
\usepackage{stmaryrd}
\usepackage{accents}
\usepackage{pict2e}
\usepackage[normalem]{ulem}
\usepackage{float}

\newcommand{\dbtilde}[1]{\accentset{\approx}{#1}}

\newcommand*\widefbox[1]{\fbox{\hspace{2em}#1\hspace{2em}}}

\newcommand{\p}{\partial}

\makeatletter
\DeclareRobustCommand{\loplus}{\mathbin{\mathpalette\dog@lsemi{+}}}
\DeclareRobustCommand{\lotimes}{\mathbin{\mathpalette\dog@lsemi{\times}}}
\DeclareRobustCommand{\roplus}{\mathbin{\mathpalette\dog@rsemi{+}}}
\DeclareRobustCommand{\rotimes}{\mathbin{\mathpalette\dog@rsemi{\times}}}

\newcommand{\dog@rsemi}[2]{\dog@semi{#1}{#2}{-90,90}}
\newcommand{\dog@lsemi}[2]{\dog@semi{#1}{#2}{270,90}}
\newcommand{\dog@semi}[3]{
  \begingroup
  \sbox\z@{$\m@th#1#2$}
  \setlength{\unitlength}{\dimexpr\ht\z@+\dp\z@\relax}
  \makebox[\wd\z@]{\raisebox{-\dp\z@}{
    \begin{picture}(1,1)
    \linethickness{\variable@rule{#1}}
    \roundcap
    \put(0.5,0.5){\makebox(0,0){\raisebox{\dp\z@}{$\m@th#1#2$}}}
    \put(0.5,0.5){\arc[#3]{0.5}}
    \end{picture}
  }}
  \endgroup
}
\newcommand{\variable@rule}[1]{
  \fontdimen8  
  \ifx#1\displaystyle\textfont3\else
    \ifx#1\textstyle\textfont3\else
      \ifx#1\scriptstyle\scriptfont3\else
        \scriptscriptfont3\relax
  \fi\fi\fi
}
\makeatother

\DeclareMathOperator\artanh{artanh}
\newcommand{\tl}[1]{\langle #1 \rangle}
\newcommand{\tldd}[2]{D_{\langle #1}D_{#2\rangle}}
\newcommand{\tluu}[2]{D^{\langle #1}D^{#2\rangle}}

\newcommand*\xbar[1]{
  \hbox{
    \vbox{
      \hrule height 0.5pt
      \kern0.3ex      
      \hbox{
        \kern-0.0em    
        \ensuremath{#1}
        \kern-0.0em 
      }
    }
  }
} 

\newcommand{\hyp}{\mathbf{h}}
\newcommand{\bceq}{\stackrel{\scriptscriptstyle\wedge}{=}}
\newcommand{\eomeq}{\doteq}
\newcommand{\bceomeq}{\stackrel{\raisebox{-1ex}{$\scriptscriptstyle\wedge$}}{\doteq}}

\begin{document}

\title{A Covariant Formulation of Logarithmic Supertranslations at Spatial Infinity}
\author{Florian Girelli${}^1$, Simon Langenscheidt${}^2$, Giulio Neri${}^{3,4}$, Christopher Pollack${}^{1,2}$, Céline Zwikel${}^{5}$}
\affiliation{$^1$ Department of Applied Mathematics, University of Waterloo,200 University Avenue West, Waterloo, Ontario, Canada, N2L 3G1
\\
$^2$ Perimeter Institute for Theoretical Physics, 31 Caroline Street North, Waterloo, Ontario, Canada N2L 2Y5\\
$^3$ SISSA, International School for Advanced Studies, 
via Bonomea 265, 34136 Trieste, Italy\\
$^4$ INFN, Sezione di Trieste,
via Valerio 2, 34127 Trieste, Italy\\
$^5$ Coll\`ege de France, 11 place Marcelin Berthelot, 75005 Paris, France\\
}

\emailAdd{florian.girelli@uwaterloo.ca}
\emailAdd{slangenscheidt@perimeterinstitute.ca}
\emailAdd{gneri@sissa.it}
\emailAdd{cajpolla@uwaterloo.ca}
\emailAdd{celine.zwikel@college-de-france.fr}

\abstract{
We investigate the asymptotic symmetries of asymptotically flat spacetimes at spatial infinity. We propose a new symplectic structure and conservative boundary conditions in a polyhomogeneous Beig–Schmidt expansion.
The asymptotic symmetries extend the BMS algebra by abelian sectors, notably incorporating regular log-translations and log-supertranslations.
The associated charges are finite and conserved, and we show that their algebra admits a central extension between supertranslations and log-supertranslations, and between the singular translations and regular log-translations. Our analysis is compatible with, and extends, both the work of~\cite{Compere:2011ve} and~\cite{Fuentealba:2022xsz}: it extends the former by incorporating log-supertranslations, and the latter by allowing both parities of the log-supertranslations in the same phase space. 
These newly identified symmetries at spatial infinity encode novel physical information that has not been revealed in other regions of asymptotically flat spacetimes, thereby opening the door to new observables to consider at null and timelike infinity.
}

\maketitle

\section{Introduction}
Penrose’s compactification of asymptotically flat spacetimes yields a conformal boundary composed of five regions: future and past timelike infinity (the endpoints of timelike geodesics), future and past null infinity (the endpoints of null geodesics), and spatial infinity, where spacelike radial geodesics terminate. These regions provide a natural framework for discussing scattering amplitudes of massive or massless bodies, as they describe the initial and final states of such processes, as well as global properties of the spacetime.

Null infinities {in particular} have been extensively studied, as they play a central role in describing gravitational radiation and massless scattering. The seminal work of Bondi, van der Burg, Metzner, and Sachs~\cite{Bondi:1962px, Sachs:1962zza} showed that the asymptotic symmetry algebra of asymptotically flat spacetimes is an infinite-dimensional extension of the Poincar\'{e} group by supertranslations, thereby called the BMS group. Since then, several generalizations have been proposed by relaxing the boundary conditions or fall-off behaviors, leading to larger symmetry algebras~\cite{Barnich:2009se,Barnich:2016lyg,Barnich:2011mi,Barnich:2010eb,Campiglia:2015yka,Campiglia:2020qvc,Campiglia:2014yka,Flanagan:2015pxa,Compere:2018ylh,Freidel:2021fxf,Geiller:2022vto,Geiller:2024amx,Geiller:2025dqe}.

An important aspect of describing the full boundary of spacetime is understanding how its different regions are connected~\cite{Regge:1974zd, Ashtekar:1978zz, Friedrich1998}. For example, a null generator entering spatial infinity from past null infinity would exit towards future null infinity with reversed spatial direction. Hence to glue past null infinity to future null infinity through spatial infinity, one needs to  use some \textit{antipodal matching} on the celestial sphere~\cite{Ashtekar:1978zz}. Equivalently, in a phase space analysis~\cite{Regge:1974zd, Fuentealba:2022xsz}, this matching is encoded by demanding that the different variables satisfy some parity conditions. 

In the context of graviton scattering, Strominger extended the antipodal matching between future and past null infinity across spatial infinity for each BMS generator so that one consistently reproduces Weinberg's soft theorems~\cite{Strominger:2013jfa}, see also~\cite{Troessaert:2017jcm,Prabhu:2019fsp,Prabhu:2021cgk,Capone:2022gme}.
More recently, a unified framework has been developed in which the BMS group is realized at each region (spatial, time, null) of the boundary~\cite{Compere:2023qoa}.

\medskip

In this work, we focus on the covariant description of spatial infinity.
In Penrose’s compactification, spatial infinity corresponds to a single point; however, fields are not single-valued there, as different directions of approach to spatial infinity may support independent asymptotic data. To account for this fact, Ashtekar and Hansen proposed a blow-up of spatial infinity that provides a four-dimensional description of this point~\cite{Ashtekar:1978zz}. A particularly useful approach to spatial infinity was developed by Beig and Schmidt~\cite{Beig:1982ifu,Beig1984Integration}, whose coordinate chart is asymptotically compatible with the Ashtekar–Hansen construction. This coordinate chart has been used to construct charges in the covariant phase space formalism~\cite{Ashtekar:1990gc,Compere:2011ve,Troessaert:2017jcm,Compere:2017knf}.

It is well-known from the study of (partial) differential equations that their solutions, expanded around (regular) singular points, should also depend on logarithmic contributions~\cite{Erdelyi1956, Hinch1991, Melrose1995}. The solutions are then written in \textit{polyhomogeneous series}, involving power laws and powers of logarithms.
The analysis of Chrusciel et al~\cite{Chrusciel1992Asymptotic, ChruscielMacCallumSingleton1994, ChruscielDelay2000} emphasizes that smoothness is not a general feature and that logarithmic terms are produced by generic initial data. Since they are naturally present, it is  interesting to see how the symmetry structure at null (see~\cite{Geiller:2024ryw}), spatial, and timelike infinity is affected by such logarithmic terms.
Beig and Schmidt~\cite{Beig:1982ifu}, Beig~\cite{Beig1984Integration}, and Friedrich~\cite{Friedrich1998} also discussed the appearance of such logarithmic contributions at spatial infinity. 

As recalled by Ashtekar~\cite{Ashtekar:1985aa}, Bergmann~\cite{PhysRev.124.274} had already pointed out that certain coordinate transformations involving logarithmic terms do not spoil the asymptotic flatness of the metric. These transformations are generated by the so-called logarithmic translations. Comp\`{e}re and Dehouck (CD)~\cite{Compere:2011ve} later incorporated these transformations within the covariant phase space framework.
Logarithmic translations also play an important role in the logarithmic soft theorem~\cite{Boschetti:2025tru} and in the matching of BMS charges between null and spatial infinity~\cite{Compere:2023qoa}.

On another front, spatial infinity has been explored in the Hamiltonian framework. In their seminal paper, Regge and Teitelboim proposed consistent boundary conditions yielding a non-trivial Hamiltonian, given by a surface integral at spatial infinity, realizing the Poincar\'{e} transformations~\cite{Regge:1974zd}. 
By relaxing these conditions, it was shown that the BMS algebra can also be consistently realized at spatial infinity~\cite{Henneaux:2018cst,Henneaux:2018hdj,Fiorucci:2024ndw}. 

Recently, Fuentealba, Henneaux, and Troessaert (FHT) introduced a new set of boundary conditions, extending the BMS algebra by an abelian sector generated by logarithmic supertranslations (log-supertranslations)~\cite{Fuentealba:2022xsz}. They obtained a well-defined Hamiltonian generator (finite and integrable) associated with such transformations and the asymptotic symmetry algebra. We will refer to the resulting algebra as the \textit{FHT log-BMS algebra}. Log-supertranslations generalize logarithmic translations by allowing the transformation parameter to depend arbitrarily on the sphere coordinates, directly analogous to the enhancement of translations to supertranslations in the BMS algebra. An important feature of the FHT log-BMS algebra is the existence of a central extension between the charges of supertranslations and log-supertranslations.  As a result, FHT demonstrated that Lorentz generators can be redefined in such a way that they commute with supertranslations, providing a BMS-frame-invariant definition of angular momentum. Their construction relied on parity conditions which were used to regularize the theory. 

The log-supertranslations also appear in the recent work~\cite{Mishra:2025nmd}. In the spirit of the Ashtekar-Hansen construction, they relaxed the usual requirement of smoothness of the expansion of the metric in the radial direction.
They derive that supertranslations and log-supertranslations are singled out as the transformations that preserve their relaxed notion of spatial infinity, while preserving finiteness of the spatial infinity mass aspect. Their analysis was performed on a purely geometric level and did not consider a specific theory.

\medskip

Motivated by these recent results, we investigate the symmetries at spatial infinity in the Beig–Schmidt chart, allowing for logarithmic terms in the radial expansion of the metric in order to accommodate the action of log-supertranslations, and without imposing parity conditions.
Parity conditions are often assumed from the outset, typically to ensure compatibility with a smooth null infinity or because one has specific classes of solutions in mind. However, in order to keep the analysis as general as possible, and bearing in mind that non-smooth solutions are also physically relevant, we refrain from imposing such conditions. 

Our main result is the realization of a log-BMS algebra within the covariant phase space framework. In this construction, no parity conditions need to be imposed on the field components. The algebra we obtain is an extension of the FHT log-BMS algebra which, upon imposing appropriate parity conditions, recovers the former.

In order to realize a log-BMS algebra in the phase space, we need to address two main challenges: (i) construct a set of finite charges, (ii) ensure their conservation.
In the Hamiltonian framework (i) is typically achieved through the imposition of parity conditions and/or some of the canonical constraints. Within the covariant phase space approach, however, it was shown that one can always use the inherent ambiguities of the formalism~\cite{Iyer:1994ys} to renormalize the symplectic structure~\cite{Andrade:2006pg,McNees:2023tus,McNees:2024iyu} without imposing parity conditions (or even boundary conditions). This ensures that all charges are finite. Note that CD likewise derived finite charges without imposing parity conditions within the covariant phase space framework, though using a different prescription~\cite{Compere:2008us} which is based on the construction of appropriate boundary Lagrangians.

Since radiation does not reach spatial infinity, there must exist a set of boundary conditions that ensure (ii). This is a delicate requirement as we want boundary conditions that do not break log-supertranslations.
Such boundary conditions amount to requiring that the Weyl tensor admits a purely polynomial asymptotic expansion in $1/\rho$ where $\rho$ is the radial coordinate, with polyhomogeneous terms appearing only at subsubleading order.

\subsubsection*{Organization of the paper}

In \S\ref{sec:BSgauge}, we use metrics in Beig-Schmidt form to probe the region around spatial infinity, by introducing fall-offs of the line element compatible with log-supertranslations and relevant fields in the asymptotic expansion. We discuss the solution space through equations of motion, give expressions for the Weyl tensor, and compute the residual symmetry algebra preserving these fall-offs, including the transformations of the fields. 
In \S\ref{sec:symplstructure} we regularize the symplectic form. Leveraging the finite ambiguities of the covariant phase space formalism, we propose a symplectic form and boundary conditions compatible with the action of log-supertranslations, as well as the associated restrictions on the residual symmetry algebra. 
We compute the charges associated with the residual symmetries in \S\ref{sec:charges}. We evaluate the algebra satisfied by the charges: it is given by the semi-direct sum of the global Lorentz algebra acting on regular translations and three Heisenberg algebras involving the supertranslations and log-supertranslations.  We also present a redefinition of the Lorentz generators such that the Poincar\'{e} algebra becomes an ideal of the log-BMS algebra. We end the section with a consistent restriction of our boundary conditions to connect with smooth null infinity and recover the algebra of FHT in this sector. 
We end the main part of the paper by an extended discussion and outlook in \S\ref{sec:outlook}. 

The main text is supplemented with a number of appendices.
Appendix \ref{app:EoM} discusses the structure of the solution space in more detail, giving a radial 3+1 ADM decomposition, explicit solutions to relevant wave equations, and important results about decompositions of tensors on the hyperboloid. Appendix
\ref{app:subsubleadresidualsymm} presents the derivation of the residual symmetry algebra of our space of fields, including the subsubleading field transformations. Appendix \ref{app:FHTDetails} describes how to make comparisons between our results and FHT, as well as the effect of FHT's parity conditions on our space of fields. Appendix \ref{App:Redef} gives details on performing the redefinition of the Lorentz charge. Finally, Appendix \ref{app:idhyp} provides useful identities on scalars, vectors, and tensors on the hyperboloid that are commonly employed within this work.

\subsubsection*{Notation and convention}\label{Notation}

We here report the notations and set the conventions used throughout this work. 
The notation $\langle\,\rangle$ denotes the symmetric and traceless part of a rank two-tensor with a factor $1/2$, for instance $\bar \tau_{\tl{ab}}=\frac12(\bar \tau_{ab}+\bar \tau_{ba})-\frac13\hyp_{ab} \bar\tau$.
The symbol $\bceq$ indicates that the relation holds for field configurations satisfying the boundary conditions (see~\eqref{eq:newBC}), while $\eomeq$ denotes equality on-shell of the (leading and subleading) equations of motion. The symbol $\bceomeq$ is a combination of the two.
We employ the notation $A\overset{\leftrightarrow}{D_a}B=A D_a B-B D_a A$ for the Wronskian derivative. 

We choose units for which the speed of light $c=1$, while we explicitly show Newton's constant $G$ and set $\kappa^2 = 8\pi G$.  

\section{Beig-Schmidt form with a radial polyhomogenous expansion}\label{sec:BSgauge}
We consider asymptotically flat spacetimes near spatial infinity in the hyperbolic representation. We generalize the Beig–Schmidt form of the metric to make it compatible with the action of logarithmic supertranslations. For this purpose, we allow for a polyhomogeneous expansion in the radial coordinate $\rho$. 

In this section we present the fall-offs and boundary conditions for our line element and set our notation. We then analyze the asymptotic equations of motion, compute the Weyl tensor, and derive the residual symmetries that preserve our fall-offs and boundary conditions.

\subsection{Line element}
To formulate the fall-off conditions close to spatial infinity we use Beig-Schmidt coordinates as in~\cite{Beig:1982ifu, Troessaert:2017jcm, Compere:2011ve,Compere:2023qoa}, 
\begin{equation}
\label{BSLineElement}
    d s^2=\sigma^2\,\dd\rho^2+2\Sigma_a\dd\rho\,\dd x^a+\rho^2 h_{ab}\dd x^a\dd x^b,
\end{equation}
where $\rho$ is a radial coordinate such that $\rho\to\infty$ corresponds to the asymptotic boundary, and $x^a$ are the (2+1)-dimensional Lorentzian coordinates.  
When compared to previous works, our approach differs by allowing for the following more generic polyhomogeneous expansion\footnote{In setting the leading term in $\sigma$ (which can be interpreted as a radial lapse function) to $1$, we implicitly set a scale for the spacetime as a whole. Our coordinate $\rho$ is then measured in terms of this scale. If we were to change this from $1$ to some constant $\sigma^{(0)}$, the leading on-shell metric would be a conformal rescaling of Minkowski space and $\sigma^{(0)}$ would appear in various expressions as a relative length scale.} 
\begin{subequations}
\label{BSExpansion}
\begin{align}
    \sigma&= 1  +\frac{1}{\rho}\pqty{\bar\sigma+\log\rho\, \tilde\sigma}+\frac{1}{\rho^2}\pqty{\bar\sigma^{(2)}+\log\rho\,\tilde\sigma^{(2)}+\log^2\rho\, \dbtilde\sigma^{(2)}}+o
    \pqty{\rho^{-2}},\\ \label{SigmaA}
    \Sigma_a&= \frac{1}{\rho}\pqty{\bar\Sigma_a+\log\rho\, \tilde\Sigma_a+\log^2\rho\dbtilde\Sigma_a }     +o \pqty{\rho^{-1}}
   ,
    \\
    h_{ab}&= h_{ab}^{(0)}+\frac{1}{\rho}\pqty{\bar h_{ab}+\log\rho\, \tilde h_{ab}}+\frac{1}{\rho^2}\pqty{ \bar h_{ab}^{(2)}+\log\rho\,\tilde h_{ab}^{(2)}+\log^2\rho\, \dbtilde h_{ab}^{(2)}}+ o
    \pqty{\rho^{-2}},
    \end{align}
\end{subequations}
which can be seen as an extension and generalization of the expansion taken in~\cite{Compere:2011ve} to now include further subsubleading contributions, $\log$ contributions, and a non-trivial expansion of the vector part of the metric (by allowing non-zero $\bar\Sigma_a$, $\tilde\Sigma_a$, $\tilde{h}_{ab},\tilde\sigma, \bar \sigma^{(2)}$, and $\dbtilde{h}^{(2)}_{ab}$). For the sake of readability we omit the superscript ${}^{(1)}$ from subleading components as they will appear most frequently in this work. Each function appearing in this expansion depends arbitrarily on $x^a$. In the following $D_a$ will denote the covariant derivative compatible with $h^{(0)}_{ab}$ and $h_0$ its determinant. 

This generalization is motivated as follows:  (i) logarithmic terms are typically generated dynamically in general relativity, even when starting from smooth initial data~\cite{Chrusciel1992Asymptotic, ChruscielMacCallumSingleton1994, ChruscielDelay2000, Beig:1982ifu}, (ii) in~\cite{Compere:2011ve}, it was shown that such terms are required at second order in the radial expansion to ensure the consistency of the equations of motion when parity conditions are not imposed, and (iii) allowing logarithmic terms at first order is necessary for compatibility with log-supertranslations, as already noted in~\cite{Compere:2011ve}. 

For convenience, we introduce the tensors
\begin{subequations}
\label{def tau}
\begin{alignat}{2}
\bar{\tau}_{ab}& :=\bar{h}_{ab} -\tilde{h}_{ab} - h^{(0)}_{ab}(\bar{h}-\tilde{h}+4\bar\sigma)\,, 
\qquad &&  \bar \tau=-2(\bar h-\tilde h+6\bar\sigma)\,, \\
\tilde{\tau}_{ab}& := \tilde{h}_{ab} - h^{(0)}_{ab} \tilde h \,,
    \qquad 
    &&\tilde \tau= -2\tilde h\,,
\end{alignat}
\end{subequations}
where $\bar{h}=h^{ab}_{(0)}\bar{h}_{ab}$, $\tilde{h}=h^{ab}_{(0)}\tilde{h}_{ab}$, $\bar\tau=h^{ab}_{(0)}\bar \tau_{{ab}}$, and $\tilde\tau=h^{ab}_{(0)}\tilde\tau_{{ab}}$. When restricting to CD boundary conditions, the field $k_{ab}$ of~\cite{Compere:2011ve} is related to $\bar\tau_{ab}$ by $\bar\tau_{ab}=k_{ab}-k h^{(0)}_{ab}$.

We consider field configurations falling off as \eqref{BSExpansion} supplemented by boundary conditions
\begin{equation}\label{eq:BC}
 h_{ab}^{(0)}\bceq \hyp_{ab} \,,\qquad   \bar\tau \bceq 0 \bceq\tilde\tau \,,\qquad \epsilon_{a}{}^{cd}D_c\tilde\tau_{db}\bceq 0 \,,
\end{equation}
where the tensor $\hyp_{ab}$ is the metric of the unit three-dimensional boundary hyperboloid $\mathcal H$,\footnote{In terms of flat spherical coordinates $(t,r,x^A)$, where the Minkowski metric takes the form $-\dd t^2+\dd r^2+r^2\dd\Omega^2$, the change of coordinates is $(t,r,x^{A})\mapsto(\rho \sinh\tau,\rho \cosh\tau,x^A) $ with inverse (for $\rho>0$) given by  $(\tau,\rho,x^{A}) \mapsto (\artanh(t/r),\sqrt{r^{2}-t^{2}},x^{A})$.}
\begin{equation}\label{boundaryhyp}
 \hyp_{ab}\dd x^a\dd x^b=   -\dd \tau^2+\cosh^2\tau\, \dd\Omega^2\,, \quad \dd\Omega^2=\dd\theta^2+\sin^2\theta\dd\varphi^2.
\end{equation}
The differential operator acting on $\tilde\tau_{ab}$ in the third condition  is the curl operator.

We leave the motivations for the choices \eqref{eq:BC} to subsequent sections. 
When computing charges \`a la Iyer–Wald~\cite{Iyer:1994ys}, it will be necessary to consider field configurations that do not satisfy the boundary conditions. For this reason, we will be careful about when these are imposed. We will use the terminology on-shell of the boundary conditions or off-shell of the boundary conditions for field configurations where \eqref{eq:BC} holds and do not hold respectively. 

With some foresight from the structure of the residual symmetries we will also impose 
\begin{equation}\label{tildesigma}
    \tilde\sigma \bceq 0
\end{equation}
for convenience (and without loss of generality) from \S\ref{sec:symplstructure} onward, which in turn allows us to henceforth set $\dbtilde{\Sigma}_{a} = 0$ (c.f. \S\ref{sec:residualsymm} and Appendix \ref{app:subsubleadresidualsymm} for a complete justification). Lastly, we will require that $\tilde \tau_{ab}$ does not have singular modes (see discussion around \eqref{BConbetasing}). 

As an example of spacetimes that admit a similar polyhomogehenous expansion at spatial infinity, let us consider Kerr-Taub-NUT spacetimes. In Beig-Schmidt coordinates~\cite{Virmani:2011gh}, they are characterized by the following non-vanishing function
\begin{equation}\label{KerrTN}
   \begin{split}
   \bar    \sigma &= G M \cosh 2 \tau \sech \tau\,,\qquad \bar{\tau}_{ab} = - 2 G N k_{(0)ab}
   \end{split}
\end{equation}
where $G$ is Newton's constant, $M$ is the black hole mass, and $N$ its NUT charge. Here, $k_{(0)ab}$ is a symmetric tracefree background tensor on the boundary hyperboloid, whose non-vanishing components are
\begin{equation}\label{Backgroundktensor}
    k_{(0)\phi \tau } = 2 \sech(\tau ) (c-\cos (\theta )),\quad k_{(0)\phi \theta }  =\frac{1}{2} \csc (\theta ) \sinh (\tau ) (-4 c \cos (\theta )+\cos (2 \theta )+3), 
\end{equation}
with $c=\pm 1$ in the north and south patch respectively. The information about the angular momentum is encoded in subsubleading terms such as $\bar{h}^{(2)}_{ab}$ and $\tilde{h}^{(2)}_{ab}$.

\subsection{Equations of motion}
The line element~\eqref{BSLineElement} can be viewed as a radial 3+1 decomposition, where $\sigma$ and $\Sigma_a$ play the roles of radial lapse and shift vector on the $\rho = \text{const}$ hypersurfaces (see appendix~\ref{app:exgeometry} for details). Using this split, the Einstein equations separate in three sets: a scalar, a vector, and a traceless tensor with respect to the hyperbolic metric. We analyze each set of equations of motion asymptotically, order by order in the radial coordinate.
\begin{itemize}
    \item 
The leading equations of motion are 
 \begin{equation}\label{leadingEq}\mathcal R_{ab}[h_{ab}^{(0)}]\eomeq 2h_{ab}^{(0)}, \end{equation}
where $\mathcal R_{ab}$ is the Ricci tensor of $h_{ab}^{(0)}$. Thus we see that the first boundary condition of \eqref{eq:BC} ensures that after a suitable rescaling, for $\rho \to \infty$, the line element~\eqref{BSLineElement} approaches  the line element of flat spacetime in hyperbolic coordinates
\begin{equation}
  \lim_{\rho\to\infty} \frac{d s^2}{\rho^2}=\hyp_{ab}\dd x^a \dd x^b.
\end{equation}

\item At subleading order, we find the following set of equations
    \begin{subequations} \label{eomsublead}
    \begin{alignat}{2}
        & (D^2+3)\bar \sigma\eomeq -\frac{1}{4}\tilde\tau,\qquad && (D^2+3)\tilde\sigma \eomeq 0, \\
        & D^a\bar \tau_{ab}\eomeq 0,\qquad && D^a\tilde \tau_{ab}\eomeq 0, \\
        \label{tracelesstensorEE}
        & (D^2-3)\bar \tau_{\tl{ab}}\eomeq -\frac{1}{2}\tldd{a}{b}(\bar\tau-4\tilde\sigma),\qquad && (D^2-3)\tilde \tau_{\tl{ab}}\eomeq -\frac{1}{2}\tldd{a}{b}\tilde\tau,
    \end{alignat}
    \end{subequations}
    where we used~\eqref{def tau} and the notation $\langle\,\rangle$ to denote the symmetric and traceless part of a rank two-tensor with a factor $1/2$, for instance $\bar \tau_{\tl{ab}}=\frac12(\bar \tau_{ab}+\bar \tau_{ba})-\frac13\hyp_{ab} \bar\tau$.

\item At subsubleading order, there are three (tensorial) equations for each set of equations. In particular, the vectorial one with no logarithms ($\sim 1/\rho^2$) ensures the existence of a symmetric two-tensor that is conserved under suitable boundary conditions
    \begin{equation}
        D_aT^{ab}\bceomeq 0\,,
    \end{equation}
where $\bceomeq$ indicates that the relation holds when both the field equations and the boundary conditions are satisfied.

We present this tensor in~\eqref{def:stresstensor} after showing that it arises as the variable canonically conjugate to the leading boundary metric $h^{(0)}_{ab}$. For this reason, we will refer to it as a stress tensor. 
\end{itemize}

\subsection{Weyl tensor}\label{app:weyl}
Given the boundary hyperboloid $\mathcal H$, we can decompose the Weyl tensor $C_{\mu\nu\rho\sigma}$ into its electric and magnetic components
\begin{equation}
E_{ab}:= C_{a\mu b \nu } n^\mu \, n^\nu\,,\qquad      B_{ab}:=\frac{1}{2}\epsilon_{a\mu\rho \sigma}C^{\rho\sigma}_{\quad  b\nu}n^\mu \, n^\nu,
\end{equation}
where $n^\mu$ is the normal vector to $\mathcal H$, and both $E_{ab}$ and $B_{ab}$ are symmetric and traceless.
Given the expansion~\eqref{BSExpansion} and the leading equation of motion~\eqref{leadingEq}, the polyhomogeneous expansion of these tensors starts at order $\rho^{-1}$
 \begin{equation}
 E_{ab} = \frac1\rho (\bar E_{ab}+\tilde E_{ab}\ln\rho)+o(\rho^{-1}) \,, \qquad   B_{ab}= \frac1\rho (\bar B_{ab}+\tilde B_{ab}\ln\rho)+o(\rho^{-1}).
 \end{equation}
On-shell of all equations of motion, one finds

\begin{subequations}\label{onshellWeyl}
\begin{align}
& \bar E_{ab}\eomeq -\tldd{a}{b}\bar\sigma+\frac{1}{2}\tilde\tau_{\tl{ab}},\qquad &&  \bar B_{ab}\eomeq-\frac12\epsilon_{a}{}^{cd}D_c\bar\tau_{db}-\frac14 \epsilon_{abc} D^c\bar\tau\,,\\
& \tilde E_{ab}\eomeq -\tldd{a}{b}\tilde\sigma\,, \qquad   && \tilde B_{ab}\eomeq -\frac12\epsilon_{a}{}^{cd}D_c\tilde\tau_{db}-\frac14 \epsilon_{abc} D^c\tilde\tau. 
\end{align}
\end{subequations}
Notice that the second term in $\bar B_{ab}$ (resp.~$\tilde B_{ab}$) compensates the fact that the curl of $\bar\tau_{ab}$ (resp.~$\tilde\tau_{ab}$) is not symmetric.

When imposing the boundary conditions \eqref{eq:BC} and \eqref{tildesigma}, we set both $\tilde\sigma$ and the curl of $\tilde\tau_{ab}$ to zero, which is equivalent to setting $\tilde E_{ab}=0=\tilde B_{ab}$. This implies that both the electric and magnetic components of the Weyl tensor decay as $1/\rho$, with logarithmic terms appearing only at the next order.

For the Kerr-Taub-NUT  black hole, $\bar E_{ab}$ encodes the mass parameter while $\bar B_{ab}$ carries the NUT charge. The angular momentum arises at the next order in the $\rho$ expansion.

\subsection{Residual symmetries}\label{sec:residualsymm}
We derive the residual symmetries that preserve the line element~\eqref{BSLineElement} together with the fall-offs~\eqref{BSExpansion}. We compute how the fields transform under these symmetries.
We then derive the additional constraints that the symmetry parameters need to satisfy to preserve the boundary conditions \eqref{eq:BC}. Lastly we derive the corresponding algebra.
A more detailed derivation and analysis of these vector fields and their algebra, up to $\mathcal{O}(\log^{2}\rho/\rho^{2})$, is included in Appendix~\ref{app:subsubleadresidualsymm}.

\subsubsection*{Residual symmetries and algebra}

The diffeomorphisms preserving the line element~\eqref{BSLineElement} with the prescribed fall-offs~\eqref{BSExpansion} are generated by asymptotic vector fields $\xi$ with components
\begin{subequations}\label{AKV}
\begin{align}
    \xi^\rho&=  \omega + \log\rho\,H+\frac1\rho \Big(D^{a}\omega D_{a}\bar{\sigma} -\omega \bar{\sigma}\Big)  +o\left(\rho^{-1}\right)\\
    \xi^a&=\mathcal Y^a + \frac1\rho \Big(D^a\omega +(1+\log\rho)D^{a}H \Big)+o\left(\rho^{-1}\right)
\end{align}
\end{subequations}
where $\omega$, $H$ and $\mathcal Y^a$ are arbitrary functions on the boundary hyperboloid.
The vector field $\mathcal Y^a$ generates diffeomorphisms of the boundary hyperboloid. The vector field parametrized by $\omega$ are the SPI-translations~\cite{Ashtekar:1978zz} that includes translations and supertranslations, while the one parametrized by $H$ corresponds to their logarithmic version and thus we will talk about log-translations and log-supertranslations.

The algebra of residual symmetries is given by the modified Lie bracket, which takes into account the dependence of the vector fields on the field configuration,
\begin{equation}
    \llbracket \xi_1 , \xi_2\rrbracket:=[\xi_1,\xi_2]-\delta_1\xi_2 +\delta_2\xi_1 \equiv \xi_{12},
\end{equation}
where we shorthand $\xi_i=\xi{(\omega_i,H_i,\mathcal Y_i)}$. Assuming that the parameters $\omega$, $H$, and $\mathcal Y^a$ are field-independent, the vector field $\xi_{12}$ generates another residual symmetry, with parameters
\begin{align}
\label{AKVBracketParams}
    \omega_{12}=
    \mathcal Y_1^a\partial_a\omega_2 -\mathcal Y_2^a\partial_a\omega_1 \,,\quad H_{12}=
    \mathcal Y_1^a\partial_aH_2 -\mathcal Y_2^a\partial_aH_1\,, \quad
    \mathcal Y_{12}^a= 
    \mathcal Y_1^b\partial_b 
    \mathcal Y_2^a- \mathcal Y_2^b\partial_b 
    \mathcal Y_1^a. 
\end{align}
Therefore, the algebra of residual symmetries is a semi-direct sum of the boundary hyperboloid diffeomorphisms, generated by $\mathcal Y^a$, with two abelian sectors, generated by $\omega$ and $H$
\begin{equation}\label{algebrabeforeBC}
    \mathfrak X(\mathcal H)\loplus\pqty{\mathbb R_\omega\oplus \mathbb R_H}^{\mathcal H},
\end{equation}
where $\mathfrak X(\mathcal H)$ is the space of vector fields on $\mathcal H$, $\pqty{\mathbb R_\omega\oplus \mathbb R_H}^{\mathcal H}$ denotes the set of functions from $\mathcal{H}$ to $\mathbb R_\omega\oplus \mathbb R_H$, and the semi-direct action is that of vector fields on functions. 

\subsubsection*{Field transformations}
It is useful to derive how the fields transform under the residual symmetries~\eqref{AKV}. At leading order, the metric of the boundary hyperboloid transforms as 
\begin{equation}\label{tranfmetric}
    \delta_\xi h_{ab}^{(0)}=\mathcal L_{\mathcal Y} h_{ab}^{(0)}.
\end{equation}
At subleading order, the fields transform as
\begin{subequations}\label{tranffield}
\begin{align}
   &\delta_{\xi}\bar\sigma= \mathcal{L}_{\mathcal{Y}}\bar\sigma+ H\,, && \delta_{\xi}\tilde\sigma=\mathcal{L}_{\mathcal{Y}}\tilde\sigma\,,\\
  &\delta_\xi \bar \tau_{\langle ab\rangle }= \mathcal{L}_{\mathcal{Y}}\bar \tau_{\langle ab\rangle }+2D_{\langle a} D_{b\rangle}\omega\,,&& \delta_\xi \bar \tau=\mathcal{L}_{\mathcal{Y}}\bar \tau-4(D^2+3)\omega\,, 
\\
 &\delta_\xi \tilde \tau_{\langle ab\rangle }= \mathcal{L}_{\mathcal{Y}}\tilde \tau_{\langle ab\rangle }+2D_{\langle a} D_{b\rangle} H\,, &&\delta_\xi \tilde \tau=\mathcal{L}_{\mathcal{Y}}\tilde \tau-4(D^2+3)H.
 \label{tranffieldtildetau}
\end{align}
\end{subequations}
Since $\tilde \sigma$ transforms homogeneously, we can consistently set it to zero,
\begin{equation}\label{eq:settildesigmatozero}
    \tilde \sigma=0,
\end{equation}
which as mentioned we will do from \S\ref{sec:symplstructure} onward. As detailed in Appendix \ref{app:subsubleadresidualsymm}, if $\tilde{\sigma} =0$ then $\dbtilde{\Sigma}_{a}$ is similarly transforming homogeneously, so we can also consistently set it to zero,
\begin{align}
    \dbtilde{\Sigma}_{a} = 0,
\end{align}
thereby justifying the statement following \eqref{tildesigma}. In fact, 
we refer the reader to  Appendix~\ref{app:subsubleadresidualsymm} to see how subsubleading terms and the vector parts of the metric transform under a residual symmetry transformation, as well as a derivation of the residual symmetry vector fields given above.

Note that the (on-shell) electric and magnetic parts of the Weyl tensor~\eqref{onshellWeyl} are invariant under $\omega$ and $H$ due to precise cancellation. They transform 
\begin{subequations}
\begin{align}
    &\delta_\xi \bar E_{ab}=\mathcal{L}_{\mathcal{Y}}\bar E_{ab},\quad \delta_\xi \tilde E_{ab}=\mathcal{L}_{\mathcal{Y}}\tilde E_{ab},\\
    &\delta_\xi \bar B_{ab}=\mathcal{L}_{\mathcal{Y}}\bar B_{ab},\quad \delta_\xi \tilde B_{ab}=\mathcal{L}_{\mathcal{Y}}\tilde B_{ab}.
\end{align}
\end{subequations}
This indicates that the effect of residual symmetries generated by $\omega$ and $H$ is not to alter the Weyl tensor, but rather to act on the non-radiative data of the phase space at spatial infinity.

\subsubsection*{Residual symmetries and algebra after enforcing the boundary conditions}
We now derive the additional constraints on the residual symmetries required to preserve the boundary conditions~\eqref{eq:BC}.

Preserving the boundary condition \eqref{eq:BC} on the leading metric $h^{(0)}_{ab}$ restricts the vector fields $\mathcal Y^a$ to be exact Killing vectors of the hyperboloid, i.e. the solutions to $D_a \mathcal Y_b+D_b \mathcal Y_a=0$. We know there are 6 of these. Using a spherical harmonic decomposition, they are given by three rotations ($\phi,\theta$ are standard coordinates on the unit two-sphere)
\begin{align}\label{rotation}
M_1&=\partial_\phi\,, \quad M_2=\cos\phi\,\partial_\theta-\sin\phi\cot\theta\,\partial_\phi \,,\quad M_3= -\sin\phi\,\partial_\theta -\cos\phi\,\cot\theta\,\partial_\phi
\end{align}
and three boosts
\begin{align}\label{boost}
N_1&=-2\sqrt{\frac{\pi}3}  B_{0}\,,\quad N_2=-i\sqrt{\frac{2\pi}3} (   B_{1}+  B_{-1}) \,,\quad N_3=\sqrt{\frac{2\pi}3} (B_{1}-  B_{-1}) .
\end{align}
where 
\begin{equation}
   B_{m}=Y_{1,m}\partial_\tau+\tanh\tau \mathcal D^A Y_{1,m}\partial_A,\qquad  m=-1,0,1
\end{equation}
where $\mathcal D_A$ is the covariant derivative on the unit two-sphere.
These 6 vector fields satisfy the Lorentz algebra $\mathfrak{so}(1,3)$, as one can check from their commutation relations
\begin{equation}
[M_i,M_j]=\epsilon_{ijk}M_k\,, \quad [M_i,N_j]=\epsilon_{ijk}N_k\,, \quad [N_i,N_j]=-\epsilon_{ijk}M_k\,.
\end{equation}

The next boundary conditions are to set subleading traces to zero
\begin{equation}\label{eq:BCtau}
         \tilde \tau=0=\bar \tau.
\end{equation}
In order to preserve these, the parameters $\omega$ and $H$ of the residual symmetries need to satisfy a second-order hyperbolic differential equation
\begin{equation}\label{eqforomegaH}
(D^2+3)\omega = 0, \qquad (D^2+3)H = 0.
\end{equation}
The solutions to this equation can be labeled according to their angular momentum and parity, as we review in Appendix~\ref{app:D2+3=0}. The parity we refer to is with respect to the involution $ \Upsilon_{\mathcal H}$ on the boundary hyperboloid, which combines time reversal with the antipodal mapping on the sphere
\begin{equation}
    \Upsilon_{\mathcal H} f(\tau,x^A)=f(-\tau,\pi-\theta,\phi+\pi)\,.
\end{equation}
We use the common terminology ``regular'', ``singular'', and ``super'' to classify functions $f$ which are solutions to $(D^2+3)f=0$. 
\begin{itemize}
    \item \textit{Regular functions} satisfy
    \begin{equation}\label{def:reg-odd-(log)translation}
    (D_aD_b+\hyp_{ab})f=0\,.
\end{equation}
On the hyperboloid, there are only four linearly independent functions that satisfy these equations. A convenient basis is given by
\begin{equation}\label{eq:regodd}
\zeta^{\text{odd}}_{0,0}=\sinh\tau\, Y_{0,0}(x^A) \,,\qquad \zeta^{\text{odd}}_{1,m}=\cosh\tau\,Y_{1,m}(x^A) \,,\quad m=-1,0,1,
\end{equation}
which contains only odd functions. We denote regular functions with a `reg' subscript: for instance,  $\omega_{\text{reg}}$ will parametrize the regular translations, and $H_{\text{reg}}$ the regular log-translations. For instance, the vector $\partial_t$ in the flat spherical coordinate system $(t,r,\theta,\phi)$ corresponds to $\omega=-\sinh\tau = -2\sqrt{\pi} \zeta^{\text{odd}}_{0,0}$ in hyperboloic coordinates.

\item \textit{Singular functions} are similarly given in terms of $\ell=0,1$ harmonics, but do not satisfy~\eqref{def:reg-odd-(log)translation}. A basis of singular functions is given by the four functions $ \zeta^{\text{even}}_{\ell=0,1,m}$
\begin{equation}\label{eq:evensing}
\begin{split}
\zeta^{\text{even}}_{0,0}&=\frac{\cosh (2 \tau )}{\cosh (\tau )} Y_{0,0}(x^A) \,,\quad\\ \zeta^{\text{even}}_{1,m}&=\left(2 \sinh (\tau )+\frac{\tanh (\tau )}{\cosh (\tau )}\right) Y_{1,m}(x^A)\,,\quad m=-1,0,1
\end{split}
\end{equation}
which are even. We denote singular functions as $f_{\text{sing}}$.

\item \textit{Super functions} correspond to higher harmonics ($\ell\geq2$) and can be either odd or even. These can be written in the following basis
\begin{align}
\zeta^{\text{even}}_{\ell\geq2,m}&=  \frac{1}{2\cosh{\tau}}\,P^2_\ell(\tanh{\tau})\,Y_{\ell, m}(x^A)\,,\\
\zeta^{\text{odd}}_{\ell\geq2,m}&=\frac{1}{2\cosh{\tau}}\,Q^2_\ell(\tanh{\tau})\,Y_{\ell, m}(x^A)\,,
\end{align}
where $P^2_\ell$, $Q^2_\ell$ are the associated Legendre polynomials and functions.
We denote super functions as $f_{\text{super}}$. A function written only in the basis spanned by $\zeta^{\text{even}}_{\ell\geq2,m}$ and $\zeta^{\text{odd}}_{\ell\geq2,m}$ will be denoted $f^{\text{even}}_{\text{super}}$ and $f^{\text{odd}}_{\text{super}}$ respectively. 
\end{itemize}
Therefore, when~\eqref{eqforomegaH} holds, the SPI-translations $\omega$ reduce to $\omega_{\text{reg}}$, $\omega_{\text{sing}}$, and $\omega_{\text{super}}$. Likewise, the otherwise arbitrary function $H$ can be divided in $H_{\text{reg}}$, $H_{\text{sing}}$, and $H_{\text{super}}$.
Importantly, the fact that we do not fix $\bar\sigma$ nor $\tilde \tau_{ab}$ allows both log-translations \textit{and} log-supertranslations to be part of the asymptotic symmetry algebra. In fact, if one sets $\tilde \tau_{ab}=0$ as in~\cite{Compere:2011ve}, the only allowed functions $H$ are the regular ones, because we have the property
\begin{equation}\label{relhyp}
    (D_a D_b +\hyp_{ab}) H=D_a D_b H-\frac{1}{3}\hyp_{ab} D^2 H +\frac{1}{3}\hyp_{ab} (D^2+3) H=\tldd{a}{b}H+\frac{1}{3}\hyp_{ab}(D^2+3)H
\end{equation}
which vanishes as a consequence of preserving the conditions $\tilde \tau_{\tl{ab}}=0$ and $\tilde \tau=0$, see equation ~\eqref{tranffieldtildetau}.
In the following, we will make extensive use of the relation ~\eqref{relhyp}, see also Appendix~\ref{app:idhyp} for more identities on the boundary hyperboloid. 

Upon the imposition of the boundary conditions \eqref{eq:BC}, the residual symmetry algebra thus schematically reduces to
\begin{equation}\label{algebraafterBC}
         \mathfrak{so}(1,3)\loplus\pqty{\bigoplus_{\ell} \,\mathbb R^\omega_{\ell; \text{odd}}\oplus \mathbb R^\omega_{\ell;\text{even}}\oplus \mathbb R^H_{\ell; \text{odd}}\oplus \mathbb R^H_{\ell;\text{even}}}^{S^2}\,.
\end{equation}
where $S^2$ is the $2-$sphere on the boundary hyperboloid. 
To describe the semi-direct action in~\eqref{algebraafterBC}, we write the commutation relations~\eqref{AKVBracketParams} in the basis of functions:  $\zeta^{\text{even}}_{\ell,m}$, and $\zeta^{\text{odd}}_{\ell,m}$.
Since Lorentz transformations (both rotations and boosts) do not mix parities, we can consider the two sectors separately. We use a condensed notation as the exact form of the structure constants is not particularly illuminating.\footnote{Note however than the structure constants in the odd and even sectors are identical except for $c_i$ and $\hat c_i$.} for the odd case we have
\begin{subequations}\label{oddLorentzalgebra}
\begin{align}
    &M^a_i \partial_a \zeta^{\text{odd}}_{\ell, m}= \sum_{m'}\pqty{r_i}_m^{m'} \zeta^{\text{odd}}_{\ell, m'}\,,\\
    &N^a_i \partial_a \zeta^{\text{odd}}_{\ell\le 1, m}= \sum_{\ell' m'}\pqty{b_i}_{\ell \, m}^{\ell ' m'} \zeta^{\text{odd}}_{\ell'\le 1, m'}\,,\\ \label{boostodd}
    &N^a_i \partial_a \zeta^{\text{odd}}_{\ell\ge 2, m}= \sum_{\ell' m'}\pqty{g_i}_{\ell \, m}^{\ell ' m'} \zeta^{\text{odd}}_{\ell'\ge 2, m'}+\delta_{\ell,2}\sum_{m'}\pqty{ c_i}^{m'}_m\zeta^{\text{odd}}_{1, m'}\,,
\end{align}
\end{subequations}
and for the even case,
\begin{subequations}\label{evenLorentzalgebra}
\begin{align}
    &M^a_i \partial_a \zeta^{\text{even}}_{\ell, m}= \sum_{m'}\pqty{r_i}_m^{m'} \zeta^{\text{even}}_{\ell, m'}\,,\\
    &N^a_i \partial_a \zeta^{\text{even}}_{\ell\le 1, m}= \sum_{\ell'm'}\pqty{b_i}_{\ell \, m}^{\ell ' m'} \zeta^{\text{even}}_{\ell'\le 1, m'}+\delta_{\ell,1}\sum_{m'}\pqty{\hat c_i}^{m'}_m\zeta^{\text{even}}_{2, m'}\,,\\
    &N^a_i \partial_a \zeta^{\text{even}}_{\ell\ge 2, m}= \sum_{\ell' m'}\pqty{g_i}_{\ell \, m}^{\ell ' m'} \zeta^{\text{even}}_{\ell'\ge 2, m'}\,.
\end{align}
\end{subequations}
We see that, under rotations $M_i$, modes with different orbital angular momenta $\ell$ do not mix. On the other hand, boosts \textit{do} transform an $\ell$-mode into a combination of $(\ell-1)$-modes and $(\ell+1)$-modes.
This implies that low-$\ell$ modes ($\ell=0,1$) can be boosted to high-$\ell$ modes ($\ell\ge 2$) and vice-versa. However, this type of mixing occurs only for odd super functions (which can be boosted down) and even singular functions (which can be boosted up). Odd regular functions, on the other hand, form a closed subalgebra, as expected from the existence of the Poincar\'{e} algebra.

Another interesting property of these commutation relations is the following. Let $f$ be a function containing regular, singular, and super modes of both parities. One may consistently set the singular and/or the odd-super modes to zero. Indeed, acting with a Lorentz transformation on the remaining non-vanishing modes does not generate singular and/or odd-super modes. This property does not hold for the other components. For instance, if one sets the regular modes to zero, such modes are generated by acting on the odd-super modes (see equation~\eqref{boostodd}).
This property is key in showing that we can exclude $H$ singular from the asymptotic symmetries, as is motivated by the next paragraph.   

In this decomposition in modes, the Kerr-Taub-NUT black hole \eqref{KerrTN} is described in terms of even functions. For instance,
\begin{equation}
\bar{\sigma}
= \hat{\sigma}^E_{0,0}\,\zeta^{\text{even}}_{0,0},
\qquad
\text{with} \qquad
\hat{\sigma}^E_{0,0} = \sqrt{4\pi}
G M .
\end{equation}
In particular, a $\ell=0$ singular log-translation would arbitrarily shift the mass of the black hole. A subset of such transformations are therefore unphysical for the following reason: 
One can always shift a configuration with these transformations to have negative mass, which we wish to exclude at least on the classical level on which we are working. 
This means that the only (field-independent) transformations among $H^E_{\text{sing}}$ which are allowed are those with
\begin{equation}
    \hat{H}^E_{0,0}\geq 0
\end{equation}
which at most form a monoid. 
This in particular implies that the $H$ parameters do not form a vector space (instead just a convex cone), and thus not a Lie algebra.\footnote{The set of these transformations is not covariant under Lorentz transformations, so must further restrict $H^E_{l<2}$. Converting the four modes in it to a 4-vector $H^\mu$, we need it to be in some one-sheeted hyperboloid in $\mathbb{R}^{1,3}$, schematically $H_\mu H^\mu \geq 0$, so that the space of valid parameters is the forward cone in the space of $H$'s. Thus the set is actually not even smooth at the identity, and one cannot attach a Lie algebra to the identity element $H=0$.}
While these transformations are technically permissible, we choose to exclude them in what follows as they do not generate an algebra.
In \S\ref{sec:charges}, we will provide consistent boundary conditions yielding $H_{\text{sing}}=0$.

\section{Finite symplectic structure at spatial infinity}\label{sec:symplstructure}

We set up a phase space on the class of metrics~\eqref{BSExpansion} and propose new boundary conditions such that: (i) the symplectic potential is finite, (ii) it vanishes when the boundary conditions are enforced, and (iii) the boundary conditions do not restrict the angular dependence of supertranslations and log-supertranslations. This will yield finite and conserved charges, which we derive in \S\ref{sec:charges}.

To address (i), the regularization of the symplectic potential, we adopt the prescription of~\cite{McNees:2023tus,McNees:2024iyu} which has the advantage of not relying on specific boundary conditions but only on the fall-offs of the fields near the boundary. Moreover, since this prescription guarantees that the symplectic current can always be rendered finite, it is not necessary to eliminate divergent terms by imposing parity conditions on the fields, unlike in~\cite{Fuentealba:2022xsz}.

Property (ii) ensures that the charges derived from this symplectic form are conserved, which is physically expected since no gravitational flux reaches spatial infinity. By exploiting the remaining finite ambiguities of the covariant phase space formalism, we are able to propose the boundary conditions~\eqref{eq:BC} without restricting the angular dependence of (log-)~supertranslations   (iii).

\subsection{Symplectic form regularisation}
The first variation of the Einstein-Hilbert Lagrangian produces a boundary term, the symplectic potential $\Theta$, that generically diverges when pulled back on an asymptotic boundary. In order to get a well-defined symplectic form, we must regularize the first variation of the action.

We use the prescription of~\cite{McNees:2023tus,McNees:2024iyu}, which has the advantage of producing a finite symplectic form current\footnote{We follow the convention that the juxtaposition $\delta A\delta B$ denotes the exterior product of field-space forms. We also use the convention $\omega=\delta \Theta$ is $\omega[\delta_1,\delta_2]=\delta_2\Theta[\delta_1]-\delta_1\Theta[\delta_2]$.} $\omega=\delta\Theta$ without relying on the specificity of boundary conditions or boundary Lagrangians. Henceforth for readability we will refer to $\omega$ simply as the ``symplectic form'', wherein the passing from a symplectic form current to a symplectic form via integration is understood contextually. We briefly review  the logic of this prescription. First, consider the symplectic form $\bar\omega^\mu$, derived from the unregularized potential $\bar\Theta^\mu$. This form is conserved on-shell
\begin{equation}
 \partial_\mu\bar \omega^\mu +\delta E^{\mu\nu}\delta g_{\mu\nu} =0.
\end{equation}
Second, consider the ambiguities in the definition of the symplectic potential: both $\Theta^\mu$ and $\Theta^\mu+\partial_\nu Y^{\mu\nu}$ are compatible with the same bulk equations of motion. The prescription uses such freedom to obtain a finite symplectic form. We introduce a cut-off $\Lambda$ at large $\rho$ and the limit $\Lambda \to \infty$ is taken at the end of the calculation. We define 
\begin{equation}\label{Yprescription}
    \Theta^\mu=\bar\Theta^\mu+\partial_\nu Y^{\mu\nu}
\,,\quad   Y^{\rho a}:=-\int^\Lambda d\rho\, \bar \Theta^a +o(\rho^{-1})\,,
\end{equation}
and the other components of $Y$ are irrelevant for the argument. 
On-shell, the new symplectic form satisfies
\begin{equation}
    \partial_\rho \omega^\rho=-\partial_a\omega^a=o(\Lambda^{-1})
\end{equation}
and fall-offs as
\begin{equation}
    \omega^a=o(\rho^{-1})\,,\quad 
    \omega^\rho=\bar\omega^\rho+\partial_a\int^\Lambda d\rho\, \delta\bar\Theta^a +o(\Lambda^{-1}).
\end{equation}
This implies that $\omega$ is finite (or, equivalently, that the symplectic potential is finite up to $\delta$-exact terms), thus the charges that we will extract from it in \S\ref{sec:charges} are finite as well.

As it stands, the prescription~\eqref{Yprescription} does not fix the finite part of $Y$. Rather than a shortcoming, we see it as a strength, as this procedure allows to shift the regularized symplectic potential by a corner (i.e. co-dimension two) term and to select boundary conditions yielding conserved charges without restricting the residual symmetries.  

We now turn to the construction of the phase space. Our strategy is to evaluate the asymptotic expansion of the symplectic potential and consider only the finite piece in the limit $\Lambda\to\infty$, as we just showed that the divergent pieces can be taken care of. Our starting point is the Einstein-Hilbert action plus the Gibbons-Hawking-York boundary term
\begin{equation}
    S=S_{\mathrm{EH}}+ S_{\mathrm{GHY}},
\end{equation}
where
\begin{equation}
    S_{\mathrm{EH}}=\frac{1}{2\kappa^2}\int_{\mathcal M}\dd[4]x\sqrt{-g}R \,, \qquad S_{\mathrm{GHY}}=\frac{\Lambda}{\kappa^2}\int_{\mathcal H_\Lambda}\dd[3]x\sqrt{-h}K
\end{equation}
and we use $\kappa^2=8\pi G$. Here, $K=h^{ab}K_{ab}$ is the trace of the extrinsic curvature of $\mathcal H_\Lambda$ and we accounted for the leading power of $\Lambda$ coming from the induced metric. We use this term because it is compatible with the Dirichlet boundary conditions we took on the asymptotic boundary metric~\eqref{eq:BC}.
The first variation of the total action on-shell is 
\begin{equation}\label{GHpot}
\delta S=-\frac{\Lambda}{2\kappa^2} \int_{\mathcal H_\Lambda}\dd[3]x\sqrt{-h} \bqty{\left( K^{ab} -
h^{ab} \,K\right)\delta h_{ab} + \mathcal D_a c^a},
\end{equation}
where where $\mathcal D$ is the covariant derivative compatible with $h_{ab}$ and $\mathcal D_a c^a$ is a well-known corner term, see for instance~\cite{McNees-Useful}.
Up to corner terms, the only relevant component of the symplectic potential is  
\begin{equation}
\label{divergentpotential}
    -\frac{\Lambda}{2\kappa^2}\sqrt{-h}\left( K^{ab} -h^{ab} \,K\right)\delta h_{ab}
\end{equation}
The finite term, $\Theta^\rho_0$, arises from those products that collectively provide a $1/\Lambda$ suppression, balancing the prefactor $\Lambda$.
Since variations in the logarithmic components of the metric come with a $\log\Lambda$ divergence that cannot be compensated, the only variations that can contribute to the finite potential are $\delta h_{ab}^{(0)}$, $\delta \bar h_{ab}$, and $\delta \bar h_{ab}^{(2)}$. Let us schematically write the result as
\begin{equation}
\label{FiniteSymPot}
    \Theta^\rho_0=\frac{\sqrt{-h_0}\,}{2\kappa^2}\pqty{ \bar{\mathcal T}^{(1)ab}\delta h^{(0)}_{ab}+\bar{\mathcal T}^{(0)ab}\delta \bar h_{ab}+\bar{\mathcal T}^{(-1)ab}\delta \bar h^{(2)}_{ab}}.
\end{equation}
where $h_0$ is the determinant of $h_{ab}^{(0)}$ and $\bar{\mathcal T}^{(n)}$ are some expansion coefficients.

As explained above, the symplectic potential~\eqref{FiniteSymPot} yields finite charges, but not necessarily conserved ones. Indeed, charges are conserved if $\omega_0=\delta\Theta_0$ vanishes for the chosen boundary conditions. Enforcing the leading boundary metric boundary condition of \eqref{eq:BC} sets the first term in~\eqref{FiniteSymPot} to zero, but not the remaining ones. In the following, we will propose additional boundary conditions, compatible with the action of both supertranslations and logarithmic supertranslations, which ensure conserved and finite charges.

\subsection{Choice of boundary conditions}
As no radiation reaches spatial infinity, there exist boundary conditions that close the system, in the sense that the symplectic form vanishes on-shell when such boundary conditions are enforced. The leading metric boundary condition $h^{(0)}_{ab} = \hyp_{ab}$ is not sufficient. We will therefore supplement it with additional conditions that lead to a vanishing symplectic potential compatible with log-supertranslations and supertranslations. 
An important point to keep in mind is that we are interested in the expression of the symplectic potential before enforcing the boundary conditions. (Corner) charges will be extracted from the symplectic form contraction as equation in~\eqref{def:chargede} and only their final expression will be evaluated on-shell of the boundary conditions.

For field configurations that admit the expansion~\eqref{BSExpansion}, the regularized symplectic potential reads\footnote{Note that $\delta \bar\tau_{\tl{ab}}$ is the variation of $\bar\tau_{\tl{ab}}$ and not the traceless part of $\delta\bar\tau_{ab}$.} (we take $\tilde \sigma=0$ from now on)
\begin{equation} \label{OriginalHypSympPot}
    \Theta^\rho_0=\frac{\sqrt{-h_0}\,}{2\kappa^2}\bqty{t^{ab}\delta h_{ab}^{(0)}+\frac{\tilde\tau+12\bar\sigma}{12}(\delta\bar\tau+\delta\tilde\tau+12\delta\bar\sigma)-\frac{3\bar\tau^{\tl{ab}}+4\tilde\tau^{\tl{ab}}}{2}(\delta\bar\tau_{\tl{ab}}+\delta\tilde\tau_{\tl{ab}})},
\end{equation}
where $t^{ab}$ comes from $\bar{\mathcal T}^{(1)}$ but includes contributions from both $\bar{\mathcal T}^{(0)}$ and $\bar{\mathcal T}^{(-1)}$ in~\eqref{FiniteSymPot}.\footnote{For example, one notices that $\bar{\mathcal T}^{(-1)ab}$ only depends on $h^{(0)}_{ab}$, so that the last term of~\eqref{FiniteSymPot} can be straightforwardly integrated by parts by adding a $\delta$-exact term.} 

The first boundary condition enforces that $h_{ab}^{(0)}$ is the unit hyperboloid metric and thus sets the first term in~\eqref{OriginalHypSympPot} to zero. The second term, which is symplectically equivalent to $(\tilde\tau+12\bar\sigma)\delta\bar\tau$, can be set to zero by the condition $\bar\tau\bceq 0$ (as in~\cite{Compere:2011ve}): this restricts the symmetry parameter $\omega$ to satisfy
\begin{equation}
    (D^2+3)\omega=0.
\end{equation}
However, imposing $\bar{\tau}_{\tl{ab}} = 0$ (resp. $\tilde{\tau}_{\tl{ab}} = 0$) to take care of the last term would restrict the residual symmetries to regular translations (resp. regular log-translations), i.e. the solutions to~\eqref{def:reg-odd-(log)translation}, see the argument~\eqref{relhyp}. To avoid this restriction, we introduce a decomposition of the dynamical variables using the equations of motion, and use the ambiguities of the covariant phase space formalism to obtain a new symplectic potential that vanishes under milder conditions.
 
Using the properties of tensors on the hyperboloid, we decompose $\bar\tau_{\langle ab\rangle}$ and $\tilde\tau_{\langle ab\rangle}$ as 
\begin{equation}\label{AuxiliaryTensorDecomposition}
\begin{split}
\bar\tau_{\langle ab\rangle}&= \bar\wp_{ab}+2D_{\langle a}D_{b \rangle }\bar \beta \,,\qquad 
  \tilde\tau_{\tl{ab}}=\tilde \wp_{ab}+2D_{\langle a}D_{b \rangle }\tilde\beta \,,
    \end{split}
\end{equation}
where $\bar \wp_{ab},\tilde \wp_{ab}$ are symmetric, traceless, and divergence-free tensors that linearly depend on a function, but cannot be expressed as $\tldd{a}{b}\Phi$.
The functions $\bar\beta$ and $\tilde\beta$ satisfy 
\begin{equation}
    (D^2+3)\bar \beta+\frac1{4}\bar\tau\eomeq 0\,,\qquad 
    (D^2+3)\tilde \beta+\frac1{4}\tilde\tau\eomeq 0\,.
\end{equation}
We refer the reader to Appendix~\ref{app:symmtensor} for details.

We use this decomposition in the symplectic potential~\eqref{OriginalHypSympPot}. Adding a $\delta$-exact term and the following corner ambiguity (see \S\ref{Notation} for notation)
\begin{equation}
\boxed{\begin{aligned}
    Y^{a\rho}=\frac{\sqrt{-h_0}\,}{2\kappa^2}\biggl[&\tilde\tau^{ab}\overset{\leftrightarrow}{D_b} \delta\bar\beta-\bar\wp^{ab}\overset{\leftrightarrow}{D_b}\delta\tilde\beta+z^{ab}\overset{\leftrightarrow}{D_b}\delta h^{(0)}+z\overset{\leftrightarrow}{D_b}\delta h_{(0)}^{ab}-z\overset{\leftrightarrow}{D^a}\delta h^{(0)}\\
    &+z^{bc}\overset{\leftrightarrow}{D^a}\delta h^{(0)}_{bc}-2z_{bc}D^b\delta h_{(0)}^{ac}+2D^c z^{ab}\delta h^{(0)}_{bc}-\gamma^{abc}\delta h^{(0)}_{bc}\biggr],
\end{aligned}}
\end{equation}
with
\begin{align}
    z^{ab}&=\frac{1}{2}(D^a D^b-h_{(0)}^{ab})(\bar\beta+\tilde\beta-\bar\sigma)\,,\\
    \gamma^{abc}&=\frac{1}{2}\tilde\tau^{ab}D^c\bar\beta-\frac{1}{2}\bar\wp^{ab}D^c\tilde\beta+\frac{1}{2}\tilde\tau^{ca}D^b\bar\beta-\frac{1}{2}\bar\wp^{ca}D^b\tilde\beta-\frac{1}{2}\tilde\tau^{bc}D^a\bar\beta+\frac{1}{2}\bar\wp^{bc}D^a\tilde\beta\,,
\end{align}
we obtain the new symplectic potential\footnote{More precisely, we have shown that, on-shell of the equations of motion \textit{and} the boundary conditions, $\Theta^\rho_0= \delta \ell +D_a \vartheta^a$, with $\vartheta^a=-Y^{a\rho}$.
Therefore, we subtract the right hand side $\Theta^\rho \equiv \Theta^\rho_0 - \delta \ell - D_a \vartheta^a$ to get a symplectic potential that only contain variations of the boundary conditions. Explicitly, we used
\begin{equation}
    \ell=\frac{\sqrt{-h_0}\,}{2\kappa^2}\pqty{6\bar\sigma^2+(\bar\sigma-\bar\beta)\tilde\tau+\frac{\tilde\tau^2}{24}-\tilde\tau^{\tl{ab}}\tilde\tau_{\tl{ab}}-\frac{3}{2}\bar\tau^{\tl{ab}}\tilde\tau_{\tl{ab}}-\frac{3}{4}\bar\tau^{\tl{ab}}\bar\tau_{\tl{ab}}-\bar\beta\tilde\tau-\frac{1}{2}\tilde\tau^{\tl{ab}}\bar\wp_{ab}}
\end{equation}}
\begin{equation}\label{finiteconserved phase space}
  \boxed{   \Theta^\rho\eomeq\frac{\sqrt{-h_0}\,}{2\kappa^2}\pqty{T^{ab}\delta h_{ab}^{(0)}+\bar\sigma\delta\bar\tau+\bar\beta\delta\tilde\tau+\frac{1}{2}\bar \wp^{ab}\delta\tilde\wp_{ab}}}
\end{equation}
where we went on-shell of the leading and subleading equations of motion, and collected every term that multiplies $\delta h^{(0)}_{ab}$ into the stress-energy tensor $T^{ab}$; we will provide its expression shortly. The associated symplectic form is
\begin{equation}\label{symplcurrent}
 \omega^\rho=\delta\Theta^\rho\,.
\end{equation}

The expression~\eqref{finiteconserved phase space} makes it clear that the remaining boundary conditions can be
\begin{equation}\label{BContildetau}
    \tilde\tau\bceq 0\,, \qquad \tilde\wp_{ab}\bceq 0\,.
\end{equation}
These are equivalent to $\tilde \tau_{ab}$ having zero curl, as reported in \eqref{eq:BC}. Indeed, \eqref{eq:newBC} $\Rightarrow$ \eqref{eq:BC} as \begin{equation}
\epsilon_{a}\,^{cd}D_c{\tilde\tau}_{bd}\eomeq\epsilon_{a}\,^{cd}D_c{\tilde \wp_{ab}
   } -\frac12 \epsilon_{abc}D^c\tilde\tau\bceq0\,,
\end{equation}
and \eqref{eq:newBC} $\Leftarrow$ \eqref{eq:BC} 
by using Lemma 1 of appendix A of~\cite{Compere:2011db}.
The first condition in \eqref{BContildetau} yields the constraint
\begin{equation}
    (D^2+3)H=0 \,,
\end{equation} while the condition $\tilde\wp_{ab}=0$ does not restrict further the residual symmetries as $\tilde \wp_{ab}$ transforms homogeneously under~\eqref{AKV}, i.e.~ 
$ \delta_{\xi}\tilde\wp_{ab}=  \mathcal{L}_{\mathcal{Y}}\tilde\wp_{ab}$.
On the other hand, the scalar potential of $\tilde\tau_{ab}$ does not transform homogeneously,
$  \delta_{\xi}\tilde\beta=  \mathcal{L}_{\mathcal{Y}}\tilde\beta+H$. 
On-shell of the boundary conditions, $\tilde\beta$ is in the kernel of  $(D^2+3)$ and can be thereby decomposed in regular, singular, and super-modes. In particular, $\delta_{H}\tilde \beta_{\text{sing}}= H_{\text{sing}}$. We can therefore impose the condition 
\begin{equation}\label{BConbetasing}
\delta\tilde \beta_{\text{sing}}=0   
\end{equation}
to set $H_{\text{sing}}$ to zero, and for simplicity choose $\tilde\beta_{\text{sing}}=0$. This is the condition mentioned at the end of \S\ref{sec:residualsymm} to prevent an arbitrary shift of the black hole mass. 

We have then justified our choice of boundary conditions \eqref{eq:BC}, which we rewrite here:
\begin{equation}\label{eq:newBC}
\boxed{h_{ab}^{(0)}\bceq \hyp_{ab} \,,\qquad   \bar\tau\bceq 0 \bceq \tilde\tau \,,\qquad \tilde \wp_{ab}\bceq 0 }
\end{equation}
These conditions provide a vanishing symplectic form (and potential) $\delta\Theta^\rho  \bceq 0$.
When they hold, the subleading solution space is given by
\begin{equation}\label{tauabafterBC}
\begin{split}
\bar\sigma\,,\quad     \bar\tau_{\langle ab\rangle}&=\bar \wp_{ab}+ 2D_{\langle a}D_{b \rangle }\bar \beta\,,\quad 
 \tilde\tau_{\tl{ab}}= 2D_{\langle a}D_{b \rangle }\tilde \beta\,, \quad  \tilde\beta_{\text{sing}}=0\,,
    \end{split}
\end{equation}
where the three functions $  \bar\sigma\,, \bar\beta\,, \tilde \beta$
are in the kernel of $(D^2+3)$.  In particular, the leading part of the Weyl tensors reads 
\begin{equation}
\bar E_{ab}\bceq -\tldd{a}{b}\bar\sigma+\tldd{a}{b}\tilde \beta_{\text{super}} ,\qquad   \bar B_{ab}\bceq-\frac12\epsilon_{a}{}^{cd}D_c\bar\wp_{db}=-\tldd{a}{b} \sigma_N\,.
\end{equation}
where $\sigma_N$ is a solution of $(D^2+3)\sigma_N=0$ and is proportional to the NUT charge for the metric \eqref{KerrTN}. 

In the solution space that respects the boundary conditions, the stress-energy tensor $T^{ab}$, which involves subsubleading quantities, reads
\begin{equation}\label{def:stresstensor}
\begin{split}
    T^{ab}\bceq &\,\bar h_{(2)}^{ab}-\frac{1}{2} \tilde h_{(2)}^{ab}-\frac{1}{2} \bar\tau^{(a}_c \tilde\tau^{b)c}-\frac{1}{2} \bar\tau^{ac} \bar\tau^b_c-\frac{3}{2} \bar\beta \tilde\tau^{ab}+\frac{1}{2} D_c\bar\beta D^{(a}\tilde\tau^{b)c}+\bar\sigma \tilde\tau^{ab}+\\
    &-\frac{1}{2} D^c\tilde\beta D_c\bar\wp^{ab}+\frac{3}{2} \tilde\beta \bar\wp^{ab}+D^{(a}\bar\Sigma^{b)}-\hyp^{ab}D_c\bar\Sigma^c+\\
    &- \hyp^{ab} \pqty{\bar h^{(2)}-\frac{1}{2} \tilde h^{(2)}-\frac{3}{8}\bar\tau_{cd} \bar\tau^{cd}-4 \bar\sigma^2+2 \bar\sigma^{(2)}-\frac{1}{4}\tilde\tau^{cd} \bar\tau_{cd}-\frac{1}{4}  \tilde\tau^{cd} \bar\wp_{cd}} \,.
\end{split}
\end{equation}
From now on, we will remove the angular brackets from the indices of $\bar\tau_{ab}$ and $\tilde\tau_{ab}$ in expressions that are evaluated on-shell of the boundary conditions, as the trace of those tensors would vanish nonetheless.
The residual symmetries are the Lorentz transformations and the functions $\omega$ and $H$ satisfying
\begin{equation}\label{eq:constraintomegaH}
   (D^2 + 3)\omega =0,\qquad (D^2 + 3)H=0\,,\qquad H_{\text{sing}}=0\,.
\end{equation}

In summary, we have obtained a new finite symplectic potential~\eqref{finiteconserved phase space} for the system and proposed new boundary conditions~\eqref{eq:newBC} that close the system. These conditions amount to requiring the electric and magnetic Weyl tensors to decay as $1/\rho$ (and not as $\log(\rho)/\rho$), along with Dirichlet boundary conditions on the traces of the subleading tensors $\bar \tau_{ab}$ and $\tilde \tau_{ab}$. 
The traceless part of these tensors is expressed in terms of $\bar \beta$ and $\tilde \beta$, respectively, the first of which we identify with the Goldstone for (singular and super) translations. 
We also set the singular part of $\tilde \beta$ to zero to avoid negative mass. Note that our boundary conditions include Kerr-Taub-NUT spacetimes. With these ingredients in place, we are now equipped to compute the charges associated with the symplectic structure.

\section{Charges and asymptotic symmetry algebra}\label{sec:charges}

Associated to a symplectic potential describing a closed system, we can construct canonical charges associated with residual symmetries, which will be conserved when the boundary conditions hold. Deriving the charges will also allow us to determine which transformations among the residual symmetries are genuine transformations of the phase space, meaning that they are associated with a non-trivial canonical generator as opposed to pure gauge. We will refer to the former as large gauge transformations, whereas the latter (those with vanishing charge) are trivial and their action will be quotiented out in the construction of the asymptotic symmetry algebra.

We use the Iyer-Wald approach~\cite{Iyer:1994ys} to construct the charges. Given a residual symmetry $\delta_\xi$, we consider the contraction of the symplectic form with it, namely $\omega^\mu[\delta_\xi,\delta]$, \textit{before} imposing the boundary conditions. Such a contraction is a total derivative on-shell of the equations of motion, which means that we can write 
\begin{equation}\label{omegadk}
    \omega^\mu[\delta_\xi ,\delta]\eomeq\partial_\nu k_\xi^{\mu\nu}\,.
\end{equation}
If the transformation is integrable in the sense that $k_\xi=\delta q_\xi$, we define the canonical charge associated with the symmetry as
\begin{equation}\label{def:chargede}
    Q_\xi= \int_{C} q_\xi\,,
\end{equation}
where $C$ is a cut at the asymptotic boundary. We take the cuts to be at $\tau=\text{const}$, namely $C$ is a two-sphere on the boundary hyperboloid. 
Since our boundary conditions make the symplectic potential vanish, $q_\xi$ corresponds to the Noether charge density (assuming that $\delta\xi =0$).
Charges obtained in this way are automatically conserved on-shell of the boundary conditions. However, those boundary conditions and the ensuing restrictions on the residual symmetry parameters need to be imposed only \textit{after} the corner charge is extracted as previously emphasized.

In addition to the transformations~\eqref{tranfmetric} and~\eqref{tranffield}, the subleading degrees of freedom introduced by the decomposition of $\bar\tau_{ab}$ and $\tilde\tau_{ab}$ transform as
\begin{subequations}
\begin{align}
   &\delta_{\xi}\bar\beta=  \mathcal{L}_{\mathcal{Y}}\bar\beta +\omega
   \,,\qquad &&
   \delta_{\xi}\tilde\beta=  \mathcal{L}_{\mathcal{Y}}\tilde\beta  \,,\\
   &\delta_{\xi}\bar\wp_{ab}=  \mathcal{L}_{\mathcal{Y}}\bar\wp_{ab}\,,\qquad && \delta_{\xi}\tilde\wp_{ab}=  \mathcal{L}_{\mathcal{Y}}\tilde\wp_{ab},.
\end{align}
\end{subequations}
The stress-energy tensor --- off-shell of the boundary conditions but on-shell of the equations of motion --- transforms as
\begin{equation}\label{SETtransformation}
\begin{split}
    \delta_\xi T^{ab}&\eomeq \mathcal L_{\mathcal Y} T^{ab}+\frac{1}{2}(\omega \tilde\tau +H \bar\tau)h_{(0)}^{ab}-2D_c(\bar\sigma D^c \omega+\bar\beta D^c H) h_{(0)}^{ab}\\
    &\quad+4D^{(a} \omega D^{b)} \bar\sigma+4D^{(a} H D^{b)} \bar\beta+W^{ab}_1(\omega)+W^{ab}_2(H)\,,
\end{split}
\end{equation}
where $W^{ab}_i$, $i=1,2$, are tensors linear in their argument such that $W^{ab}_i\delta h^{(0)}_{ab}\eomeq D_c V_i^c$ with $V_i^c\bceq 0$, so that they do not contribute to any charge. The transformation of the subsubleading components which enters the stress-energy tensor are reported in Appendix~\ref{app:subsubleadresidualsymm}.

In the remainder of this section we will compute the Iyer-Wald charges first associated to $\omega$ and $H$ before tackling the Lorentz charge. This will allow us to compute the asymptotic symmetry algebra and show that there is a non-vanishing central extension between $\omega$ and $H$. This can be used to redefine the Lorentz generators such as to make the Poincar\'{e} subalgebra an ideal of the asymptotic symmetry algebra. In the final subsection we restrict our boundary conditions and compare with the literature on spatial infinity.

\subsection{Charges}
\subsubsection*{(Super)translation and log-(super)translation charges}

Since translations and log-translations have a very similar action on phase space, we compute the corresponding charges simultaneously by contracting the symplectic form~\eqref{symplcurrent} by $\omega$ and $H$. As those transformations do not change $h^{(0)}_{ab}$, $\bar\wp_{ab}$, and $\tilde\wp_{ab}$, we obtain
\begin{equation}\label{sutracontraction}
\begin{split}
    \omega^\rho[\delta_{(\omega,H)},\delta]\eomeq -\frac{1}{2\kappa^2}\biggl(&\sqrt{-h_0}\,\delta_{(\omega,H)} T^{ab}\delta h_{ab}^{(0)}+ \sqrt{-h_0}\,(H \delta\bar\tau+\omega\delta\tilde\tau)+\\&+4\delta (\sqrt{-h_0}\,\bar\sigma)(D^2+3)\omega+4\delta (\sqrt{-h_0}\,\bar\beta)(D^2+3)H\biggr)\,.
\end{split}
\end{equation}
Integrating by parts the second line, we obtain the corner term
\begin{equation}\label{sutraaspect}
\begin{split}
    \partial_a k_\xi^{\rho a}&=-\frac{2\sqrt{-h_0}\,}{\kappa^2}D_a\pqty{\pqty{\frac{1}{2}\bar\sigma\delta h^{(0)}+\delta \bar\sigma}\overset{\leftrightarrow}{D^a}\omega+\pqty{\frac{1}{2}\bar\beta\delta h^{(0)}+\delta \bar\beta}\overset{\leftrightarrow}{D^a}H}\\ 
    &\bceq- \frac{2\sqrt{-\hyp}}{\kappa^2}D_a\pqty{\delta \bar\sigma\overset{\leftrightarrow}{D^a}\omega+\delta \bar\beta\overset{\leftrightarrow}{D^a}H}\,,
    \end{split}
\end{equation}
where $\delta h^{(0)}= 2\delta \ln\sqrt{-h_0}=h^{ab}_{(0)}\delta h^{(0)}_{ab}\,$, plus another term in the bulk of the hyperboloid which, evaluated for linear perturbations around the equations of motion, gives\footnote{
We use that two functions $x,y$ satisfy
\begin{equation}
    x[D^2,\delta]y+\frac{1}{2}xy D^2\delta h^{(0)}+x D^a y D_a\delta h^{(0)}=\frac{1}{4}C^{ab}\delta h^{(0)}_{ab}+\frac{1}{4}D_c U^c.
\end{equation}
with
\begin{equation}
    C^{ab}=2 h^{ab}_{(0)} D_c(y D^c x)-4 D^{(a} x D^{b)} y,\quad U^c= 2y x D^c\delta h^{(0)}-2\delta h^{(0)} yD^c x-4x h^{ca}_{(0)} \delta h_{ab}^{(0)} D^by\,.
\end{equation}
}
\begin{equation}
\begin{split}
    &-\frac{2\sqrt{-h_0}\,}{\kappa^2}\pqty{\omega(D^2+3)\pqty{\delta \bar\sigma+\frac{1}{2}\delta h^{(0)}\bar\sigma}+H(D^2+3)\pqty{\delta \bar\beta+\frac{1}{2}\delta h^{(0)}\bar\beta}}\\
    &\eomeq-\frac{2\sqrt{-h_0}\,}{\kappa^2}\biggl[-\frac{1}{4}\omega \delta\tilde\tau+\omega [D^2,\delta]\bar\sigma+\frac{1}{2}\omega\bar\sigma D^2\delta h^{(0)}-\frac{1}{8}\omega \tilde\tau \delta h^{(0)}+\omega D^a\bar\sigma D_a \delta h^{(0)}+\\
    &\qquad\qquad\quad -\frac{1}{4}H \delta\bar\tau+H [D^2,\delta]\bar\beta+\frac{1}{2}H\bar\beta D^2\delta h^{(0)}-\frac{1}{8}H \bar\tau \delta h^{(0)}+H D^a\bar\beta D_a \delta h^{(0)}\biggr]\\ 
    &=\frac{\sqrt{-h_0}\,}{2\kappa^2}\pqty{\omega\delta \tilde\tau+H\delta \bar\tau+\frac{1}{2}\omega\tilde\tau\delta h^{(0)}+\frac{1}{2}H\bar\tau\delta h^{(0)}-C^{ab}\delta h^{(0)}_{ab}-D_c U^c}.
\end{split}
\end{equation}
Considering the last equation, the terms in $\delta\bar\tau$ and $\delta\tilde\tau$ cancel with the analogous ones in the first line of~\eqref{sutracontraction}, the terms in $\delta h^{(0)}_{ab}$ cancel with the stress-energy tensor transformation~\eqref{SETtransformation}, and the corner term coming from $U^c$ vanishes on-shell of the boundary condition $\delta h^{(0)}_{ab}=0$.

The corner term~\eqref{sutraaspect} is integrable on field space assuming $\delta \omega=0=\delta H$. We then integrate it on a sphere $S^2\subset\mathcal H$ at fixed $\tau$ to get canonical (super)translation and log-(super)translation charges 
\begin{equation}
\label{chargesG}
\boxed{Q_{(\omega,H)}=\frac{2}{\kappa^2}\oint_{S^2}  \cosh^2\tau\,s_a \bqty{\omega D^a\bar\sigma-\bar\sigma D^a \omega+H D^a\bar\beta-\bar\beta D^a H}}
\end{equation}
where $\oint_{S^2}\equiv \oint_{S^2} \dd^2\Omega$, and $s_a$ is the unit normal to the sphere in $\mathcal H$, which is a timelike normal. These results match with~\cite{Compere:2011ve} upon restricting $H$ to regular log-translations.  

Since the parameters of allowed (log-)translations $\omega$, $H$, and the conjugated variables $\bar\sigma$, $\bar \beta$ are in the kernel of $(D^2+3)$, we can use the terminology introduced in \S\ref{sec:residualsymm} and 
decompose each function in spherical harmonics as in Appendix~\ref{app:D2+3=0} (remembering that $H_{\text{sing}}=0$)
\begin{alignat}{2}
    &\bar\sigma
    =\sum_{\ell, m} (\hat \sigma^E_{\ell, m}\zeta^{\text{even}}_{\ell,m}+\hat \sigma^O_{\ell, m}\zeta^{\text{odd}}_{\ell,m})\,,\qquad &&\bar\beta
        =\sum_{\ell, m} (\hat \beta^E_{\ell, m}\zeta^{\text{even}}_{\ell,m}+\hat \beta^O_{\ell, m}\zeta^{\text{odd}}_{\ell,m})\,,\\
    &\omega
    =\sum_{\ell, m} (\hat \omega^E_{\ell, m}\zeta^{\text{even}}_{\ell,m}+\hat \omega^O_{\ell, m}\zeta^{\text{odd}}_{\ell,m})\,,\qquad && H
    =\sum_{\ell\geq2, m} \hat H^E_{\ell, m}\zeta^{\text{even}}_{\ell,m}+\sum_{\ell, m} \hat H^O_{\ell, m}\zeta^{\text{odd}}_{\ell,m}\,.
\end{alignat}
The charges read
\begin{align}
\label{ChargeModeExpansionLargegauge}
Q_{(\omega, H)}&=Q_\omega+Q_H\,,\\
    Q_{\omega}&=-\frac{2}{\kappa^2}\sum_{\ell, m}(\hat \sigma^E_{\ell, m}\hat \omega^O_{\ell,-m}-\hat \sigma^O_{\ell, m}\hat \omega^E_{\ell,-m})\mathcal C_\ell\equiv -\frac{2}{\kappa^2}\langle \bar\sigma,\omega\rangle\,,\\
    Q_{H}&=-\frac{2}{\kappa^2}\sum_{\ell, m}\hat \beta^E_{\ell, m}\hat H^O_{\ell,-m}\mathcal C_\ell-\frac{2}{\kappa^2}\sum_{\ell\geq2, m}\hat \beta^O_{\ell, m}\hat H^E_{\ell,-m}\mathcal C_\ell \equiv-\frac{2}{\kappa^2}\langle \bar\beta,H\rangle\,,
\end{align}
where $\mathcal C_\ell$ is a constant given in \eqref{ClDefinition} and we also used the bracket notation defined in \eqref{KernelInnerProduct}. In addition to the usual translations, which are generated by $\hat\omega^O_{\ell=0,1}$, the large gauge transformations are generated by the supertranslations and log-supertranslations of both parities, as well as a nonzero charge for even and odd $\ell=0,1$ translations and odd $\ell=0,1$ logarithmic translations.

It is interesting to stress that we have found a charge associated to the regular log-translations which are usually treated as pure gauge. 
There may appear to be some tension between our results and those of FHT~\cite{Fuentealba:2022xsz}, who report having proper gauge transformations instead. However, a comparison at this level is too naive and requires careful clarification, which we provide in Appendix~\ref{app:FHTDetails}. Furthermore, we will shortly see that imposition of supplementary parity conditions compatible with a smooth null infinity will render them pure gauge.  

Let's focus on the regular translations, which are generated by $\hat\omega^O_{\ell=0,1}$. The charge can be written in terms of the electric part of the Weyl tensor. Using the property 
\begin{equation}
-D_{\langle a}D_{b\rangle}\bar \sigma \,D^a\omega+2D^a(D_{[a}\omega D_{b]} \bar\sigma) =2\bar \sigma \overset{\leftrightarrow}{D_b} \omega
\end{equation}
and the fact that $D^a(D_{[a}\omega D_{b]} \bar\sigma)$ is zero when integrating over the sphere by
\begin{equation}\label{propertyoninthyp}
    \oint_{S^2}\cosh\tau^2s_a D_bM^{[ab]}=0
\end{equation}
for any antisymmetric tensor $M^{[ab]}$, we have 
\begin{equation}\label{chargeomegareg}
    Q_{\omega_{\text{reg}}}= 
-\frac{2}{\kappa^2}\oint_{S^2} \cosh^2\tau \,s_a \, \bar \sigma \overset{\leftrightarrow}{D_b} \omega_{\text{reg}}= -\frac{1}{\kappa^2}\oint_{S^2} \cosh^2\tau \,s_a \bar E^{ab}\,D_b\omega_{\text{reg}}\,.
\end{equation}
Similarly we have for $H_{\text{reg}}$, 
\begin{equation}\label{chargeHreg}
Q_{H_{\text{reg}}}= 
-\frac{2}{\kappa^2}\oint_{S^2} \cosh^2\tau \,s_a \, \bar \beta \overset{\leftrightarrow}{D_b} H_{\text{reg}}= \frac{1}{\kappa^2}\oint_{S^2} \cosh^2\tau \,s_a \,D_{\langle a}D_{b\rangle}\bar\beta\,D_b H_{\text{reg}}\,.
\end{equation}
Note that this is not the magnetic part of the Weyl tensor, for that we would need the curl of that expression. 

For Kerr-Taub-NUT (KTN) spacetimes \eqref{KerrTN}, the non-vanishing charges \eqref{chargesG} are associated to the regular time-translation $\omega=-\sinh\tau$, 
\begin{equation}
 Q_{\omega}^{\text{KTN}}=M,
\end{equation}
whence we recover that the regular time-translation gives the ADM mass $M$. 

\subsubsection*{Lorentz charges}

Let us now compute the charge associated with Lorentz transformations. To do so, we contract the symplectic form with $\xi=\mathcal Y^a \partial_a=\mathcal Y$. Assuming $\delta\mathcal Y^a=0$ and recalling that any field in our phase space transforms under Lorentz transformations as $\delta_\mathcal Y (\bullet)=\mathcal L_{\mathcal Y} (\bullet)$, the contribution from each term in the symplectic form~\eqref{finiteconserved phase space} can be schematically written as
\begin{equation}\label{Lorentzcontribution}
    \delta_{\mathcal Y}(\sqrt{-h_0}\, X)\delta Y-\delta(\sqrt{-h_0}\, X)\delta_{\mathcal Y} Y=\delta_\mathcal Y(\sqrt{-h_0}\, X \delta Y)-\delta(\sqrt{-h_0}\, X\delta_{\mathcal Y} Y)\,,
\end{equation}
where both $X$ and $Y$ are tensorial quantities on the hyperboloid. Since $X\delta Y$ is a scalar, the first term on the r.h.s.~can written as a total derivative
\begin{equation}
    \sqrt{-h_0}\,\mathcal L_{\mathcal Y}(X\delta Y)+\sqrt{-h_0}\, X\delta Y D_c \mathcal Y^c=\partial_c\pqty{\sqrt{-h_0}\, X\delta Y \mathcal Y^c}.
\end{equation}
Considering the second term in \eqref{Lorentzcontribution}, we see that it contains derivatives of $\mathcal Y$ as soon as $X$ and $Y$ are not scalars. In these cases, we integrate by parts to remove such derivatives from the bulk of the hyperboloid. Explicitly
\begin{equation}
    \frac{1}{2}\sqrt{-h_0}\, T^{ab}\delta_\mathcal Y h^{(0)}_{ab}=\sqrt{-h_0}\, T^{ab} D_a \mathcal Y_b=\partial_a (\sqrt{-h_0}\, T^{ab}\mathcal Y_b)-\sqrt{-h_0}\, D_a T^{ab}\mathcal Y_b\,,
\end{equation}
and
\begin{equation}
    \frac{1}{2}\sqrt{-h_0}\, \bar\wp^{ab}\delta_\mathcal Y \tilde\wp_{ab}=\partial_a(\sqrt{-h_0}\, \bar\wp^{ab} \tilde\wp_{bc} \mathcal Y^c)+\sqrt{-h_0}\, \mathcal Y^c \pqty{\frac{1}{2}\bar\wp^{ab} D_c \tilde\wp_{ab}-D_a (\bar\wp^{ab}  \tilde\wp_{bc})}\,.
\end{equation}
Collecting all the terms, the resulting expression is 
\begin{align}
    \omega^\rho[\delta_{\mathcal Y},\delta]&\eomeq\partial_a  k^{\rho a}_{\mathcal Y}\,,
\end{align}
where
\begin{align} \label{superrotation}
    k^{\rho a}_{\mathcal Y}&=\delta\bqty{\frac{\sqrt{-h_0}\,}{2\kappa^2} \pqty{2T^{a}_b+\bar\wp^{ac}\tilde\wp_{bc}}\mathcal Y^b}
    -\frac{\sqrt{-h_0}\,}{2\kappa^2} \, \left( T^{bc}\delta h_{bc}^{(0)} + \bar\sigma \delta \bar\tau +\bar\beta\delta\tilde\tau+\frac12\bar\wp^{bc}\delta\tilde\wp_{bc}\right) \mathcal Y^a\\
    &\bceq  \delta\pqty{\frac{\sqrt{-\hyp}}{\kappa^2} T^{a}_b\mathcal Y^b}\,,
\end{align}
follows from imposing the equations of motion for the stress tensor 
\begin{equation}\label{Wardidentity}
    D_a T^{ab}\eomeq  \frac{1}{2}   \bar\sigma D^b \bar\tau +\frac{1}{2} \bar\beta D^b \tilde\tau -\frac{1}{2} D^a(\bar\wp_{ac}\tilde\wp^{cb})+\frac{1}{4} \bar\wp^{ac}D^b\tilde\wp_{ac}\,
\end{equation}
off-shell of the boundary conditions.

As a side remark, we note that this equation can be interpreted as the Ward identity associated with boundary diffeomorphism invariance of a putative dual holographic theory. Indeed, in holographic duality, the symplectic potential corresponds to the first variation of the boundary action, written schematically in the form ``expectation value'' times ``source''. We have contracted this expression with a boundary diffeomorphism and obtained two contributions: one localized at the corner, corresponding to the charge, and one on the boundary. From the perspective of the boundary theory, the latter term represents the Ward identity associated with boundary diffeomorphism invariance and should vanish (unless there is an anomaly). In our case, we can show that this term vanishes identically upon using the bulk equations of motion.   

Since $\delta\mathcal Y^a=0$, the corner term can be easily integrated to get the standard expression for the charge associated to diffeomorphisms (see for instance the derivation in~\cite{McNees:2025acf})
\begin{equation}
 \boxed{   Q_{\mathcal{Y}}=\frac{1}{\kappa^2}\oint_{S^2} \cosh^2\tau\,s_a \,T^{ab}\mathcal Y_b\,.}
\end{equation}
When plugging in the stress tensor \eqref{def:stresstensor} and restricting to $\tilde \tau_{ab}=0$, we recover the results of~\cite{Compere:2011ve}. In particular, for the Kerr black hole family of solutions, the charge associated with rotations  around the $z$-axis $\mathcal{Y}=M_1$ measures the angular momentum, $Q_{M_1}= M a$.

Note that our derivation also yields a finite infinitesimal charge associated to arbitrary diffeomorphisms of the hyperboloid $\mathcal Y$ \eqref{superrotation}. It would be interesting to further investigate the properties of this charge and clarify how it relates to the superrotation charge discussed in \cite{Fiorucci:2024ndw}.

\subsection{Charge algebra}
In the previous section we have established that the large gauge transformations are generated by 
\begin{equation}
\xi=\xi(\omega(\omega_{\text{reg}},\omega_{\text{sing}},\omega_{\text{super}}),H(H_{\text{reg}},H_{\text{super}}),\mathcal Y_i)   
\end{equation}
where $\mathcal Y_i$ are the Lorentz generators and the symmetry parameters $\omega$ and $H$ satisfy~\eqref{eq:constraintomegaH} but $H$ does not have singular modes.  They satisfy the algebra
\begin{equation}\label{largeAKValgebra}
     \mathfrak{so}(1,3)\loplus\pqty{\mathbb R_{\omega_{\text{reg}}}\oplus \mathbb R_{H_{\text{reg}}}\oplus R_{\omega_{\text{sing}}} \oplus R_{\omega_{\text{super}}}\oplus \mathbb R_{H_{\text{super}}}}.
\end{equation}
The total charges are then expressed as 
\begin{equation}
\boxed{Q_\xi=\frac{2}{\kappa^2}\oint_{S^2}  \cosh^2\tau\,s_a \bqty{\omega D^a\bar\sigma-\bar\sigma D^a \omega+H D^a\bar\beta-\bar\beta D^a H+\frac12T^{ab}\mathcal Y_b  } . }
\end{equation}

 According to the representation theorem~\cite{Brown:1986nw,Brown:1986ed} (see also~\cite{Barnich:1991tc, Barnich:2001jy, Barnich:2007bf} for the covariant formulation), the charges~\eqref{chargesG} satisfy the same algebra~\eqref{largeAKValgebra} of the large gauge transformations up to a central term
\begin{equation}\label{reptheorem}
\{ Q_{\xi_1}, Q_{\xi_2} \}=\delta_{\xi_2}Q_{\xi_1} = Q_{\llbracket\xi_1,\xi_2 \rrbracket}+K_{(\xi_1,\xi_2)}\,.    
\end{equation}
Now we will proceed to compute the algebra and show that it is well represented and exhibits central extensions. 
\begin{itemize}
    \item 
First we compute $\delta_{\xi_2}Q_{\xi_1}$ for $\xi_1=\xi(\omega_1,H_1)$ and $\xi_2=\xi(\omega_2,H_2)$. Using the fact that $\llbracket\xi_1,\xi_2 \rrbracket=0$, we reproduce \eqref{reptheorem} with the central term
\begin{equation}\label{CentralTermExpansion}
\begin{split}
  K_{((\omega_1,H_1),(\omega_2,H_2))}&=\frac{2}{\kappa^2}\left(\langle\omega_1,H_2\rangle-\langle\omega_2,H_1\rangle\right)=\\
  &=\frac{2}{\kappa^2}\sum_{\ell\geq2, m}((\hat\omega_1)^E_{\ell, m}(\hat H_2)^O_{\ell,-m}- (\hat\omega_1)^O_{\ell, m}(\hat H_2)^E_{\ell,-m})\mathcal C_\ell \\
  &\quad+\frac{2}{\kappa^2}\sum_{\ell=0,1, m} (\hat\omega_1)^E_{\ell, m}(\hat H_2)^O_{\ell,-m}\mathcal C_\ell -(1\leftrightarrow2)\,,
\end{split}
\end{equation}
where we use the notation introduced in~\eqref{KernelInnerProduct}.
The expression is manifestly antisymmetric in $1\leftrightarrow 2$. There is a central extension between supertranslations and log-supertranslations and between the regular log-translations and singular translations. Importantly, there is \textit{no} central extension involving the regular translations, which makes it straightforward to recover Poincar\'e as a subalgebra.

We can thus write compactly
\begin{equation}
    \{Q_\omega,Q_H\}=  K_{(\omega,H)}=\frac{2}{\kappa^2}\langle\omega,H\rangle\,,
\end{equation}
from which we deduce the Poisson algebra of the fields: 
\begin{equation}\label{centralchargeinmodecomposition}
\begin{split}
\{\hat\sigma^O_{\ell,m},\hat\beta^E_{\ell',m'}\}&=-\frac{\kappa^2}{2 \mathcal C_\ell}\delta_{\ell,\ell'}\delta_{m,-m'}\,,\qquad
\{\hat\sigma^E_{\ell\geq2,m},\hat\beta^O_{\ell'\geq2,m'}\}=\frac{\kappa^2}{2 \mathcal C_\ell}\delta_{\ell,\ell'}\delta_{m,-m'}\,,
\end{split}
\end{equation}
with the other commutation relations between $\bar\sigma$ and $\bar\beta$ being zero.
\item Then we consider $\delta_{\xi_2}Q_{\xi_1}$ for $\xi_1=\xi(\omega,H)$ and $\xi_2=\mathcal Y$. Recalling that $\llbracket\xi_1,\xi_2 \rrbracket=\xi(-\mathcal L_{\mathcal Y} \omega,-\mathcal L_{\mathcal Y} H,0)$, we have 
\begin{align}
   \delta_{\mathcal Y}Q_{(\omega,H)}&=-\frac{2}{\kappa^2}\langle\mathcal L_{\mathcal Y}\bar\sigma, \omega\rangle -\frac{2}{\kappa^2}\langle\mathcal L_{\mathcal Y}\bar\beta, H\rangle=\frac{2}{\kappa^2}\langle\bar\sigma, \mathcal L_{\mathcal Y}\omega\rangle +\frac{2}{\kappa^2}\langle\bar\beta, \mathcal L_{\mathcal Y} H\rangle+\nonumber\\
   &\quad+\frac{2}{\kappa^2}\oint_{S^2}\cosh^2\tau s^a\,\bqty{D_b(\mathcal Y^b(\bar\sigma\overset{\leftrightarrow}D_a \omega+\bar\beta\overset{\leftrightarrow}D_a H))-D^b\mathcal Y_a (\bar\sigma\overset{\leftrightarrow}D_b \omega+\bar\beta\overset{\leftrightarrow}D_b H) } \nonumber \\
  & \eomeq Q_{(-\mathcal L_{\mathcal Y}\omega,-\mathcal L_{\mathcal Y}H)} +\frac{4}{\kappa^2}\oint_{S^2}\cosh\tau^2s_a D_b(\mathcal Y^{[b}  \bar\sigma {\overset{\leftrightarrow}{D}{}^{a]}} \omega )
\end{align}
In the second equality, we integrated by parts to have the Lorentz transformation act on the parameters. In the third equality, we went on-shell of the equations of motion and recognized the form of the (super)translation and log-(super)translation charge. Since the very last term vanishes using~\eqref{propertyoninthyp}, we conclude that there is no central extension between $\omega,H$ and the Lorentz sector.

\item We cross-check the previous computation by evaluating the opposite bracket, i.e.~$\delta_{\xi_2}Q_{\xi_1}$ for $\xi_1=\mathcal Y$ and $\xi_2=\xi(\omega,H)$. For that, we use the stress-energy tensor transformation~\eqref{SETtransformation} that we report here
\begin{equation}
\begin{split}
    \delta_{(\omega,H)} T^{ab}&\eomeq -2D_c(\bar\sigma D^c \omega+\bar\beta D^c H) h_{(0)}^{ab}+4D^{(a} \omega D^{b)} \bar\sigma+4D^{(a} H D^{b)} \bar\beta+\\
    &\quad+\frac{1}{2}(\omega\tilde\tau+H\bar\tau)h_{(0)}^{ab}+W^{ab}_1(\omega)+W^{ab}_2(H)\,.
\end{split}
\end{equation}
Using the identity
\begin{equation}
    -2D_c(xD^c y)\mathcal Y^a+4 D^{(a}x D^{b)} y\,\mathcal Y^b=2(\mathcal Y^b D_b y) \overset{\leftrightarrow}{D}{}^a x+4D_b(\mathcal Y^{[b} x D^{a]} y)\,
\end{equation}
valid for any $x,y$ and for a Killing vector field $\mathcal Y$, we can show that
\begin{align}
    \delta_{(\omega,H)}Q_{\mathcal Y}&\eomeq \frac{2}{\kappa^2}\oint_{S^2} \cosh^2\tau\,s_a  \bqty{\mathcal L_{\mathcal Y} \omega \overset{\leftrightarrow}{D}{}^a \bar\sigma+\mathcal L_{\mathcal Y} H \overset{\leftrightarrow}{D}{}^a \bar\beta+2D_b(\mathcal Y^{[b} \bar\sigma D^{a]} \omega+\mathcal Y^{[b} \bar\beta D^{a]} H)}+ \nonumber \\
    &\quad +\frac{1}{\kappa^2}\oint_{S^2} \cosh^2\tau\,s_a  \bqty{\frac{1}{2}(\omega\tilde\tau+H\bar\tau)\mathcal Y^a+(W_1^{ab}+W_2^{ab})\mathcal Y_b}=\nonumber\\
    &= Q_{(\mathcal L_{\mathcal Y}\omega,\mathcal L_{\mathcal Y}H)} +\frac{1}{2\kappa^2}\oint_{S^2} \cosh^2\tau\,s_a  \bqty{(\omega\tilde\tau+H\bar\tau)\mathcal Y^a+(V_1^{a}+V_2^{a})}\,.
    \end{align}
Since the last term vanishes on-shell of the boundary conditions, it confirms that the algebra
\begin{equation}
    \{Q_\mathcal Y,Q_{(\omega,H)}\}=Q_{(\mathcal L_{\mathcal Y}\omega,\mathcal L_{\mathcal Y}H)}
\end{equation}
has no central terms.

\item Last, we consider $\delta_{\xi_2}Q_{\xi_1}$ for $\xi_1=\mathcal Y$ and $\xi_2=\mathcal Y'$,
\begin{equation}
\begin{split}
    \delta_{\mathcal Y'}Q_{\mathcal Y}&=\frac{1}{\kappa^2}\oint_{S^2} \cosh^2\tau\,s_a \mathcal L_{\mathcal Y'} T^{ab}_{BY}\mathcal Y_b= \\
    & \eomeq \frac{1}{\kappa^2}\oint_{S^2} \cosh^2\tau\,s_a T^{ab}_{BY} (\mathcal L_{\mathcal Y}\mathcal Y')_b + \frac{1}{\kappa^2}\oint_{S^2}\cosh^2\tau\,s_a D_c\pqty{T^{c[a}\mathcal Y'^{b]}\mathcal Y_c}\\
    &= Q_{\mathcal L_{\mathcal Y}\mathcal Y'}=Q_{\llbracket\mathcal Y,\mathcal Y'\rrbracket}\,.
\end{split}
\end{equation}

\end{itemize}

Summing up the results of this section, the charge algebra is well represented, with a central extension between singular translations and regular log-translations, as well as between the supertranslations and log-supertranslations: 
\begin{equation}\label{chargealg}
 \boxed{    \mathfrak{so}(1,3)_{\mathcal Y}\loplus\left[\mathbb R_{\omega_{\text{reg}}} \oplus {\mathfrak{h}_3}_{(\omega_{\text{super}}^{\text{odd}},H_{\text{super}}^{\text{even}})}
     \oplus{ \mathfrak{h}_3}_{(\omega_{\text{super}}^{\text{even}},H_{\text{super}}^{\text{odd}})} 
     \oplus
 {\mathfrak{h}_3}_{(\omega_{\text{sing}},H_{\text{reg}})}
    \right]}
\end{equation}
where $\mathfrak{h}_3$ is the three-dimensional Heisenberg algebra.

A feature of this algebra is that the vacuum is degenerate along independent directions in phase space, parametrized by 
singular translation, supertranslation, regular log-translation and log-supertranslation charges. Indeed, acting with either of these transformations preserves the energy of the system, since the corresponding charges commute with time translations. As such, the fields $\bar\sigma$ and $\bar\beta$ can be interpreted as Goldstone modes associated with these symmetries. 

There are a few interesting subalgebras. First, there is the Poincar\'{e} subalgebra, generated by $\mathcal Y$ and $\omega_{\text{reg}}$, which is not centrally extended. 
Including $\omega^{\text{odd}}_{\text{super}}$, we have the BMS algebra, which instead presents a central extension with the $H^{\text{even}}_{\text{super}}$ transformations. Finally, we have
FHT-log BMS algebra which does not have the two last terms
\begin{equation}\label{FHTlogBMSalgebra}
 \text{FHT log-BMS:}\qquad     \mathfrak{so}(1,3)_{\mathcal Y}\loplus\left[\mathbb R_{\omega_{\text{reg}}} \oplus {\mathfrak{h}_3}_{(\omega_{\text{super}}^{\text{odd}},H_{\text{super}}^{\text{even}})}
    \right]\,.
\end{equation}

\paragraph{Redefinition of the Lorentz generators}
In~\cite{Fuentealba:2022xsz,Fuentealba:2023hzq} the authors perform a redefinition of the Lorentz generators such that the semi-direct sum in the charge algebra \eqref{FHTlogBMSalgebra} becomes a direct sum.
We show that a similar procedure can be done for the algebra~\eqref{chargealg}, whence it can be written as 
\begin{equation}
\mathfrak{iso}(1,3)_{(\mathcal Y,\omega_{\text{reg}})}\oplus
{\mathfrak{h}_3}_{(\omega_{\text{sing}},H_{\text{reg}})}\oplus {\mathfrak{h}_3}_{(\omega_{\text{super}}^{\text{even}},H_{\text{super}}^{\text{odd}})} \oplus {\mathfrak{h}_3}_{(\omega_{\text{super}}^{\text{odd}},H_{\text{super}}^{\text{even}})}
\end{equation}
after redefining the Lorentz generators. The main advantage of this construction is that the charges associated to $\mathcal Y$ (which include rotation charges) are not transforming under the supertransformations. We can then evaluate the charges without specifying the superframe in which we are considering them. 
For instance, acting on a Kerr black hole with a (log)-supertranslation modifies the value of the original charge $Q_{\mathcal{Y}}$, but not that of the redefined charge $\tilde{Q}_{\mathcal{Y}}$, which continues to capture the Kerr angular momentum. In this sense, the redefined charge represents the `center-of-mass angular momentum', and thus provides a preferred notion of angular momentum in asymptotically flat spacetimes. 

 We define new Lorentz generators schematically as
\begin{equation}
\tilde{Q}_{\mathcal{Y}}= Q_{\mathcal{Y}} -Q_\omega\, K^{-1}_{(\omega,H)} \{ Q_{\mathcal{Y}}  ,Q_H \} \,,
\end{equation}
where we introduced $K^{-1}_{(\omega,H)}$ as the inverse of the central charge between $\omega$ and $H$ in the sense that
\begin{equation}
    K_{(\omega,H)}^{-1}K_{(\omega',H)}=\delta_{\omega,\omega'},\qquad K_{(\omega,H)} K^{-1}_{(\omega,H')}=\delta_{H,H'}\,.
\end{equation}
In these expressions, we see $\omega$ and $H$ as placeholders for a collection of indices (parity and angular momentum) and $K_{(\omega,H)}$ as elements of a matrix, which we can formally invert.
We can then formally verify that
\begin{subequations}
    \begin{equation}
    \begin{split}
        \{\tilde{Q}_{\mathcal Y},Q_{\omega'}\}&=\{ Q_{\mathcal Y},Q_{\omega'}\}-Q_{\omega}K^{-1}_{(\omega,H)}\{\{ Q_{\mathcal Y},Q_{H}\},Q_{\omega'}\}\\
        &=Q_{\mathcal L_{\mathcal Y}\omega'}-Q_{\omega}K^{-1}_{(\omega,H)}\{\{ Q_{\mathcal Y},Q_{\omega'}\},Q_H\}\\
        &=Q_{\mathcal L_{\mathcal Y}\omega'}-Q_{\omega}K^{-1}_{(\omega,H)}K_{(\mathcal L_{\mathcal Y}\omega',H)}=0\,,\\
        \end{split}
    \end{equation}
    \begin{equation}
        \begin{split}   
        \{\tilde{Q}_{\mathcal Y},Q_{H'}\}&=\{ Q_{\mathcal Y},Q_{H'}\}-\{Q_{\omega},Q_{H'}\}K^{-1}_{(\omega,H)}\{ Q_{\mathcal Y},Q_{H}\}\nonumber\\
        &=Q_{\mathcal L_{\mathcal Y}H'}-K_{(\omega,H')}K^{-1}_{(\omega,H)}Q_{\mathcal L_{\mathcal Y}H}=0\,,
    \end{split}
    \end{equation}
\end{subequations}
for any symmetry parameters $\omega'$ and $H'$. In the first relation, we used the commutativity between (super)translations and the Jacobi identity to go to the second line. In the second relation, we used that $ \{ Q_{\mathcal{Y}}  ,Q_H \}$ is another log-(super)translation charge and hence commutes with any log-(super)translation. A formalization of this reasoning is made precise by working in a mode decomposition, which we detail in Appendix \ref{App:Redef}.

\subsection{Imposing parity boundary conditions}
So far, our analysis did not require to impose any parity conditions on the fields or the symmetry parameters. However, regularity conditions and compatibility conditions for solutions are typically imposed in the analysis of spatial infinity. We want to stress that these are additional requirements that the charge analysis at spatial infinity itself does not force us to take. 

In the literature, boundary conditions have been imposed to ensure the regularity of the solution space in the limit $|\tau|\to\infty$, so as to attach \textit{smooth} past and future null infinity to spatial infinity. Such boundary conditions take the form of parity conditions imposed on the fields in order to recover asymptotic simplicity; see~\cite{Troessaert:2017jcm,Capone:2022gme}. It would be interesting to investigate whether opposite parity conditions, which do not necessarily lead to a smooth null infinity, can arise dynamically in general relativity and how they relate to null infinity. We leave this for future work.

We start by commenting on the parity conditions imposed in~\cite{Compere:2023qoa}, as they include previous boundary conditions in the literature.
The first requirement is to impose $\bar\sigma$ to be even for $\ell\geq2$. This implies that $H_{\text{super}}$ has to be even. In terms of the Weyl tensor, this amounts to requiring the leading electric part of the Weyl tensor to be even for $\ell\geq2$. 

Then, they further require that $\bar\tau_{ab}$ derives from a scalar potential (which, for us, is $\bar\beta$). Comparing with~\eqref{tauabafterBC}, it requires to set $\bar\wp_{ab}$ to zero, which excludes a NUT charge. They further impose the potential $\bar\beta$ to be odd, and to not have $\ell=0,1$ harmonics. This implies that $\omega$ has to be an odd function. 

Using only these two restrictions, the charges are now 
\begin{align}
\label{ChargeModeExpansionLargegaugeparitycond}
 Q_{(\omega,H)}&=-\frac{2}{\kappa^2}\sum_{\ell\geq2, m}(\hat \sigma^E_{\ell, m}\hat \omega^O_{\ell,-m}-\hat \beta^O_{\ell, m}\hat H^E_{\ell,-m})\mathcal C_\ell -\frac{2}{\kappa^2}\sum_{\ell=0,1, m}(\hat \sigma^E_{\ell, m}\hat \omega^O_{\ell,-m})\mathcal C_\ell
\end{align}
With these conditions, $H_{\text{reg}}$ becomes pure gauge, as its charge vanishes. In~\cite{Compere:2023qoa}, they use this pure gauge transformation to obtain finite, when taking the limit $\tau\to\pm\infty$, singular $\bar\sigma$.
The central charge is now only among the super modes:
\begin{equation}\label{CentralTermExpansionparitycond}
    K_{(\omega,H)}=\frac{2}{\kappa^2}\sum_{\ell\geq2, m}(\hat\omega_1)^O_{\ell, m}(\hat H_2)^E_{\ell,-m}\,\mathcal C_\ell  -(1\leftrightarrow2)
\end{equation}
At this stage, we recover the structure of the log-BMS algebra found by FHT in the canonical setting. We will comment more on the matching conditions in Appendix \ref{app:FHTDetails}. 

The last condition we need to impose to match those of~\cite{Compere:2023qoa} is $ \tilde \tau_{\tl{ab}}=0$. This restricts $H$ to be a regular log-translation, which is pure gauge, see~\eqref{ChargeModeExpansionLargegaugeparitycond}. In that case, the asymptotic symmetry algebra reduces to global BMS. Regge-Teitelboim~\cite{Regge:1974zd} further considered $\bar \tau_{\tl{ab}}$ to be even. This disallows odd supertranslations, and therefore reduces the asymptotic symmetry algebra to the Poincar\'{e} algebra. 

We now discuss the parity conditions imposed in~\cite{Ashtekar:1990gc}. They require $\bar\sigma$ to be even in all harmonics. As a consequence, the residual symmetries are generated only by even $H$, which in particular excludes the regular logarithmic translation. Furthermore, they impose $\tilde \tau_{ab}=0$, which eliminates the remaining $H$ transformations altogether.
To further reduce the symmetry group to Poincar\'{e}, they impose the condition $\bar\tau_{\tl{ab}}=0$. As we have demonstrated, this choice can be reached by a large gauge transformation and can therefore be interpreted as a restriction of the phase space.

Lastly, we briefly comment on the connection between the solution space at spatial infinity and that of null infinity. We use the change of coordinates between spatial and null infinity~\cite{Compere:2023qoa} to interpret the fields $\bar\sigma,\bar\beta$ in terms of null quantities. To achieve this, the parity conditions on $\bar\sigma,\bar\tau_{ab}$ are necessary for consistency. We have
\begin{equation}
    \bar\sigma \propto m_B,\qquad \bar\tau_{AB}\propto C_{AB}\implies \bar\beta \propto C
\end{equation}
where $m_B$ denotes the Bondi mass aspect, $C_{AB}$ the Bondi shear and $C$ its scalar potential for the electric part. The latter transforms, just like $\bar\beta$, linearly under supertranslations, and therefore both can be seen as parametrising the orbit of these transformations. In this sense, both of them are "Goldstone fields" for supertranslations. Conversely, $\bar\sigma$ is then a Goldstone field for log-supertranslations.
It would be interesting to study the action of such transformations at null infinity. 

\section{Summary and outlook}\label{sec:outlook}
In this work, we performed a symmetry analysis at spatial infinity. We reviewed the different aspects of our results and compared them with those of CD (Comp\`{e}re–Dehouck)~\cite{Compere:2011ve} and FHT (Fuentealba–Henneaux–Troessaert)~\cite{Fuentealba:2022xsz}, who explored similar questions using different approaches, before presenting interesting questions to explore in the future.

We started by generalizing the Beig–Schmidt description of spatial infinity by allowing a polyhomogeneous radial expansion of the metric, including logarithmic terms at subleading orders. We recall that these logarithmic contributions are natural from the perspective of solving some (partial) differential equations using a power-law expansion. The logarithmic contributions are the result of resonance effects when solving these equations order by order~\cite{Chrusciel1992Asymptotic, ChruscielMacCallumSingleton1994, ChruscielDelay2000}. 
The metric \eqref{BSExpansion} generalizes the one of CD~\cite{Compere:2011ve}. In particular, this enabled us ultimately to switch on an extra set of symmetries with respect to CD: the log-supertranslations. These symmetries were identified in a different coordinate chart by FHT and studied, in the Hamiltonian picture, imposing parity conditions.

We solved Einstein's equations asymptotically, identifying the freely specifiable data and determining the asymptotic structure of the Weyl tensor. The residual diffeomorphisms preserving the fall-offs were derived and shown to form an enlarged asymptotic symmetry algebra containing Lorentz transformations, translations, supertranslations, logarithmic translations, and logarithmic supertranslations, \textit{without} imposing parity conditions.

We then constructed a finite and conserved symplectic structure for the enlarged phase space. Using the results of~\cite{Andrade:2006pg,McNees:2023tus,McNees:2024iyu}, the symplectic form was renormalized without specifying boundary conditions and without adding boundary terms to the action. Boundary conditions were then identified in a second stage to ensure conservation of the charges while \textit{preserving} logarithmic supertranslations. Some of these conditions are formulated in terms of the asymptotic behavior of the Weyl tensor and do not rely on parity assumptions. Our phase space has the feature to include Kerr-Taub-NUT spacetimes.

The symplectic potential we obtained is written off-shell of the boundary conditions in the form of ``expectation value'' times variations of the ``source''.  We use this form to compute the Ward identities associated to the boundary diffeomorphism invariance of the putative dual theory, see equation \eqref{Wardidentity}. We found that the Ward identity precisely corresponds to the bulk equations of motion and that there is no anomaly associated to boundary diffeomorphisms. 

The specific details behind our approach are therefore distinct from CD's and FHT's works which relied respectively on a boundary action~\cite{Compere:2008us} or parity conditions to renormalize the infinities arising from the asymptotic limit.   
Before the work of CD, strong emphasis was put on the use of parity conditions to make the theory well-defined. Our results   
are therefore  in agreement with CD~\cite{Compere:2011ve} as they emphasize again that the standard reliance on parity conditions is not forced by consistency at spatial infinity in the covariant phase space, but rather should reflect an additional choice motivated by matching to a smooth null infinity~\cite{Troessaert:2017jcm}. This disentangles the intrinsic structure of spatial infinity from assumptions about global regularity and opens the door to a systematic treatment of asymptotically flat spacetimes that does not satisfy peeling or asymptotic simplicity. 

All the above results enabled us to compute the finite surface charges associated with the residual symmetries and to evaluate their Poisson algebra. We used the Iyer-Wald method~\cite{Iyer:1994ys} of contracting the symplectic form \textit{off-shell} of the boundary conditions to extract the co-dimension two objects defining the charges.  We obtained non-vanishing charges associated with Lorentz transformations, regular translations, singular log-translations, supertranslations, and log-supertranslations (where the supertransformations contain both parities). The Lorentz charges encode information about the angular momentum of the spacetime. The charges associated with the four regular translations measure the electric part of the leading Weyl tensor. For instance, the charge associated with time translation is the ADM mass. The interpretation of what is measured by the singular log-translation is still an open question.\footnote{We can however make heuristic statements: if we have a stress-energy matter source $T^{\mathrm{mat}}_{\mu\nu}$ with the form
    \begin{equation}
        T^{\mathrm{mat}}_{a\rho} = \frac{1}{\rho} D^b \bar S_{ab}+\cdots\implies D^b(\bar\tau_{ab}+2\kappa^2 \bar S_{ab})\eomeq 0\implies (D^2+3)\bar\beta + \frac{\bar\tau}{4}\eomeq \kappa^2 \bar S_{ab}\hyp^{ab},
    \end{equation}
    then even when $\bar\tau=0$, there is a source for the log-supertranslation charges. For this, however, one needs superleading fields, i.e. a massless scalar with decay $\phi \sim \bar\phi^0 + \log(\rho) \tilde\phi^0+\mathcal{O}(\frac{1}{\rho})$.}

The orbits generated by singular translations, regular log-translations, and (log-) supertranslations preserve the energy of the system, as they commute with time translations, yet they connect physically inequivalent states. The vacuum is therefore degenerate, characterized by such charges. Furthermore, the log-supertranslation charges at spatial infinity are related to the Goldstone mode of supertranslations at null infinity via the coordinate transformation constructed in~\cite{Compere:2023qoa}.

The resulting asymptotic symmetry algebra is a log-BMS algebra containing the Lorentz algebra and several Heisenberg-type subalgebras, i.e.~non-trivial central extensions between abelian sub-algebras, see equation \eqref{chargealg}. Here lies the main result of our work: \textit{in the covariant phase space formalism, we obtained a more general charge algebra structure than CD, since we do realize a log-supertranslations sector, just like FHT, but without relying on parity conditions and the Hamiltonian formalism as FHT do.} As a consistency check, upon imposition of such parity conditions, we recover a charge algebra isomorphic to FHT's.  

Using a similar argument as~\cite{Fuentealba:2022xsz,Fuentealba:2023hzq}, namely that log-supertranslations and supertranslations form an Heisenberg algebra, we construct new Lorentz generators that make the Poincar\'{e} algebra an ideal of the log-BMS algebra. This shows that the realization of log-supertranslations at spatial infinity allows us to define an angular momentum that does not depend on a BMS frame. At null infinity, this was accomplished by using supertranslation Goldstone fields, which does not appear as a Noether charge at null infinity, in order to obtain such a redefinition of the angular momentum~\cite{Compere:2019gft,Javadinezhad:2022hhl,Javadinezhad:2023mtp}. An advantage of working at spatial infinity is that these fields originate from a symmetry principle, namely the charge generator associated with logarithmic supertranslations.

\medskip

This project opens new interesting directions to explore which we detail now. As we have emphasized, our work recovers several aspects of both the CD and FHT approaches. It would be worthwhile to develop a more systematic dictionary between these frameworks and ours. In particular, it would be useful to clarify whether the regularization of the symplectic form adopted here can be related to the boundary Lagrangian prescription of CD.
Recall that in the prescription used here~\cite{Andrade:2006pg,McNees:2023tus,McNees:2024iyu}, one obtains a finite symplectic potential up to $\delta$-exact terms, whose contributions cancel when constructing the symplectic form, and hence do not affect the charges. The prescription in CD follows the Comp\`{e}re–Marolf approach~\cite{Compere:2008us}, which employs boundary Lagrangians to renormalize the symplectic potential. In the context of asymptotically locally AdS$_4$ spacetimes, these boundary Lagrangians are precisely the ones that render the action itself finite. On the other hand, it was shown in~\cite{McNees:2025acf} using the prescription~\cite{McNees:2023tus,McNees:2024iyu} that the remaining divergent $\delta$-exact terms in the symplectic potential are canceled by the same boundary Lagrangians.
It would be interesting to establish an analogous result at spatial infinity and, in doing so, to generalize the CD construction of boundary Lagrangians to a setting allowing log-supertranslations.

As also explained in Appendix \ref{app:FHTDetails}, the different realizations of the charge algebra due to  using different routes to build the symplectic form becomes manifest when attempting to compare our results with those of FHT at the level of explicit charge expressions. Indeed, the obstruction does not arise from the symmetry algebra itself, but from the fact that the charges are defined with respect to different coordinate charts and in terms of canonical variables.
Understanding whether these constructions are related by a change of variables or correspond to genuinely distinct phase spaces would be an interesting question to explore. 

Our starting point, while more general than previous analyses, still allows for further gauge relaxations. For instance, by relaxing the leading-order metric $h_{ab}^{(0)}$, we expect to enlarge our symmetry algebra to include superrotations as shown in~\cite{Fiorucci:2024ndw}, although obtained in the Hamiltonian framework. We have already demonstrated that the symplectic form remains finite in that case, and provide the expression of the infinitesimal charge associate with arbitrary diffeormophisms of the hyperboloid.
We have not yet analyzed all the charges and their algebra when $h_{ab}^{(0)}$ is left unfixed. It would therefore be interesting to compute that algebra and determine whether the central term we found between our supertransformations remains central or instead receives non-trivial modifications in the presence of superrotations.

Nonetheless, the enhancement of symmetry at spatial infinity naturally raises the question of whether a similar structure can be realized in other regions of asymptotically flat spacetime, namely at null and timelike infinities. Timelike infinity is related to spatial infinity by a Wick rotation (see, for instance,~\cite{Compere:2023qoa}); one may therefore expect a closely analogous mathematical construction to hold there. The physical interpretation, however, is substantially different, and we defer a detailed investigation of this case to future work.
The relation to null infinity is more subtle, as it requires a involved coordinate transformation. Progress along these lines has already been achieved for scalar fields~\cite{Fuentealba:2024lll}, for electromagnetism~\cite{Fuentealba:2025ekj}, and for general relativity without log-supertranslations~\cite{Troessaert:2017jcm,Prabhu:2019fsp,Prabhu:2021cgk,Capone:2022gme,Compere:2023qoa,Compere:2025bnf}. 

An important open question is how to establish a holographic correspondence for asymptotically flat spacetimes. Current proposals, such as celestial holography (see for instance \cite{Raclariu:2021zjz}) or Carrollian holography (see for instance \cite{Ruzziconi:2026bix}), rely primarily on the structure at null infinity to construct the dual theory from the bulk perspective (see \cite{Nguyen:2021ydb}, however, for a connection to spatial infinity). Since we have identified new symmetries beyond those existing at null infinity, it would be particularly interesting to understand how they are encoded within flat holography proposals.

Another interesting avenue concerns the Ward identities and infrared structure implied by the enlarged symmetry algebra. Since BMS Ward identities are known to be equivalent to the leading and subleading soft graviton theorems, the log-BMS algebra realized at spatial infinity suggests the existence of additional identities associated with logarithmic supertranslations. While log-translations were shown to be related to universalities in scattering amplitudes~\cite{Boschetti:2025tru}, it would be worthwhile to connect our results with their treatment and to investigate the effect of log-supertranslations on soft theorems. 

A main feature of our analysis is that we did not assume any connection of spatial infinity with smooth null infinity, thereby not imposing parity for instance.  When one seeks to establish a connection with smooth null infinity, by which we mean spacetimes satisfying the peeling property~\cite{Sachs:1961zz}, this is done by imposing parity conditions on the fields thereby reducing the asymptotic symmetry algebra to the FHT-log BMS algebra. 
However, it is now better appreciated that there exist solutions of the Einstein equations that do not generally lead to a smooth null infinity~\cite{Friedrich:1983vx,PhysRevD.19.3483,PhysRevD.19.3495,doi:10.1098/rspa.1981.0101,Andersson:1993we,Valiente-Kroon:2002xys,Kroon:2004me,Kehrberger:2021uvf,Kehrberger:2021vhp,Kehrberger:2021azo,Kehrberger:2024clh,Kehrberger:2024aak,Gajic:2022pst,Kehrberger:2023btg,Bieri:2023cyn,Geiller:2024ryw}. The parity conditions may therefore be too restrictive to capture the full spectrum of allowed fall-offs. It will be interesting to probe the non-smooth sector starting from our analysis at spatial infinity.

\section*{Acknowledgments}
The authors thank Geoffrey Compère, Laurent Freidel, Marc Henneaux, and Beniamino Valsesia for discussions. The authors GN and SL would also like to thank the Perimeter Institute for Theoretical Physics for its hospitality during the initial stages of this project.
Research at the Perimeter Institute is supported in part by the Government of Canada through NSERC and by the Province of Ontario through MEDT.

\appendix

\section{Equations of motion}\label{app:EoM}
In this appendix, we give the necessary detail to understand the linearised solution space of metrics we consider. We begin with a radial 3+1 decomposition around spatial infinity, after which we solve the most important equation exactly. Then, we show how the remaining equations can be systematically solved using decompositions of tensors in analogy to the Helmholtz decomposition.  \subsection{3+1 split and equations of motion}
\label{app:exgeometry}

The choice of Beig-Schmidt coordinates is adapted to a foliation of spacetime in hypersurfaces sitting at constant $\rho$ that we denote as $\mathcal H_\rho$. This 3+1 split is the radial equivalent of the ADM decomposition~\cite{danieliADMFormalismHamiltoniana}, where we use timelike hypersurfaces instead of spacelike ones. These surfaces are implicitly characterised to be such that Lorentz transformations preserve them individually.

On each $\mathcal H_\rho$, we induce the metric
\begin{align}
   d s^2|_{\mathcal{H}_{\rho}}=  \rho^{2}h_{ab}(\rho,x^c)\dd x^a \dd x^b.
\end{align}
Until we consider the effects of the asymptotic expansion in~\eqref{leadeom}, Latin indices are raised with $h^{ab}$, with no factors of $\rho$.
If we let $n_\mu$ be the unit 1-form normal to $\mathcal H_\rho$, we can introduce a radial lapse $N$ and a radial shift $N^a$ by analogy with the ADM decomposition
\begin{equation}
    n_\mu \dd x^\mu=N 
    \dd\rho\implies n^\mu\partial_\mu=\frac{1}{N}\pqty{\partial_\rho-N^a\partial_a}.
\end{equation}
This implies that the full metric (and its inverse) can also be written as
\begin{equation}
    g_{\mu\nu}=\pmqty{N^2+\rho^2 N^a N_a \; & \rho^2 N_a\\
    \rho^2 N_b & \rho^2 h_{ab}},\qquad g^{\mu\nu}=\pmqty{\frac{1}{N^2} & -\frac{N^a}{N^2}\\
    -\frac{N^b}{N^2} &\; \rho^{-2}h^{ab}+\frac{N^a N^b}{N^2}}
\end{equation}
so that, comparing with BS coordinates, we deduce that
\begin{equation}
    N^2=\Sigma^2-\frac{1}{\rho^2}\Sigma^a\Sigma_a,\qquad N_a=\frac{1}{\rho^2}\Sigma_a.
\end{equation}
Given the normal, we can introduce the extrinsic curvature of the $\mathcal H_\rho$ hypersurface by the standard formula (with a slight abuse of notation, we write $h$ with mixed indices to indicate the projector on the hypersurface)
\begin{equation}
    K_{ab}=h^\mu_a h^\nu_b \nabla_\mu n_\nu=\frac{\rho^2}{2N}\pqty{\partial_\rho h_{ab}+\frac{2}{\rho}h_{ab}-\mathcal D_a N_b-\mathcal D_b N_a},
\end{equation}
where $\mathcal D$ is the covariant derivative compatible with $h_{ab}$; this is different from $D$, which is compatible with $\hyp_{ab}$.

Given this split, we can rewrite the Einstein equations as a set of three equations: a scalar, a vector, and a tensor on $\mathcal H_\rho$. Using Gauss, Codazzi and Ricci relations, we can write
\begin{subequations}
\begin{align}
\label{EinsteinEqs}
    H&\equiv -2\rho^2 G_{\mu\nu}n^\mu n^\nu=\mathcal R+\frac{K_{ab}K^{ab}-K^2}{\rho^2},\\
    F_{a}&\equiv 2\rho R_{\mu\nu}h^\mu_a n^\nu=\frac{2}{\rho}(\mathcal D_b K^b_a-\mathcal D_a K),\\
    F_{ab}&\equiv N R_{\mu\nu}h^\mu_a h^\nu_b= 
    N\mathcal R_{ab}-\partial_\rho K_{ab}-\mathcal D_a\mathcal D_b N-\frac{N}{\rho^2} K K_{ab}+\frac{2N}{\rho^2} K_{ac}K^c_b,
\end{align}
\end{subequations}
where $\mathcal R_{ab}$ is the Ricci tensor of the induced metric on $\mathcal H_\rho$.
Since the $\rho=\text{const}$ hypersurfaces are timelike, all the equations are dynamical, but there will be some constraints given by the Bianchi identity. Note that in the above, we raise indices $a,b$ with the induced inverse metric $\rho^{-2}h^{ab}$, meaning every upper index brings a factor $\rho^{-2}$ compared to CD~\cite{Compere:2011ve}.

In addition to this split, the $1/\rho$ polyhomogeneous expansion~\eqref{BSExpansion} induces an analogous expansion on the Ricci tensor, allowing us to separate equations~\eqref{EinsteinEqs} into an infinite set. At order $\rho^{-n}$, we will find a partial differential equation that is linear in the $n$-th coefficient of the metric expansion, with possible non-linear contributions (starting from $n\ge 2$) involving the lower order coefficients.
For our purposes, we are interested only in the leading, subleading and subsubleading terms in this expansion. Let us analyze them carefully, with the reminder that, from now on, latin indices are raised with the leading order metric.
\begin{itemize}

\item At leading order $n=0$, the equations of motion are
\begin{subequations}\label{leadeom}
\begin{align}
\bar H^{(0)}&=\bar{\mathcal R}^{(0)}-6,\\
\bar F^{(0)}_a&=0,\\
\bar F^{(0)}_{ab}&=\bar{\mathcal R}^{(0)}_{ab}-2h^{(0)}_{ab}.
\end{align}
\end{subequations}
These equations, together with the vanishing of the Weyl tensor in 3d, imply that the leading order metric is that of a negatively curved, maximally symmetric Einstein spacetime with $\mathcal R=6$, \textit{i.e.} the unit hyperboloid. Thus it is consistent to require $h^{(0)}_{ab}=\hyp_{ab}$.
    
\item At subleading order $n=1$, we have two sets of equations, coming from terms in $1/\rho$ and $\log\rho/\rho$. Explicitly, they are
\begin{subequations}
\label{subleadpolyeom}
\begin{align}
\bar H^{(1)}&=D^a D^b \bar h_{ab}-D^2\bar h+12\bar\sigma-2\tilde h,\\
\bar F^{(1)}_a&=D_a(\bar h-\tilde h)-D^b(\bar h_{ab}-\tilde h_{ab})+4D_a\bar\sigma,\\
\begin{split}
    \bar F^{(1)}_{ab}&=-\frac{1}{2}D_a D_b \bar h+D^c D_{(a}\bar h_{b)c}-\frac{1}{2}(D^2+3) \bar h_{ab}+\frac{1}{2}(\bar h-\tilde h)\hyp_{ab}+\\
    &\quad-D_a D_b \bar\sigma+\hyp_{ab}(3\bar\sigma+\tilde\sigma),
\end{split}
\end{align}
\end{subequations}
and
\begin{subequations}
\label{subleadlogeom}
\begin{align}
\tilde H^{(1)}&=D^a D^b \tilde h_{ab}-D^2\tilde h+12\tilde\sigma,\\
\tilde F^{(1)}_a&=D_a\tilde h-D^b\tilde h_{ab}+4D_a\tilde\sigma,\\
\begin{split}
    \tilde F^{(1)}_{ab}&=-\frac{1}{2}D_a D_b \tilde h+D^c D_{(a}\tilde h_{b)c}-\frac{1}{2}(D^2+3) \tilde h_{ab}+\frac{1}{2}\tilde h\hyp_{ab}+\\
    &\quad-D_a D_b \tilde\sigma+3\hyp_{ab}\tilde\sigma.
\end{split}
\end{align}
\end{subequations}
We immediately notice that the trace of the tensorial equations gives the scalar one, in both the polynomial and the logarithmic sector. Rewriting these equations in terms of the auxiliary tensors in~\eqref{def tau} and separating the trace from the traceless components gives back~\eqref{eomsublead} -- notice that the trace of the tensorial equations is not independent from the other equations, since $4\bar F^{(1)}=3\bar H^{(1)}+\tilde H^{(1)}-D^a \bar F^{(1)}_a$, and similarly $4\tilde F^{(1)}=3\tilde H^{(1)}-D^a \tilde F^{(1)}_a$. The logarithmic sector~\eqref{subleadlogeom} evolves independently and behaves as a source for the polynomial sector~\eqref{subleadpolyeom}. The coupling of the two sectors originates from $\tilde\sigma$ and the trace $\tilde{\tau}$: if these vanish, the two sectors decouple and we simply get a doubling of our kinematical variables.

\item At subsubleading order $n=2$, there are three sets of equations, corresponding to $1/\rho^2$, $\log\rho/\rho^2$, and $\log^2\rho/\rho^2$.
We, however, only require a specific combination of them to check the conservation of the boundary stress-energy tensor. This combination is equivalent to the mixed component $\bar G_{\rho a}$, explicitly:
\begin{equation}
    \begin{aligned}
       \bar G_{\rho a}= &\frac{1}{2} \bar h^{bc} D_a \tilde h_{bc}
    +\frac{1}{4} \tilde h^{bc} D_a \bar h_{bc} 
    +\frac{1}{4} \tilde h_a^b D_b\bar h
    -\frac{1}{2} \bar h^{bc} D_c\tilde h_{ab}\\
    &-\frac{1}{2} \tilde h_a^b D_c\bar h_b^c
    -\frac{3}{4} \bar h^{bc} D_a\bar h_{bc}
    -\frac{1}{4} \bar h_a^b D_b\bar h
    +\frac{1}{2} \bar h^{bc} D_c\bar h_{ab}\\
    &+\frac{1}{2} \bar h_a^b D_c\bar h_b^c
    +\frac{1}{2} \tilde h D_a\bar\sigma
    -\frac{1}{2} \tilde h_{ab} D^b\bar\sigma
    -\frac{1}{2} D_a\dbtilde{h}^{(2)}\\
    &+\frac{1}{2} D_b\dbtilde{h}^{(2)b}_a
    -\frac{1}{2} \bar h D_a\bar\sigma
    +\frac{1}{2} \bar h_{ab} D^b\bar\sigma\\
    &
    +D_a\bar{h}^{(2)}-D_b\bar{h}^{(2)b}_a
    -2 \bar\sigma D_a\bar\sigma
    +2 D_a\bar\sigma^{(2)} \\
    &
    -\frac{1}{2} D^2\bar\Sigma_a+\frac{1}{2} D_b D_a\bar\Sigma^b-2 \bar\Sigma_a
    \eomeq 0
    \end{aligned}
\end{equation}


\end{itemize}

\subsection{$(D^2+3)F = 0$}\label{app:D2+3=0}
As it appears generically in our equations of motion, we need to solve the equation 
\begin{align}\label{eomf}
    (D^2+3)F = 0\, ,
\end{align}
for a scalar $F$ on the unit hyperboloid $\mathcal H$ with line element
\begin{equation}
    \hyp_{ab}\dd x^a \dd x^b=-\dd\tau^2+\cosh^2\tau \dd\Omega^2,\quad\qquad \dd\Omega^2=\gamma_{AB}\dd x^A\dd x^B=\dd\theta^2+\sin^2\theta\dd\varphi^2.
\end{equation}
We follow the derivation and convention of~\cite{Compere:2023qoa} and~\cite{Troessaert:2017jcm}. 
It is convenient to organize the solutions in terms of their properties under the parity operator
\begin{align}\label{parity}
    \Upsilon_\mathcal{H} (\tau,\theta,\phi) = (-\tau, \pi - \theta, \phi + \pi)\,.
\end{align}
We include the subscript $\mathcal{H}$ to distinguish this operator from that on celestial sphere
\begin{align}\label{parity-sphere}
    \Upsilon(\theta,\phi) = (\pi - \theta, \pi + \phi)\,.
\end{align}
We say that a tensor $X$ has definite (even or odd) parity when $\Upsilon_\mathcal{H}X = \pm X$, with the upper/lower sign corresponding to even/odd parity, and similarly for tensors on the sphere. In particular, the spherical harmonics have the following parity
\begin{align}
    \Upsilon Y_{\ell, m} = (-1)^\ell Y_{\ell, m}.
\end{align}
We assume our spherical harmonics to be normalised as in our references, $\oint Y_{\ell, m}Y_{\ell'm'}=\delta_{\ell,\ell'}\delta_{m,-m'}$. Here and in what follows, we omit the unit round sphere area element in all integrals over 2-spheres.\\
Starting with the ansatz 
\begin{align}\label{eqKGhyp}
 F_{\ell, m}(\tau, x^A) = \frac{1}{2\cosh{\tau}}f_\ell(\tanh{\tau}) Y_{\ell, m}(x^A)\,,
\end{align}
we can turn the original equation into Legendre equation by a change of variable $s=\tanh\tau$ (recall also that $\partial_\tau=(1-s^2)\partial_s$)
\begin{align}\label{eomfx}
    (1-s^2)\partial_s^2f_\ell(s)- 2s \partial_sf_\ell + \left(\ell(\ell+1)- \frac{4}{1-s^2}\right)f_\ell(s) = 0\, .
\end{align}
The solutions of this equation are well-known 
\begin{align}
f^E_\ell(s) &= \begin{cases}2 \frac{1+s^2}{1-s^2}, & \ell = 0\,  \\ 
\frac{2 s \left(s^2-3\right)}{s^2-1}\, & \ell = 1\,  \\ 
P_\ell^2(s)& \ell \geq 2 \, \end{cases}\,,\quad\qquad 
f_\ell^O(s)=Q_\ell^2(s)\, ,
\end{align}
where $P_\ell^2$ and $Q_\ell^2$ are the associated Legendre functions. We notice that the only regular solutions at $s=\pm 1$ are $f^E_{\ell\ge 2}$, and they vanish there.
The notation $f^E$ and $f^O$ comes from the fact that, since $P^2_\ell(-s)=(-1)^{\ell}P^2_\ell(s)$ and $Q^2_\ell(-s)=(-1)^{\ell+1}Q^2_\ell$(s), we have
\begin{align} \Upsilon_\mathcal{H}[f_\ell^{E}(s)Y_{\ell, m}(x^A)] & = f_\ell^{E}(s)Y_\ell^m(x^A)\, \qquad
\Upsilon_\mathcal{H}[f_\ell^O(s)Y_{\ell, m}(x^A)] = - f_\ell^O(s)Y_{\ell, m}(x^A)\,,
\end{align}
and so the two families have definite hyperboloid parities, even and odd, respectively.\\
The general solution of~\eqref{eomf} is then a sum over different modes,
\begin{equation}
\label{modedecomp}
F
=\frac{1}{2\cosh\tau} \sum_{\ell, m} (\hat F^E_{\ell, m}f^E_\ell(\tanh\tau)+\hat F^O_{\ell, m}f^O_{\ell }(\tanh\tau))Y_{\ell, m}(x^A).
\end{equation}
In terms of the basis introduced in \S\ref{sec:residualsymm}, it reads similarly
\begin{align}
 F
= \sum_{\ell, m} (\hat F^E_{\ell, m}\zeta^{\text{even}}_{\ell,m}+\hat F^O_{\ell, m}\zeta^{\text{odd}}_{\ell,m}). 
\end{align}

Given the existence of a conserved product in the space of solutions to Legendre equation given by $(1-s^2)(f\partial_s g-g\partial_s f)$, we can construct a conserved inner product in the kernel of $(D^2+3)$ as well,
\begin{equation}
\label{KernelInnerProduct}
\langle F,G\rangle:=\oint_{S^2} \cosh^2\tau \, (F\partial_\tau G-G\partial_\tau F).
\end{equation}
Explicitly, using again $s=\tanh\tau$ and recalling that $\partial_\tau=(1-s^2)\partial_s$, one can show that
\begin{equation}
\begin{split}
\langle F,G\rangle &=\frac{1-s^2}{4}\oint_{S^2}\sum_{\ell, m,\ell'm'} (\hat F^E_{\ell, m}f^E_\ell+\hat F^O_{\ell, m}f^O_{\ell })\overset{\leftrightarrow}{\partial_s}  (\hat G^E_{\ell, m}f^E_\ell+\hat G^O_{\ell, m}f^O_{\ell })Y_{\ell, m} Y_{\ell' m'}=\\
&=\frac{1-s^2}{4}\sum_{\ell, m} (\hat F^E_{\ell, m}f^E_\ell+\hat F^O_{\ell, m}f^O_{\ell })\overset{\leftrightarrow}{\partial_s}  (\hat G^E_{\ell, - m}f^E_\ell+\hat G^O_{\ell, -m}f^O_{\ell })=\\
&=\frac{1-s^2}{4}\sum_{\ell, m} (\hat F^E_{\ell, m}\hat G^O_{\ell,-m}-\hat F^O_{\ell, m}\hat G^E_{\ell,-m})(f^E_\ell \partial_s f^O_\ell-f^O_\ell \partial_s f^E_\ell).
\end{split}
\end{equation}
We can evaluate explicitly the product in the last line
\begin{equation}\label{ClDefinition}
    \mathcal C_\ell\equiv \frac{1-s^2}{4}(f^E_\ell \partial_s f^O_\ell-f^O_\ell \partial_s f^E_\ell)=\begin{cases} 1, & \ell = 0\,  \\ 
    -3, & \ell = 1\,  \\ 
    \frac{(\ell-1)\ell(\ell+1)(\ell+2)}{4},\quad & \ell\ge 2 \end{cases}
\end{equation}
and see that it is indeed a constant. These constants are relevant in the Poisson brackets of the charges in the main text, see~\eqref{CentralTermExpansion}.

\subsection{Decomposition of symmetric tensors on the hyperboloid}\label{app:symmtensor}

In this appendix, we provide an on-shell decomposition for the tensors $\bar\tau_{ab}$ and $\tilde\tau_{ab}$. As the line of arguments is the same for both, we are going to consider here $\bar\tau_{ab}$ only.
The subleading equations of motion, written in terms of $\bar\tau_{ab}$ are
\begin{equation}
    D^a\bar\tau_{ab}=0\,, \quad (D^2-3)\bar\tau_{\langle ab \rangle}=-\frac12 D_{\langle a}D_{b\rangle}\bar\tau\,.
\end{equation}

In Appendix A of~\cite{Compere:2011db} it is proven that any regular symmetric, divergenceless and trace-free tensor $X_{ab}$ satisfying $(D^2-3)X_{ab}=0$ can be constructed as a linear combination of three terms
\begin{equation}\label{SDT}
\begin{split}
X_{ ab}&= (D_a D_b+\hyp_{ab})\Phi_{\text{sing}}+T_{ab}^{(I)}(\Phi_{\text{super}}^{(1)})+\tldd{a}{b}\Phi_{\text{super}}^{(2)}
    \end{split}
\end{equation}
where all the scalars $f=(\Phi_{\text{sing}}, \Phi_{\text{super}}^{(1)}, \Phi_{\text{super}}^{(2)})$ satisfy $(D^2+3)f=0$. In particular, given a Lorentz frame, $\Phi_{\text{sing}}$ can be chosen to have only even $\ell=0,1$ modes. The addition of regular modes $\Phi_{\text{reg}}$ has no effect on $X_{ab}$ because, by definition, $(D_a D_b+\hyp_{ab})\Phi_{\text{reg}}=0$. The tensor $T_{ab}^{(I)}(\Phi)$ is symmetric, divergenceless and trace-free
\begin{equation}
    T^{(I)}_{[ab]}(\Phi)=0\,\quad D^a T^{(I)}_{[ab]}(\Phi)=0\,\quad T^{(I)}(\Phi)=0\,,
\end{equation}
and it satisfies
\begin{equation}
    \epsilon_{a}{}^{cd} D_cT_{db}^{(I)}(\Phi)=D_{\langle a}D_{b \rangle }\Phi\,.
\end{equation}
The precise expression of this tensor can be found in~\cite{Compere:2011db}. 

The tensor $\bar\tau_{\tl{ab}}$, although symmetric and trace-fee, it is neither divergenceless nor in the kernel of $(D^2-3)$. However, we can define a new tensor $\bar\tau_{\tl{ab}}-2\tldd{a}{b}\bar\beta$ which satisfies all the required assumptions to be of the form~\eqref{SDT} if the scalar $\bar\beta$ satisfies
\begin{equation}
    (D^2+3)\bar\beta=-\frac{1}{4}\bar\tau\,.
\end{equation}
Therefore, we can write
\begin{equation}
    \bar\tau_{\tl{ab}}=(D_a D_b+\hyp_{ab})\Phi_{\text{sing}}+T_{ab}^{(I)}(\Phi_{\text{super}}^{(1)})+\tldd{a}{b}\Phi_{\text{super}}^{(2)}+2\tldd{a}{b}\bar\beta\,,
\end{equation}
or
\begin{equation}
    \bar\tau_{\tl{ab}}=T_{ab}^{(I)}(\Phi_{\text{super}}^{(1)})+2\tldd{a}{b}\bar\beta\,,
\end{equation}
after the redefinition $\bar\beta\to \bar\beta-\frac{1}{2}\Phi_{\text{sing}}-\frac{1}{2}\Phi^{(2)}_{\text{super}}$. 

This accounts for the decomposition of a regular symmetric, and trace-free tensor $\bar\tau_{\tl{ab}}$. However, solutions like those in the Kerr-Taub-NUT class~\eqref{KerrTN} do not have regular $\bar\tau_{ab}$, instead carrying terms proportional to~\eqref{Backgroundktensor}. These tensors have singularities on the sphere. In order to include Kerr-Taub-NUT in our phase space, we will allow a non-regular term. 
Such terms have the property to not be written in terms of a singular scalar potential but their the curl does. In particular, for Kerr-Taub-NUT, the relevant tensor $k_{(0)ab}$ satisfies $\curl k_{(0)ab}\propto D_{\langle a}D_{b\rangle}\zeta_{0,0}^{\text{even}}$. The other three tensors can be found in Appendix D of~\cite{Compere:2011ve}, and we label them as $k_{(\mu) ab}$. 
We then have the general expansion of a symmetric, trace-free tensor
\begin{equation}\label{onshellSDT}
  \bar\tau_{\tl{ab}}=\sum_{\mu=0,...,4}N_{(\mu)}k_{(\mu)ab}+ T_{ab}^{(I)}(\Phi_{\text{super}}^{(1)})+2\tldd{a}{b}\bar\beta\, =: \bar\wp_{ab} + 2\tldd{a}{b}\bar\beta \,,   
\end{equation}
where we collected the two non-double-gradient tensors in the middle term of Eq.~\eqref{onshellSDT} into $\bar\wp_{ab}$, which is the form we use in the main text~\eqref{AuxiliaryTensorDecomposition}.

\section{Derivation of the residual symmetries and field transformations}\label{app:subsubleadresidualsymm}

Our starting point is a general ansatz for the residual symmetry vector fields, which includes the generator of (super)translations $\xi_{\omega}$, log-(super)translations $\xi_{H}$, and Lorentz transformations $\xi_{\mathcal{Y}}$:
\begin{equation}{\label{genericAKV}}
    \xi=\Xi^{\rho} \partial_\rho+\Xi^a \partial_a\,,
\end{equation}
where $\Xi^{\rho}$ and $\Xi^{a}$ are functions on the boundary hyperboloid taken to have the respective polyhomogeneous expansions,
\begin{subequations}
    \begin{align}
    \Xi^{\rho}&=\omega+ \log\rho H+\frac{1}{\rho}\Big(\bar \Xi+ \log\rho\tilde\Xi+\log^2\rho\dbtilde\Xi\Big)+o
    \pqty{\rho^{-1}},\\
    \Xi^a&=\mathcal{Y}^{a}+\frac{1}{\rho}\Big(\bar{\Xi}^{a}+\log\rho\tilde{\Xi}^{a}+\log^{2}\rho\dbtilde{\Xi}^{a}\Big)+\frac{1}{\rho^{2}}\Big(\bar \Xi^{a(2)}+ \log\rho\tilde\Xi^{a(2)}+ \log^2\rho\dbtilde\Xi^{a(2)}\Big)+o
    \pqty{\rho^{-2}}\,.
\end{align}
\end{subequations}
Our sole initial assumption is that the leading order functions in the above expansion are field independent, i.e.~$\delta\omega = \delta H = \delta\mathcal{Y}^{a} = 0$,\footnote{This assumption is inspired by the desire for the supertranslation and log supertranslation vector fields to form an Abelian algebra. The consistency of this assumption is then captured in \S\ref{appsubsec:ResSymConstraints} by the resulting constraints at $\mathcal{O}(\log\rho)$ and $\mathcal{O}(1)$.} whereas generic field dependence is a priori permitted for all non-leading order functions in the vector field expansion given above.

Note that the vector field associated to Lorentz transformations is already determined and simply takes the form
\begin{align}{\label{vecLorentz}}
   \boxed{ \xi_{\mathcal{Y}} = \mathcal{Y}^{a}\partial_{a}\,.}
\end{align}
Therefore, we focus on deriving the subleading components for the vector fields $\xi_{\omega}$ and $\xi_{H}$~\eqref{AKV} for (super)translations and log-(super)translations used throughout the manuscript.

We start in \S\ref{appsubsec:ResSymConstraints} by providing an exposition of the necessary constraints on~\eqref{genericAKV} that make $\xi_{\omega}$ and $\xi_{H}$ compatible with the polyhomogeneous expansion~\eqref{BSExpansion} of the  metric. Following this derivation, in \S\ref{appsubsec:fieldtransformations}, we provide a solution to the constraints and list the resulting transformation rules of all relevant fields, including subsubleading components.

\subsection{Constraints from the Lie brackets and the field space actions}\label{appsubsec:ResSymConstraints}

We determine $\xi_{\omega}$ and $\xi_{H}$ by imposing several consistency and algebraic conditions. The first consistency condition comes from noticing that there is no leading $\mathcal{O}(1)$ component in our expansion~\eqref{BSExpansion} of the vector part $g_{\rho a}$ of the metric. Preserving this condition we find that
\begin{align}
    D_{a}\omega = \bar{\Xi}_{a}-\tilde{\Xi}_{a}\quad\text{for $\xi_\omega$},\qquad 0 = \bar{\Xi}_{a}-\tilde{\Xi}_{a}\quad\text{for $\xi_H$}\,.
\end{align}
Imposing these constraints on the vector fields leads us to the polyhomogeneous forms
\begin{subequations}\label{polylogwH}
\begin{align}
    \xi_{\omega}&=\Big(\omega+\frac{1}{\rho}(\bar \omega+ \log\rho\tilde \omega+\log^2\rho\dbtilde \omega)+o\pqty{\rho^{-1}}\Big)\partial_{\rho}+\frac{1}{\rho}\bigg(\bar{\omega}^{a}+\log\rho(\bar{\omega}^{a}-D^{a}\omega)+\log^{2}\rho\dbtilde{\omega}^{a}+\nonumber\\
    &\qquad+\frac{1}{\rho}(\bar \omega^{a(2)}+ \log\rho\tilde \omega^{a(2)}+ \log^2\rho\dbtilde \omega^{a(2)})+o\pqty{\rho^{-1}}\bigg)\partial_{a}\,,\\
     \xi_{H}&= \Big(\log\rho H+\frac{1}{\rho}(\bar H+ \log\rho\tilde{H}+\log^2\rho\dbtilde H)+o\pqty{\rho^{-1}}\Big)\partial_{\rho}+\frac{1}{\rho}\bigg((1+\log\rho) \bar{H}^{a}+\log^{2}\rho\dbtilde{H}^{a}+\nonumber\\
    &\qquad+\frac{1}{\rho}(\bar H^{a(2)}+ \log\rho\tilde H^{a(2)}+ \log^2\rho\dbtilde H^{a(2)})+o\pqty{\rho^{-1}}\bigg)\partial_{a}\,.
    \end{align}
\end{subequations}

We determine the rest of the component functions by imposing the following conditions: (i) the (possibly field dependent) Lie bracket of the vector fields is Abelian, and (ii) the resulting field space action is a homomorphism.
The field dependent Lie bracket is given by $\llbracket\xi_{1},\xi_{2}\rrbracket := \comm{\xi_{1}}{\xi_{2}}-\delta_{1}\xi_{2}+\delta_{2}\xi_{1}$.
Requiring that it
vanishes order by order for all combinations of $\xi_{1},\xi_{2}\in\{\xi_{\omega},\xi_{H}\}$ is equivalent to imposing
\begin{align}{\label{vfcond1}}
    \comm{\xi_{1}}{\xi_{2}} \overset{!}{=} \delta_{1}\xi_{2}-\delta_{2}\xi_{1}\,.
\end{align}
The second condition amounts to imposing that $\comm{\delta_{1}}{\delta_{2}} = -\delta_{\llbracket \xi_{1},\xi_{2}\rrbracket}$ for all fields in the polyhomogeneous expansion of the metric. Because of condition~\eqref{vfcond1}, this is equivalent to imposing
\begin{align}{\label{vfcond2}}
    \comm{\delta_{1}}{\delta_{2}} \overset{!}{=}0\,,
\end{align}
for (super)translations and log-(super)translations.

Beginning with the condition on the algebra, we first compute the mixed Lie bracket of $\xi_{\omega}$ and $\xi_{H}$:
\begin{subequations}
\begin{align}
    \comm{\xi_{\omega}}{\xi_{H}}^{\rho} &= \frac{1}{\rho}\Big(H\omega - \bar{H}^{a}D_{a}\omega\Big)+\frac{\log\rho}{\rho}\Big(\bar{\omega}^{a}D_{a}H-\bar{H}^{a}D_{a}\omega\Big)\nonumber\\
    &\quad+\frac{\log^{2}\rho}{\rho}\Big(\bar{\omega}^{a}D_{a}H-D_{a}\omega D^{a}H-\dbtilde{H}^{a}D_{a}\omega\Big)\,,\\
    \comm{\xi_{\omega}}{\xi_{H}}^{a}&=\frac{1}{\rho^{2}}\Big(\bar{\omega}^{c}D_{c}\bar{H}^{a}-\bar{H}^{c}D_{c}\bar{\omega}^{a}\Big)+\frac{\log\rho}{\rho^{2}}\Big(-\bar{H}^{a}\omega + 2\dbtilde{H}^{a}\omega + HD^{a}\omega + 2\bar{\omega}^{c}D_{c}\bar{H}^{a}\nonumber\\
    &\quad-D_{c}\omega D^{c}\bar{H}^{a}-2\bar{H}^{c}D_{c}\bar{\omega}^{a}+\bar{H}^{c}D_{c}D^{a}\omega\Big)\nonumber\\
    &\quad+\frac{\log^{2}\rho}{\rho^{2}}\Big(\bar{\omega}^{a}H-2H\dbtilde{\omega}^{a}-\dbtilde{H}^{a}\omega -HD^{a}\omega+\bar{\omega}^{c}D_{c}\bar{H}^{a}+\dbtilde{\omega}^{c}D_{c}\bar{H}^{a}\nonumber\\
    &\quad-D_{c}\omega D^{c}\bar{H}^{a}-\bar{H}^{c}D_{c}\bar{\omega}^{a}-\dbtilde{H}^{c}D_{c}\bar{\omega}^{a}+\bar{\omega}^{c}D_{c}\dbtilde{H}^{a}-\bar{H}^{c}D_{c}\dbtilde{\omega}^{a}+\bar{H}^{c}D_{c}D^{a}\omega\Big)\,.
\end{align}
\end{subequations}
Then, imposing~\eqref{vfcond1} order by order, we obtain the following constraints from $\comm{\xi_{\omega}}{\xi_{H}}^{\rho}$:
\begin{subequations}
\begin{align}
    \mathcal{O}(\log\rho):&\quad \delta_{\omega}H = 0\,,\\
    \mathcal{O}(1):&\quad \delta_{H}\omega = 0\,,\\
    \mathcal{O}(1/\rho):&\quad \delta_{\omega}\bar{H}-\delta_{H}\bar{\omega} = H\omega - \bar{H}^{a}D_{a}\omega  \,,\\
    \mathcal{O}(\log\rho/\rho):&\quad \delta_{\omega}\tilde{H}-\delta_{H}\tilde{\omega} = \bar{\omega}^{a}D_{a}H-\bar{H}^{a}D_{a}\omega\,,\\
    \mathcal{O}(\log^{2}\rho/\rho):&\quad \delta_{\omega}\dbtilde{H}-\delta_{H}\dbtilde{\omega} =\bar{\omega}^{a}D_{a}H-D_{a}\omega D^{a}H-\dbtilde{H}^{a}D_{a}\omega \,,
\end{align}
\end{subequations}
and the following from $\comm{\xi_{\omega}}{\xi_{H}}^{a}$:
\begin{subequations}
    \begin{align}
         \mathcal{O}(1/\rho):&\quad \delta_{\omega}\bar{H}^{a}-\delta_{H}\bar{\omega}^{a} = 0\,,\\
    \mathcal{O}(\log\rho/\rho):&\quad \delta_{\omega}\bar{H}^{a}-\delta_{H}(\bar{\omega}^{a}-D^{a}\omega) =0\,,\\
    \mathcal{O}(\log^{2}\rho/\rho):&\quad \delta_{\omega}\dbtilde{H}^{a}-\delta_{H}\dbtilde{\omega}^{a} =0\,,\\
    \mathcal{O}(1/\rho^{2}):&\quad \delta_{\omega}\bar{H}^{a(2)}-\delta_{H}\bar{\omega}^{a(2)} =\bar{\omega}^{c}D_{c}\bar{H}^{a}-\bar{H}^{c}D_{c}\bar{\omega}^{a}\,,\\
    \mathcal{O}(\log\rho/\rho^{2}):&\quad \delta_{\omega}\tilde{H}^{a(2)}-\delta_{H}\tilde{\omega}^{a(2)} =-\bar{H}^{a}\omega + 2\dbtilde{H}^{a}\omega + HD^{a}\omega + 2\bar{\omega}^{c}D_{c}\bar{H}^{a}\nonumber\\
    &\quad-D_{c}\omega D^{c}\bar{H}^{a}-2\bar{H}^{c}D_{c}\bar{\omega}^{a}+\bar{H}^{c}D_{c}D^{a}\omega\,,\\
     \mathcal{O}(\log^{2}\rho/\rho^{2}):&\quad \delta_{\omega}\dbtilde{H}^{a(2)}-\delta_{H}\dbtilde{\omega}^{a(2)} =\bar{\omega}^{a}H-2H\dbtilde{\omega}^{a}-\dbtilde{H}^{a}\omega -HD^{a}\omega\nonumber\\
     &\quad+\bar{\omega}^{c}D_{c}\bar{H}^{a}+\dbtilde{\omega}^{c}D_{c}\bar{H}^{a}-D_{c}\omega D^{c}\bar{H}^{a}-\bar{H}^{c}D_{c}\bar{\omega}^{a}\nonumber\\
     &\quad-\dbtilde{H}^{c}D_{c}\bar{\omega}^{a}+\bar{\omega}^{c}D_{c}\dbtilde{H}^{a}-\bar{H}^{c}D_{c}\dbtilde{\omega}^{a}+\bar{H}^{c}D_{c}D^{a}\omega\,.
    \end{align}
\end{subequations}
Similarly, imposing~\eqref{vfcond1} for the Lie bracket of two (super)translation vector fields $\xi_{\omega}$ and $\xi_{\omega'}$, we obtain
\begin{subequations}
\begin{align}
    \mathcal{O}(\log\rho):&\quad \text{No constraint},\\
    \mathcal{O}(1):&\quad \delta_{\omega}\omega' -\delta_{\omega'}\omega = 0 \,,\\
    \mathcal{O}(1/\rho):&\quad \delta_{\omega}\bar{\omega}'-\delta_{\omega'}\bar{\omega} = -\bar{\omega}'^{a}D_{a}\omega+\bar{\omega}^{a}D_{a}\omega'\,,\\
    \mathcal{O}(\log\rho/\rho):&\quad \delta_{\omega}\tilde{\omega}'-\delta_{\omega'}\tilde{\omega} =-\bar{\omega}'^{a}D_{a}\omega+\bar{\omega}^{a}D_{a}\omega' \,,\\
    \mathcal{O}(\log^{2}\rho/\rho):&\quad \delta_{\omega}\dbtilde{\omega}-\delta_{\omega'}\dbtilde{\omega}  =-\dbtilde{\omega}'^{a}D_{a}\omega+\dbtilde{\omega}^{a}D_{a}\omega'\,,
\end{align}
\end{subequations}
from the radial part, and 
\begin{subequations}
    \begin{align}
        \mathcal{O}(1/\rho):&\quad \delta_{\omega}\bar{\omega}'^{a}-\delta_{\omega'}\bar{\omega}^{a} = 0\,,\\
    \mathcal{O}(\log\rho/\rho):&\quad \delta_{\omega}(\bar{\omega}'^{a}-D^{a}\omega')-\delta_{\omega'}(\bar{\omega}^{a}-D^{a}\omega) =0 \,,\\
    \mathcal{O}(\log^{2}\rho/\rho):&\quad \delta_{\omega}\dbtilde{\omega}'^{a}-\delta_{\omega'}\dbtilde{\omega}^{a} =0\,,\\
    \mathcal{O}(1/\rho^{2}):&\quad \delta_{\omega}\bar{\omega}'^{a(2)}-\delta_{\omega'}\bar{\omega}^{a(2)} =D^{a}\omega \omega'-\omega D^{a}\omega' - \bar{\omega}'^{c}D_{c}\bar{\omega}^{a}+\bar{\omega}^{c}D_{c}\bar{\omega}'^{a}\,,\\
    \mathcal{O}(\log\rho/\rho^{2}):&\quad \delta_{\omega}\tilde{\omega}'^{a(2)}-\delta_{\omega'}\tilde{\omega}^{a(2)} = -\bar{\omega}'^{a}\omega+2\dbtilde{\omega}'^{a}\omega+\bar{\omega}^{a}\omega'-2\dbtilde{\omega}^{a}\omega'\nonumber\\
    &\quad-D^{a}\omega\omega'+\omega D^{a}\omega'-2\bar{\omega}'^{c}D_{c}\bar{\omega}^{a}+D^{c}\omega'D_{c}\bar{\omega}^{a}+2\bar{\omega}^{c}D_{c}\bar{\omega}'^{a}\nonumber\\
    &\quad-D^{c}\omega D_{c}\bar{\omega}'^{a}+\bar{\omega}'^{c}D_{c}D^{a}\omega-\bar{\omega}^{c}D_{c}D^{a}\omega'\,,\\
     \mathcal{O}(\log^{2}\rho/\rho^{2}):&\quad \delta_{\omega}\dbtilde{\omega}'^{a(2)}-\delta_{\omega'}\dbtilde{\omega}^{a(2)} =-\dbtilde{\omega}'^{a}\omega+\dbtilde{\omega}^{a}\omega'-\bar{\omega}'^{c}D_{c}\bar{\omega}^{a}-\dbtilde{\omega}'^{c}D_{c}\bar{\omega}^{a}\nonumber\\
     &\quad+D^{c}\omega'D_{c}\bar{\omega}^{a}+\bar{\omega}^{c}D_{c}\bar{\omega}'^{a}+\dbtilde{\omega}^{c}D_{c}\bar{\omega}'^{a}-D^{c}\omega D_{c}\bar{\omega}'^{a}-\bar{\omega}'^{c}D_{c}\dbtilde{\omega}'^{a}+\bar{\omega}^{c}D_{c}\dbtilde{\omega}'^{a}\nonumber\\
     &\quad+\bar{\omega}'^{c}D_{c}D^{a}\omega-D^{c}\omega'D_{c}D^{a}\omega-\bar{\omega}^{c}D_{c}D^{a}\omega'+D^{c}\omega D_{c}D^{a}\omega'\,,
    \end{align}
\end{subequations}
from the tangential part.
Lastly, imposing~\eqref{vfcond1} for the Lie bracket of two log-( super)translations $\xi_{H}$ and  $\xi_{H'}$, we find the constraints
\begin{subequations}
    \begin{align}
         \mathcal{O}(\log\rho):&\quad \delta_{H}H' -\delta_{H'}H = 0\,,\\
    \mathcal{O}(1):&\quad \text{No constraint},\\
    \mathcal{O}(1/\rho):&\quad \delta_{H}\bar{H}'-\delta_{H'}\bar{H} =0\,,\\
    \mathcal{O}(\log\rho/\rho):&\quad \delta_{H}\tilde{H}'-\delta_{H'}\tilde{H} = -\bar{H}'^{a}D_{a}H+\bar{H}^{a}D_{a}H'\,,\\
    \mathcal{O}(\log^{2}\rho/\rho):&\quad \delta_{H}\dbtilde{H}'-\delta_{H'}\dbtilde{H} =-\bar{H}'^{a}D_{a}H+\bar{H}^{a}D_{a}H'\,,
    \end{align}
\end{subequations}
and
\begin{subequations}
    \begin{align}
        \mathcal{O}(1/\rho):&\quad \delta_{H}\bar{H}'^{a}-\delta_{H'}\bar{H}^{a} = 0\,,\\
    \mathcal{O}(\log\rho/\rho):&\quad \delta_{H}\bar{H}'^{a}-\delta_{H'}\bar{H}^{a} = 0\,,\\
    \mathcal{O}(\log^{2}\rho/\rho):&\quad \delta_{H}\dbtilde{H}'^{a}-\delta_{H'}\dbtilde{H}^{a} = 0\,,\\
    \mathcal{O}(1/\rho^{2}):&\quad \delta_{H}\bar{H}'^{a(2)}-\delta_{H'}\bar{H}^{a(2)} = -\bar{H}'^{c}D_{c}\bar{H}^{a}+\bar{H}^{c}D_{c}\bar{H}'^{a}\,,\\
    \mathcal{O}(\log\rho/\rho^{2}):&\quad \delta_{H}\tilde{H}'^{a(2)}-\delta_{H'}\tilde{H}^{a(2)} = 2(-\bar{H}'^{c}D_{c}\bar{H}^{a}+\bar{H}^{c}D_{c}\bar{H}'^{a})\,,\\
     \mathcal{O}(\log^{2}\rho/\rho^{2}):&\quad \delta_{H}\dbtilde{H}'^{a(2)}-\delta_{H'}\dbtilde{H}^{a(2)} = -\bar{H}'^{a}H+\bar{H}^{a}H'-2H'\dbtilde{H}^{a}+2H\dbtilde{H}'^{a}-\bar{H}'^{c}D_{c}\bar{H}^{a}\nonumber\\
     &\quad-\dbtilde{H}'^{c}D_{c}\bar{H}^{a}+\bar{H}^{c}D_{c}\bar{H}'^{a}+\dbtilde{H}^{c}D_{c}\bar{H}'^{a}-\bar{H}'^{c}D_{c}\dbtilde{H}^{a}+\bar{H}^{c}\dbtilde{H}'^{a}\,.
    \end{align}
\end{subequations}
The last ingredient we need to solve for the component functions of the vector fields (and their possible field dependency) is to impose condition~\eqref{vfcond2}. It turns out that, even for generic functions in the vector fields, this condition only imposes a few additional constraints, most notably\footnote{The technique used to determine the transformation laws $\delta_{\omega}\tilde{\sigma}^{(2)},\delta_{H}\tilde{\sigma}^{(2)},\delta_{\omega}\dbtilde{\sigma}^{(2)},\delta_{H}\dbtilde{\sigma}^{(2)}$ of the fields $\tilde{\sigma}^{(2)}$ and $\dbtilde{\sigma}^{(2)}$, along with the transformation laws of all other fields, is presented at the beginning of \S\ref{appsubsec:fieldtransformations}.}
\begin{subequations}
\begin{align}
     \comm{\delta_{\omega}}{\delta_{H}}\tilde{\sigma}^{(2)} &=0 \iff D_{a}\omega(\bar{H}^{a}- D^{a}H-2\dbtilde{H}^{a})= 0\,, \\
    \comm{\delta_{\omega}}{\delta_{H}}\dbtilde{\sigma}^{(2)}&=0\iff -\bar{\omega}^{a}D_{a}H+\frac{1}{2}D_{a}HD^{a}\omega+\frac{1}{2}\bar{H}^{a}D_{a}\omega - \dbtilde{\omega}^{a}D_{a}H=0\,.
\end{align}
\end{subequations}
In the next section, we solve all the constraints above and list the resulting field transformation rules.

\subsection{Solving the constraints and determining the transformations of the fields}\label{appsubsec:fieldtransformations}

The constraints listed in the previous section imply a non-trivial field dependence for the symmetry vector fields. To determine this field dependence, we first determine the transformation properties of the fields under the residual symmetries and then choose the appropriate field dependencies of $\xi_{\omega}$ and $\xi_{H}$ that solve the constraints of \S\ref{appsubsec:ResSymConstraints}.

To extract the transformation rules of the fields, we compare the Lie derivative of the metric $g_{\mu\nu}$ defined by~\eqref{BSLineElement} to the variation of the expansion of the metric $g_{\mu\nu}$ defined by~\eqref{BSExpansion} and read off the field transformation rules order by order, i.e.
\begin{align}
    \mathcal{L}_{\xi_{i}}g_{\mu\nu} \overset{!}{=} \delta_{i} g_{\mu\nu}\,,
\end{align}
for $\xi_{i}\in \{\xi_{\omega}, \xi_{H}, \xi_{\mathcal{Y}}\}$. Note that, under this prescription, fields transform under $\xi_{\mathcal{Y}}$ as boundary tensors, i.e.~with a Lie derivative along $\mathcal{Y}$. For this reason we now focus only on (super)translation and log-(super)translation transformation rules. For the sake of compactness, we will use the generic expansion~\eqref{genericAKV} for a generic symmetry vector field. 

Following this prescription, we find the following transformation rules for the scalar part of the metric $g_{\rho\rho}$:
\begin{subequations}{\label{AKVScalarGenTrans}}
\begin{align}
   \delta_{\xi}\bar{\sigma}&= \mathcal{L}_{\mathcal{Y}}\bar{\sigma}+H\,,\\
   \delta_{\xi}\tilde{\sigma}&=\mathcal{L}_{\mathcal{Y}}\tilde{\sigma}\,,\\
   \delta_{\xi}\bar{\sigma}^{(2)} &=-\bar \Xi+\tilde{\Xi}-\omega(\bar{\sigma} - \tilde{\sigma})+\bar{\sigma}H +\bar{\Xi}^{a}D_{a}\bar{\sigma}+\mathcal{L}_{\mathcal{Y}}\bar{\sigma}^{(2)}\,,\\
   \delta_{\xi}\tilde{\sigma}^{(2)} &=-\tilde{\Xi}+2\dbtilde{\Xi}-\bar{\sigma}H-\tilde{\sigma}(\omega-2H)+\tilde{\Xi}^{a}D_{a}\bar{\sigma}+\bar{\Xi}^{a}D_{a}\tilde{\sigma}+\mathcal{L}_{\mathcal{Y}}\tilde{\sigma}^{(2)}\,,\\
   \delta_{\xi}\dbtilde{\sigma}^{(2)} &=-\dbtilde{\Xi}-H\tilde{\sigma}+\tilde{\Xi}^{c}D_{c}\tilde{\sigma}+\dbtilde{\Xi}^{c}D_{c}\bar{\sigma}+\mathcal{L}_{\mathcal{Y}}\dbtilde{\sigma}^{(2)}\,,
 \end{align}
 \end{subequations}
 for the vector part $g_{\rho a}$:
 \begin{subequations}{\label{AKVVectorGenTrans}}
 \begin{align}
  \delta_{\xi}\bar{\Sigma}_{a}&=\tilde{\Xi}^{(2)}_{a}-2\bar{\Xi}^{(2)}_{a}-\bar{h}_{ca}(\bar{\Xi}^{c}-\tilde{\Xi}^{c})+D_{a}\bar{\Xi}+2\bar{\sigma}D_{a}\omega+\mathcal{L}_{\mathcal{Y}}\bar{\Sigma}_{a}\,,\\
   \delta_{\xi}\tilde{\Sigma}_{a}&=-2\tilde{\Xi}^{(2)}_{a}+2\dbtilde{\Xi}_{a}^{(2)}+\bar{h}_{ca}(2\dbtilde{\Xi}^{c}-\tilde{\Xi}^{c})+\tilde{h}_{ca}(\tilde{\Xi}^{c}-\bar{\Xi}^{c})+D_{a}\tilde{\Xi}+2\bar{\sigma}D_{a}H\nonumber\\
   &\quad+2\tilde{\sigma}D_{a}\omega+\mathcal{L}_{\mathcal{Y}}\tilde{\Sigma}_{a}\,,\\
   \delta_{\xi}\dbtilde{\Sigma}_{a} &=(2\tilde{h}_{ca}-\bar{h}_{ca})\dbtilde{\Xi}^{c}-2\dbtilde{\Xi}^{(2)}_{a}-\tilde{h}_{ca}\tilde{\Xi}^{c}+2\tilde{\sigma}D_{a}H+D_{a}\dbtilde{\Xi}\,,
   \end{align}
   \end{subequations}
   and for the tensor part $g_{ab}$:
\begin{subequations}{\label{AKVTensorGenTrans}}
\begin{align}
   \delta_{\xi}h^{(0)}_{ab} &= \mathcal{L}_{\mathcal{Y}}h^{(0)}_{ab} = 0\,,\\
   \delta_{\xi}\bar{h}_{ab} &=2h^{(0)}_{ab}\omega+2D_{(a}\bar{\Xi}_{b)}+\mathcal{L}_{\mathcal{Y}}\bar{h}_{ab}\,,\\
   \delta_{\xi}\tilde{h}_{ab} &=2h^{(0)}_{ab}H+2D_{(a}\tilde{\Xi}_{b)}+\mathcal{L}_{\mathcal{Y}}\tilde{h}_{ab}\,,\\
   \delta_{\xi}\bar{h}^{(2)}_{ab} &=2h^{(0)}_{ab}\bar{\Xi}+\omega(\bar{h}_{ab} + \tilde{h}_{ab})+\bar{\Xi}^{c}D_{c}\bar{h}_{ab}+2\bar{h}_{c(a}D_{b)}\bar{\Xi}^{c}\nonumber\\
   &\quad+2D_{(a}\bar{\Xi}^{(2)}_{b)}+\mathcal{L}_{\mathcal{Y}}\bar{h}_{ab}^{(2)}\,,\\ 
   \delta_{\xi}\tilde{h}^{(2)}_{ab} &=H(\bar{h}_{ab}+\tilde{h}_{ab})+2h^{(0)}_{ab}\tilde{\Xi}+\tilde{h}_{ab}\omega+\tilde{\Xi}^{c}D_{c}\bar{h}_{ab}+2\tilde{h}_{c(a}D_{b)}\bar{\Xi}^{c}\nonumber\\
   &\quad+\bar{\Xi}^{c}D_{c}\tilde{h}_{ab}+2\bar{h}_{c(a}D_{b)}\tilde{\Xi}^{c}+2D_{(a}\tilde{\Xi}^{(2)}_{b)}+\mathcal{L}_{\mathcal{Y}}\tilde{h}^{(2)}_{ab}\,,\\
   \delta_{\xi}\dbtilde{h}^{(2)}_{ab} &=\tilde{h}_{ab}H+2h^{(0)}_{ab}\dbtilde{\Xi}+\dbtilde{\Xi}^{c}D_{c}\bar{h}_{ab}+\tilde{\Xi}^{c}D_{c}\tilde{h}_{ab}+2\bar{h}_{c(a}D_{b)}\dbtilde{\Xi}^{c}+2D_{(a}\dbtilde{\Xi}^{(2)}_{b)}\nonumber\\
   &\quad+2\tilde{h}_{c(a}D_{b)}\tilde{\Xi}^{c}+\mathcal{L}_{\mathcal{Y}}\dbtilde{h}^{(2)}_{ab}\,.
\end{align}
\end{subequations}

By a careful consideration of the above results, we may choose the functions in the vector fields which solve all of the constraints from \S\ref{appsubsec:ResSymConstraints}. We note that seemingly many choices exist, but we only present here the particular one that we used, which is
\begin{subequations}
\begin{empheq}[box=\widefbox]{align}{\label{vecST}}
\xi_{\omega}&:=\Big(\omega+\frac{1}{\rho}(D^{a}\bar{\sigma}D_{a}\omega - \bar{\sigma}\omega)+o\pqty{\rho^{-1}}\Big)\partial_{\rho}\nonumber\\
&\quad+\frac{1}{\rho}\bigg(D^{a}\omega+\frac{1}{\rho}\pqty{-\frac{1}{2}\bar{h}^{ac}D_{c}\omega-\omega D^{a}\bar{\sigma} }+\nonumber\\
    &\qquad+ \frac{\log\rho}{\rho}(-D^{a}(\bar{\sigma}\omega)-D^{c}\bar{\sigma}D_{c}D^{a}\omega -D_{c}D^{a}\bar{\sigma}D^{c}\omega)+o\pqty{\rho^{-1}}\bigg)\partial_{a}\,.
\end{empheq}
\begin{empheq}[box=\widefbox]{align}{\label{vecLogST}}
   \xi_{H}&:= (\log\rho H+o\pqty{\rho^{-1}})\partial_{\rho}+\frac{1}{\rho}\bigg((1+\log\rho) D^{a}H+\nonumber\\
   &\qquad-\frac{1}{2\rho}\bar{h}^{ac}D_{c}H-\frac{\log\rho}{\rho}\bar{h}^{ac}D_{c}H-\frac{\log^{2}\rho}{2\rho}\tilde{h}^{ac}D_{c}H+o\pqty{\rho^{-1}}\bigg)\partial_{a}\,.
   \end{empheq}
\end{subequations}
up to less l contributions.
Note that, by construction, these vector fields are such that
\begin{equation}
    \llbracket \xi_{\omega},\xi_{H}\rrbracket = 0=\comm{\delta_{\omega}}{\delta_{H}} ,\quad \llbracket \xi_{\omega},\xi_{\omega'}\rrbracket =0 = \comm{\delta_{\omega}}{\delta_{\omega'}} ,\quad \llbracket \xi_{H},\xi_{H'}\rrbracket = 0 = \comm{\delta_{H}}{\delta_{H'}} \,,
\end{equation}
at all relevant orders.

For the sake of completeness, we can list how all the fields in~\eqref{BSExpansion} transform under these vector fields (and $\xi_{\mathcal Y}$). The scalar part transforms as
\begin{subequations}
    \begin{align}
    \delta_{\xi}\bar{\sigma}&=\mathcal L_{\mathcal Y}\bar\sigma+H,\\
   \delta_{\xi}\tilde{\sigma}&=\mathcal L_{\mathcal Y}\tilde\sigma,\\
   \delta_{\xi}\bar{\sigma}^{(2)} &=\mathcal L_{\mathcal Y}\bar\sigma^{(2)}+\omega\tilde{\sigma}+\bar\sigma H+D^a H D_a\bar\sigma,\\
   \delta_{\xi}\tilde{\sigma}^{(2)} &=\mathcal L_{\mathcal Y}\tilde{\sigma}^{(2)}+D^{a}\omega D_{a}\tilde{\sigma}-\omega\tilde{\sigma}-\bar{\sigma}H+2\tilde{\sigma}H+D^{a}H D_{a}\bar{\sigma}+D^{a}H D_{a}\tilde{\sigma},\\
   \delta_{\xi}\dbtilde{\sigma}^{(2)} &=\mathcal L_{\mathcal Y}\dbtilde{\sigma}^{(2)} -H\tilde{\sigma}+D^{c}H D_{c}\tilde{\sigma}.
 \end{align}
\end{subequations}
The vector part transforms as
\begin{subequations}
     \begin{align}
     \delta_{\xi}\bar{\Sigma}_{a}&=\mathcal L_{\mathcal Y}\bar{\Sigma}_{a}-2D_{a}\bar{\sigma}\omega , \\
   \delta_{\xi}\tilde{\Sigma}_{a}&=\mathcal L_{\mathcal Y}\tilde{\Sigma}_{a}+2(D^{a}\bar{\sigma}\omega +\bar{\sigma}D^{a}\omega+D^{c}\bar{\sigma}D_{c}D^{a}\omega+D_{c}D^{a}\bar{\sigma}D^{c}\omega )+\\
   &\quad-\tilde{h}_{ca}D^{a}\omega+2\tilde{\sigma}D_{a}\omega +D^{c}H(\bar{h}_{ac}-\tilde{h}_{ac})+2\bar{\sigma}D_{a}H,\\
   \delta_{\xi}\dbtilde{\Sigma}_{a}&=\mathcal L_{\mathcal Y}\dbtilde{\Sigma}_{a}+2\tilde{\sigma}D_{a}H.
 \end{align}
\end{subequations}
And, finally, the tensor part transforms as
\begin{subequations}
    \begin{align}
   \delta_{\omega}h^{(0)}_{ab} & = 0,\\
   \delta_{\omega}\bar{h}_{ab} &=\mathcal L_{\mathcal Y}\bar h_{ab}+2( D_{a}D_{b}+h^{(0)}_{ab})\omega +2D_{a}D_{b}H,\\
   \delta_{\omega}\tilde{h}_{ab} &=\mathcal L_{\mathcal Y}\tilde h_{ab}+2(D_{a}D_{b}+h^{(0)}_{ab})H,\\
   \delta_{\omega}\bar{h}^{(2)}_{ab} &=\mathcal L_{\mathcal Y}\bar h^{(2)}_{ab}+2h^{(0)}_{ab}(D^{c}\bar{\sigma}D_{c}\omega-\bar{\sigma}\omega)+\omega(\bar{h}_{ab}+\tilde{h}_{ab})+D^{c}\omega D_{c}\bar{h}_{ab}+\bar{h}_{c(a}D_{b)}D^{c}\omega\nonumber\\
   &\quad-D_{(a}\bar{h}_{b)c}D^{c}\omega- 2D_{(a}\omega D_{b)}\bar{\sigma}-2\omega D_{a}D_{b}\bar{\sigma}+D^{c}H D_{c}\bar{h}_{ab}+\bar{h}_{c(a}D_{b)}D^{c}H\nonumber\\
   &\quad-D_{(a|}\bar{h}_{c|b)}D^{c}H,\\
   \delta_{\omega}\tilde{h}^{(2)}_{ab} &=\mathcal L_{\mathcal Y}\tilde h^{(2)}_{ab}+\tilde{h}_{ab}\omega+2\tilde{h}_{c(a}D_{b)}D^{c}\omega+D^{c}\omega D_{c}\tilde{h}_{ab}-2D_{a}D_{b}\bar{\sigma}\omega -2D_{(a} \bar{\sigma}D_{b)}\omega\nonumber\\
   &\quad-2D_{(a|}D^{c}\bar{\sigma}D_{c}D_{|b)}\omega-2D_{(a|}D_{c}D_{|b)}\bar{\sigma}D^{c}\omega+H(\bar{h}_{ab}+\tilde{h}_{ab})\nonumber\\
   &\quad+D^{c}H D_{c}\bar{h}_{ab}+D^{c}H D_{c}\tilde{h}_{ab}+2\tilde{h}_{c(a}D_{b)}D^{c}H-2D_{(a|}\bar{h}_{c|b)}D^{c}H,\\
   \delta_{\omega}\dbtilde{h}^{(2)}_{ab} &=\mathcal L_{\mathcal Y}\dbtilde h^{(2)}_{ab}+\tilde{h}_{ab}H+D^{c}H D_{c}\tilde{h}_{ab}+\tilde{h}_{c(a}D_{b)}D^{c}H-D_{(a|}\tilde{h}_{c|b)}D^{c}H,
   \end{align}\label{lasteq}
\end{subequations}
We note that $\tilde{\sigma}$ is invariant (under $\xi_{\omega}$ and $\xi_{H}$) and transforms homogeneously under $\xi_{\mathcal Y}$, thereby justifying our choice to set it to $0$. As a consequence, when $\tilde{\sigma} = 0$, $\dbtilde{\Sigma}_{a}$ also transforms homogeneously and can be set to $0$.

The above transformation laws for the metric component fields imply the following transformation laws for the tensors $\bar{\tau}_{ab}$ and $\tilde{\tau}_{ab}$ defined by~\eqref{def tau}, and their traces
\begin{alignat}{2}
    \delta_{\xi}\bar{\tau}_{ab} &= \mathcal L_{\mathcal Y}\bar\tau_{ab}-2h^{(0)}_{ab}(D^{2}+2)\omega + 2D_{a}D_{b}\omega,\quad \delta_{\xi}\bar{\tau} &&=\mathcal L_{\mathcal Y}\bar\tau-4(D^{2}+3)\omega,\\
    \delta_{\xi}\tilde{\tau}_{ab} &= \mathcal L_{\mathcal Y}\tilde\tau_{ab}-2h^{(0)}_{ab}(D^{2}+2)H + 2D_{a}D_{b} H,\quad \delta_{\xi}\tilde{\tau} &&=\mathcal L_{\mathcal Y}\tilde\tau -4(D^{2}+3)H.
\end{alignat}

\section{Comparing our results with  Fuentealba-Henneaux-Troessaert}\label{app:FHTDetails}

Fuentealba-Henneaux-Troessaert (FHT)~\cite{Fuentealba:2022xsz} constructed a consistent extension of the
BMS symmetry algebra at spatial infinity by including {logarithmic supertranslations},
while preserving the finiteness of the action and the well-definedness of the canonical
generators in the \textit{Hamiltonian framework}. It is therefore worthwhile to compare, as much as we can, our results obtained in a different framework with theirs. To achieve this,  we employ the methods previously used by Comp\`{e}re-Dehouck (CD)~\cite{Compere:2011ve} (mostly appendix A) in order to make the appropriate comparisons.

Both their work and ours demonstrate the presence of logarithmic supertranslations as asymptotic symmetries, as well as the presence of central charges. However, the precise relation between their transformations and boundary conditions with ours is subtle.

The comparison relies on identifying the correct way of going canonical as being on the $\tau=0$ slice. On this slice, we only need the first few $\tau$-derivatives at most to compare to the canonical formalism, regardless of radial expansions or not. Then, once this is done, we can \textit{further} do radial expansions in a coordinate $r$ which is asymptotically equal to $\rho$. 
The correct radial expansion coordinate on the slice is $r$ as opposed to $\rho$, though they agree on the $\tau=0$ slice. 

With this, we then end up with a comparable expansion to FHT so that  we can compare our classes of metrics, and identify how our phase space transformations and Goldstone fields parametrising the symmetry orbits are related. We can also discuss some of the parity conditions in both scenarios, but the comparison of the charges and their algebra cannot be done so easily. The construction of the symplectic form relies on very different techniques, in particular the renormalization process, which makes it hard to compare term by term. 

Out of our analysis come the following results: first, we can identify a set of field configurations which is common to both FHT and our setup. On this set, we can provide a dictionary between the different parametrisations used. Second, we can identify the set of vector fields generating our residual symmetry algebra as a subset of the one considered by FHT, and can identify the condition $\Sigma_A=\mathcal O(1/\rho)$
in our set of fields \eqref{BSExpansion} as the one restricting the symmetry algebra. Furthermore, the actual algebra of symmetry charges is isomorphic to the one found by FHT (when appropriate parity conditions hold). However, the expressions of the charges, and the central terms in particular, is different from FHT. This fact remains to be analysed in more detail, and likely can be addressed by appropriate reparametrisations, coordinate changes or choice of ambiguities in the symplectic form.

\subsection{Matching of metric components}\label{matching of components}
As appendix A of~\cite{Compere:2011ve} suggests, we do a 0th-order coordinate change from our hyberbolic-type to radial-type coordinates\
\begin{equation}
    \rho = r \sqrt{1-\frac{t^2}{r^2}} = r\bqty{1-\frac{t^2}{2r^2}+\order{\frac{t^4}{r^4}}}\,, \quad \tau = \artanh\pqty{\frac{t}{r}} = \frac{t}{r}+ \order{\frac{t^3}{r^3}}\,,
\end{equation}
with Jacobian
\begin{subequations}
\label{jacobian}
\begin{alignat}{2}
    \frac{\p \rho}{\p r} &= 1 + \frac{t^2}{2 r^2}+ \order{\frac{t^4}{r^4}}\,,
    \quad
    \frac{\p \rho}{\p t} &&= -\frac{t}{r} + \order{\frac{t^3}{r^3}}\,,
\\
    \frac{\p \tau}{\p r} &= -\frac{1}{r}\bqty{\frac{t}{r} + \order{\frac{t^3}{r^3}}}\,,
    \quad
    \frac{\p \tau}{\p t} &&= \frac{1}{r}\bqty{1 + \frac{t^2}{r^2}+\order{\frac{t^4}{r^4}}}\, ,
\end{alignat}
\end{subequations}
and go to the $\tau=0$ slice. In the radial coordinates, this means $\frac{t}{r}\rightarrow 0$, so we expand everything in powers of this fraction, before performing radial expansions. We also  need to organize the expansions in terms of $\frac{1}{r^k}$, which comes after the expansion in $\tau$. To set some notation, in direct analogy to~\cite{Compere:2011ve}, given a field F, we write its Taylor expansion in $\tau$ as
\begin{equation}
    F(\tau) = F + F^\pi \frac{t}{r} + \frac{1}{2} F^{\pi\pi} \frac{t^2}{r^2} +  \order{\frac{t^3}{r^3}},
\end{equation}
where the superscript $\pi$, in analogy to~\cite{Compere:2011ve}, denotes the $\tau$-derivative of the hyperboloid field, evaluated at $\tau=0$.
So, we keep terms up to second order in $\tau$. Next, we consider the coordinate change on the metric: Let $J$ be the Jacobian \eqref{jacobian}, then it can also be expanded
\begin{equation}
    J(\tau)^\alpha_\nu = J^\alpha_\nu + (J^\pi)^\alpha_\nu\frac{t}{r} + \frac{1}{2} (J^{\pi\pi})^\alpha_\nu \frac{t^2}{r^2}\,.
\end{equation}
which is zero in the angular components for $J^\pi,J^{\pi\pi}$, and diagonal in radial/timelike components for $J,J^{\pi\pi}$, but off-diagonal for $J^\pi$.
As for the metric, for $\mu,\nu = t,r$ versus $\alpha,\beta=\tau,\rho$,
\begin{equation}
\begin{aligned}
        g_{\mu\nu}(\tau) &= J(\tau)^\alpha_\mu J(\tau)^\beta_\nu g_{\alpha\beta}(\tau) 
\end{aligned}
\end{equation}
has expansion
\begin{equation}
    g_{\mu\nu} = J^\alpha_\mu J^\beta_\nu g_{\alpha\beta}\,,
\end{equation}
\begin{equation}
    g_{\mu\nu}^{\pi} = 
    J^\alpha_\mu J^\beta_\nu g_{\alpha\beta}^\pi
    +(J^\pi)^\alpha_\mu (J)^\beta_\nu g_{\alpha\beta}
    +(J)^\alpha_\mu (J^\pi)^\beta_\nu g_{\alpha\beta}\,,
\end{equation}
\begin{equation}
\begin{aligned}
        g_{\mu\nu}^{\pi\pi} &= 
    J^\alpha_\mu J^\beta_\nu g_{\alpha\beta}^{\pi\pi}
    +2((J^\pi)^\alpha_\mu (J)^\beta_\nu+(J)^\alpha_\mu (J^\pi)^\beta_\nu) g_{\alpha\beta}^\pi\\
    &\quad+((J^{\pi\pi})^\alpha_\mu (J)^\beta_\nu+(J)^\alpha_\mu (J^{\pi\pi})^\beta_\nu+ 2(J^\pi)^\alpha_\mu (J^\pi)^\beta_\nu) g_{\alpha\beta}\,.
\end{aligned}
\end{equation}
Now assume we have converted to radial coordinates and carried out the $\tau$-expansion. From this point on, we  only care about the expansion in $r$, and can match with FHT in that expansion, particularly on the $t=0$ coordinate slice. Even for the conjugate momenta, we only need the first $t$-derivative as long as we work on the $t=0$ slice, so we can omit $g^{\pi\pi}$ even in principle.  

With this, we  present the first matching dictionary: We can express the metric components and their $\tau$-derivatives in radial coordinates, on the $\tau=0$ slice, through our fields in Beig-Schmidt coordinates. We get explicitly:
\begin{equation}\boxed{
    \begin{gathered}
        g_{rr}= g_{\rho\rho}\,,\quad g_{rt}= \frac{1}{r}g_{\rho\tau}\,,\quad g_{tt}=\frac{1}{r^2}g_{\tau\tau}\,,\\
        g_{rA}=g_{\rho A},,\quad g_{tA}=\frac{1}{r}g_{\tau A}\,,\quad g_{AB}=g_{AB}\,,\\
        \,\\
        g^\pi_{rr}= g^\pi_{\rho\rho}-\frac{2}{r}g_{\rho\tau}\,,\quad g^\pi_{r t} = \frac{1}{r}g^\pi_{\rho\tau} - \frac{1}{r^2} g_{\tau\tau}-g_{\rho\rho}\,,\\
        g^\pi_{tt}= \frac{1}{r^2} g^\pi_{\tau\tau} -\frac{2}{r}g_{\tau\rho}\,,\quad 
        g^\pi_{rA}= g^\pi_{\rho A} - \frac{1}{r}g_{\tau A}\,,\\
        g^\pi_{tA}=\frac{1}{r}g^\pi_{\tau A} - g_{\rho A}\,, \quad g^\pi_{AB}=g^\pi_{AB}\,.
    \end{gathered}
    }
\end{equation}
We can therefore see that the $\frac{t}{r}$ and the $\frac{1}{r}$-expansion is not quite the same, and the order in which we do them matters, as orders in $\frac{1}{r}$ of different components can mix.

Now we can further go and insert our specific notations,
\begin{equation}
    g_{\rho\rho} =\sigma^2,\quad g_{\rho a}=\Sigma_a, \quad g_{ab}=\rho^2 h_{ab} = r^2 h_{ab} + \order{\frac{t^2}{r^2}},
\end{equation}
where each of the objects has a polyhomogeneous expansion
\begin{equation}
\begin{aligned}
        F 
    &= F^{(0)} + \frac{1}{\rho}(\bar{F}+ \log{\rho} \tilde{F})
    + \frac{1}{\rho^2}(\bar{F}^{(2)}+ \log{\rho} \tilde{F}^{(2)}+\log^2\rho \dbtilde{F}^{(2)}) + o\pqty{\frac{1}{\rho^2}}\,, \\
    &=F^{(0)} + \frac{1}{r}(\bar{F}+ \log{r} \tilde{F})
    + \frac{1}{r^2}(\bar{F}^{(2)}+ \log{r} \tilde{F}^{(2)}+\log^2 r \dbtilde{F}^{(2)})+ \order{\frac{t^2}{r^2}} + o\pqty{\frac{1}{\rho^2}}\,,
\end{aligned}
\end{equation}
which we rewrote using $\log\rho = \log r + \mathcal{O}(t^2/r^2) $.

We can finally compare our set of fields to the  FHT's set of fields. We only list the relations up to first order. We always express the fields appearing in FHT's phase space as functions of our fields (hence FHT is on the LHS).

\begin{itemize}
\item{\textbf{$g_{rr}$} component:} Just like FHT, we have $\sigma^{(0)}=1$ with the same leading coefficient. 
\begin{equation}\label{grr}\boxed{
    \bar{h}_{rr}= 2\bar{\sigma}
    }
\end{equation}
A first point of difference is that FHT do not have a separate $\log r/r$-term in $g_{rr}$. But in setting  $\tilde{\sigma}=0$ as we did (see \eqref{tildesigma}), we recover the same phase space variables.
\\

\item{\textbf{$g_{rA}$} component:} There are some differences.
\begin{equation}\boxed{
    \bar \lambda_A = \bar{\Sigma}^{(0)}_A\,,\quad h^{(2)}_{rA}= \bar{\Sigma}_A\,,\quad h^{\log(1)}_{rA}=   \tilde{\Sigma}_A\,,\quad h^{\log(2)}_{rA}= \dbtilde{\Sigma}_{a}
    }
\end{equation}
Here, $\bar\Sigma^{(0)}_a$ would be the leading term in \eqref{SigmaA} that we set to zero from the start.
As argued in Appendix \ref{app:subsubleadresidualsymm}, shortly after \eqref{lasteq} we have set the $h^{\log(2)}_{rA}$ contribution to zero since this term is not affected by the symmetries. 
We also restricted our $\bar{\Sigma}^{(0)}_A$ to be zero, which they do not do. The former is not problematic, as it does not show up in any of the charges. The latter is more important. {{We will see in \S\ref{sec:parity-cond} that $\bar\lambda=0$ is not an invariant condition under the full set of transformations considered by FHT, unlike for our transformations. Our set of transformations are therefore a subset of the FHT transformations which preserve the $\bar \lambda_A=0$ condition.} } Since we also managed to get the full set of asymptotic (logarithmic) symmetries with the appropriate algebra, this difference does not seem to be an issue.
\\
\item{\textbf{$g_{AB}$} component:} These terms match exactly.
\begin{equation}\boxed{
    \bar{g}_{AB}= \hyp_{AB} =: \Omega_{AB}\,,\quad \bar{h}_{AB}= \bar{h}_{AB}\,,\quad \theta_{AB}= \tilde{h}_{AB}
    },
\end{equation}
where $\Omega_{AB}$ is the unit round sphere metric.

\item{Conjugate momenta:}
We only will consider 
$\bar\pi^{rr},\bar\pi $ and $\pi^{rr}_{log} $. These are calculated from the objects $g^\pi_{ij}$, according to the formula (Eq. (4.25) of~\cite{danieliADMFormalismHamiltoniana})
\begin{equation}
    \pi^{ij} = \frac{\sqrt{|g|}}{2N}\bqty{
    (g^{ik}g^{jl}-g^{ij}g^{kl})\partial_t g_{kl}
    +2 g^{ij} g^{kl}D_k N_l - D^i N^j-D^j N^i}\,.
\end{equation}
This has the opposite sign to the convention of CD~\cite{Compere:2011ve}.
To apply this formula, we first need the expressions for Lapse $N$ and shift $N_i$. These are
\begin{gather}
    N_r = g_{tr}=\frac{1}{r}g_{\rho\tau}\,,\quad N_A = g_{tA}=\frac{1}{r}g_{\tau A}\,,\\
    N^2 = g^{ij}N_i N_j - g_{tt} = \frac{1}{r^2}( g^{rr} g_{\rho\tau}^2 + 2g^{rA}g_{\rho\tau}g_{\tau A}+ g^{AB}g_{\tau A}g_{\tau B} - g_{\tau\tau} )\,.
\end{gather}
Next, consider that the $t$-derivative of a function $F(\tau)$ is shifted by one order in $r$:
\begin{equation}
    \partial_t F(\tau) = F^\pi \frac{1}{r} + F^{\pi\pi}\frac{t}{r^2} + o\pqty{\frac{1}{r^2}}.
\end{equation}
So, a time derivative $\partial_t g_{ij}$ always starts at order $\frac{1}{r}$. Here, we are only interested in the leading order, $r^0$ component of the conjugate momenta. For this, we expand the fields to lowest order, mostly replacing them with the 0th order expressions. In particular, we use that corrections to the determinant, lapse as well as the contributions from Christoffel symbols in $D_{r}N_j$, are subleading\footnote{We have $N-1 = \Gamma^r_{rr}= \mathcal{O}(\frac{1}{r^2})$, $\Gamma^r_{rr}= \mathcal{O}(\frac{1}{r^2})$, $\Gamma^A_{rr}= \mathcal{O}(\frac{1}{r^3})$, $\Gamma^A_{rB}= \frac{1}{r}\delta^A_B+\mathcal{O}(\frac{1}{r^2})$, $\Gamma^r_{AB}= - r \Omega_{AB}+ \mathcal{O}(1)$.}.  
Then, for $\pi^{rr}$, we have the leading contribution
\begin{equation}
    \frac{1}{\sqrt{\Omega}}\pi^{rr} = -r^2\,\Omega^{AB}(\partial_t g_{AB} - \bar D_A N_B)
\end{equation}
where $\bar D$ denotes the covariant derivative on the unit sphere. One needs to be careful with the first term:
\begin{equation}
    \partial_t g_{AB} = r^2 \pqty{\partial_t \Omega_{AB} + \frac{1}{r}\partial_t h_{AB}} = r^2 \pqty{0 + \frac{1}{r}\times\frac{1}{r}h^\pi_{AB}} = h^\pi_{AB}\,,
\end{equation}
but this then yields
\begin{equation}
    \frac{1}{\sqrt{\Omega}}\bar\pi^{rr} = -\frac{1}{2}\hyp^{AB} \bar h^\pi_{AB} + \hyp^{AB} \bar D_A \bar h_{\tau B}\,,
    \quad
    \frac{1}{\sqrt{\Omega}}\pi^{rr}_{log} = -\frac{1}{2}\hyp^{AB} \tilde h^\pi_{AB} + \hyp^{AB} \bar D_A \tilde h_{\tau B}\,.
\end{equation}
Here, the log-form works in perfect analogy to the polynomial case. We also find
\begin{equation}
    \frac{1}{\sqrt{\Omega}}\bar\pi = -2\bar\sigma^\pi + \hyp^{AB} \bar h^\pi_{AB} + 2\hyp^{AB} \bar D_A \bar h_{\tau B}\,.
\end{equation}
The conjugate momenta are the same as CD found in their appendix. 
\end{itemize}

\subsection{Parity conditions and residual symmetries}\label{sec:parity-cond}
Parity conditions are essential to glue past null infinity to future null infinity through spatial infinity. From our perspective, this is the only reason why we would consider them. They did not contribute to the definition of our phase space as we focused only on spatial infinity.  Parity conditions are instead  an essential part of the FHT construction  since FHT use them to construct finite objects as variables for their phase space. As such their phase space variables satisfy some parity conditions by definition, unlike ours. In order to continue matching our framework to theirs, we need therefore to see what the parity conditions the FHT field satisfy implies for our fields. 

The FHT parity conditions  are formulated in terms of sphere parity in 2 dimensions. Therefore, we will have to bear in mind that a direct identification of the parity conditions in our covariant set up might be over reaching, since in principle we should also deal with time reversal as a transformation of interest, and have the time components of the fields to consider for this. 

We will explore the correspondence relying only on the FHT parity conditions, both for the phase space variables and the transformations. This approach is reasonable given that, as stated in Appendix A of~\cite{Compere:2011ve}, on the $t= 0$ slice the notions of parity on the hyperboloid and parity of the canonical fields are identical.

\medskip

We recall also that parity conditions can be twisted by diffeomorphisms (parameterized by $U,V, \tilde U, \tilde V, U_A$) with specific asymptotic behavior and parity ($U=U^E+U^O$, $V=V^E+V^O$, $\tilde U=\tilde U^E$, $\tilde V=\tilde V^E, U_A=U_A^O$). This means at the end of the day that these FHT parity conditions (``twisted parity condition'') actually become relations between different fields with appropriate parity, as follows (numbering is from FHT)   
\begin{align}
\bar h_{rr}
&= \bar h_{rr}^{\text{even}} + 2 \tilde U
= \text{even},
\tag{3.7}
\\
\bar\lambda_A
&= \bar D_A U^{\text{even}} + \bar D_A \tilde U - U_A
= \text{odd},
\tag{3.8}
\\
\theta_{AB}
&= 2\bigl(\bar D_A \bar D_B \tilde U + \bar g_{AB} \tilde U\bigr)
= \text{even},
\tag{3.9}
\\
\bar h_{AB}
&= \bar h_{AB}^{\text{even}}
+ 2\Bigl(\bar D_{(A} U_{B)}
+ \bar D_A \bar D_B U^{\text{odd}}
+ \bar g_{AB} U\Bigr).
\tag{3.10}
\end{align}
We also recall that the derivative $\bar D$ on the sphere provides a odd contribution. There is an abuse of notation in (3.7) and (3.10) since the even part of $\bar h_{rr}$ and 
$\bar h_{AB}$ has a contribution respectively from $\bar h_{rr}^{\text{even}}$ and $\bar h_{AB}^{\text{even}}$, but also from the rest depending on respectively $\tilde U$ and  $U$, $U_A$.

Eq.~(3.7), together with \eqref{grr}, shows  that $\bar \sigma$ has to be even, and we split like them
\begin{equation}
    2 \bar\sigma^E = 2 \bar\sigma_0^E + 2 \tilde U^E.
\end{equation}
To avoid the abuse of notation we introduce an index $0$ on the component $\bar \sigma$ corresponding to $\bar h_{rr}^{\text{even}}$.

Next, Eq.~(3.8) must vanish for us, as we set $\bar\lambda_A=0$. This implies that we must restrict
\begin{equation}
    D_A( U^E + \tilde U^E) - U^O_A = 0
\end{equation}
on their fields in order to match correctly. This provides a relation between  $U^O_A $ and $ U^E, \, \tilde U^E$\,.
This then implies the twisted parity conditions
\begin{align}
    \tilde h_{AB} &= 2(D_A D_B + \Omega_{AB})\tilde U^E\,,\\
   \bar h_{AB} &= (\bar h_{AB}^E)_0 + 2(D_A D_B + \Omega_{AB})U + 2 D_A D_B \tilde U^E\,,
\end{align}
or, in other words, the odd part of {$\bar h_{AB}$ must be a pure supertranslation,} and the even part of $\tilde h_{AB}$ a pure logarithmic supertranslation. Meanwhile, the odd part of $\tilde h_{AB}$ must vanish. The even part of $\bar h_{AB}$ may be further split
\begin{equation}
    \bar h_{AB}^E=(\bar h_{AB}^E)_0 + 2(D_A D_B + \Omega_{AB}) U^E + 2 D_A D_B \tilde U^E
\end{equation}
and so, $U^E,\tilde U^E$ parametrise the parts in $\bar h^E_{AB}$ due to supertranslations and log-supertranslations.
These parity conditions, together with $\lambda_A=0=\tilde\sigma$ and the dictionary above, define a comparable subset of our and FHT's set of fields on which we can perform comparisons directly.

Next, we turn to the parity conditions on the transformations, which are listed on p.17 of~\cite{Fuentealba:2022xsz}. 
Unlike our present work, FHT studies an in principle larger class of transformations induced by diffeomorphisms. We express them here in the spherical coordinates $(t,r,x^A)$ used by FHT, and will then express our transformations in terms of theirs. We include their parity to make the identification easier.  

First, the \textit{radial} transformations, which are comparable to ours:
\begin{equation}
    \begin{aligned}
        \xi^r &= W+ \log r \tilde{W}^E+o(1)\,,\\
        \xi^t&=o(1)\,,\\
        \xi^A&=\frac{1}{r}( D^A W + \log r D^A \tilde{W}^E )+o\pqty{\frac{1}{r}}\,,
    \end{aligned}
\end{equation}
which clearly is similar to our radial (log)supertranslations. 
Second, the \textit{timelike} transformations
\begin{equation}
    \begin{aligned}
        \xi^r &= o(1)\,,\\
        \xi^t&=T^E+  \tilde{T}^O+o(1)\,,\\
        \xi^A&=o\pqty{\frac{1}{r}}\,,
    \end{aligned}
\end{equation}
as well as the 'angular supertranslations'
\begin{equation}
    \begin{aligned}
        \xi^r &= o(1)\,,\\
        \xi^t&=o(1)\,,\\
        \xi^A&=\frac{1}{r}(I^A)^O + o\pqty{\frac{1}{r}}\,.
    \end{aligned}
\end{equation}
Our vector fields given in \eqref{AKV}, recalled now for convenience,
\begin{equation}
    \begin{aligned}
    \xi^\rho&=  \omega + \log\rho\,H+\frac1\rho \Big(D^{a}\omega D_{a}\bar{\sigma} -\omega \bar{\sigma}\Big)  +o\pqty{\frac{1}{\rho}}\,,\\
    \xi^a&=\mathcal Y^a + \frac1\rho \Big(D^a\omega +(1+\log\rho)D^{a}H \Big)q+o\pqty{\frac{1}{\rho}}\,,
    \end{aligned}
\end{equation}
can be translated just like the metric components into radial coordinates, via
\begin{equation}
    \xi^t = r \xi^\tau + \frac{t}{r}\xi^\rho + \order{\frac{t^2}{r^2}}\,,
    \qquad
    \xi^r = \xi^\rho + t \xi^\tau + \order{\frac{t^2}{r^2}}\,.
\end{equation}
The result is
\begin{equation}\label{AKVbutRadial}{
    \begin{aligned}
        \xi^r&= \omega+ \log r H +o(1)\,,\\
        \xi^t&=D^\tau\omega + (1+\log r) D^\tau H +o(1)\,,\\
        \xi^A&=\frac{1}{r} \Big(D^A\omega +(1+\log r)D^A H \Big)+o\pqty{\frac{1}{r}}\,.
    \end{aligned}
    }
\end{equation}

This shows that our (log)- supertranslations are in fact combinations of the radial, timelike and angular transformations of FHT, with parameters
\begin{equation}
\begin{gathered}
    W = \omega^{O}+\omega^E\,,\quad 
    T^E= D^\tau \omega^{O}\,,\quad
    \tilde{W}^E=H^{E}\,,\\
    \tilde{T}^O=D^\tau H^{E}\,,\quad (I^A)^O = D^A( H^{E} +\omega^E)\,.
\end{gathered}
\end{equation}
This means in particular that under our transformations, including Lorentz transformations, the condition $\bar\lambda_A=0$ is preserved. 
Moreover, our transformations are singled out as those combinations that preserve that same condition. 

The Goldstone fields $\bar\beta,\tilde \beta$ then parametrise a subspace of that encompassed by the transformations considered by FHT, parametrised by the fields $U,\tilde U$ for radial transformations, $V,\tilde{V}$ for timelike ones and $U_A$ for angular transformations. We can then solve for these fields, using the parity conditions above, in terms of $\bar\beta,\tilde \beta$:
\begin{enumerate}
    \item $\tilde U$: The parity condition on $\tilde h_{AB}$ identifies $\tilde U^E = \tilde \beta^E$ exactly.
    \item $U^O$: Similarly, this is identified with $\bar\beta^O$ up to a linear combination of odd $l=1$ spherical harmonics, corresponding to spatial translations.\footnote{If, instead, the function were even, the ambiguity would not be present, like with $U^E$.} This ambiguity is intrinsic to $U^O$. We choose to simply identify them exactly, and let $\bar\beta$ parametrise these translations as well.
    \item $U^E$: $U^E$ is again determined uniquely from the parity condition on $\bar h^E_{AB}$. \item $(\bar h^E_{AB})_0$: We choose this offset as a function of $\bar h^E_{AB}$ so that we can identify $U^E=\bar\beta^E$ exactly. 
    \item $U_A$: The condition $\lambda_A=0$ lets us solve for $U_A=D_A(U^E+\tilde U^E)$ exactly. 
    \item $\sigma^E_0$: We determine it as $\sigma^E - \tilde\beta^E$.
\end{enumerate}
By pure analogy, then, we can identify the remaining fields $\tilde V, V$ (the latter up to a constant ambiguity, corresponding to time translations) with time derivatives of $\tilde\beta,\bar\beta$. We must however again assume that $D^\tau \bar\beta^E=0$ in order for the parity of $V$ to be correct. Our final matching is then
\begin{equation}
    {U = \bar{\beta}^{O}+\bar{\beta}^{E}\,,\quad  V= D^\tau \bar{\beta}^{O}\,, \quad \tilde{U}=\tilde \beta^{E}\,, \quad \tilde{V}=D^\tau \tilde \beta^{E}\,, \quad  U_A = D_A (\tilde \beta^{E}+\bar \beta^{E})\,.
    } 
\end{equation}
This means in particular that $V$ is even, while $\tilde V, U_A$ are odd. Notice that the fields are therefore subject to the restriction $U_A = D_A(U^E + \tilde{U})$, which ensures $\lambda_A=0$.
In particular, we have the parity conditions on the Goldstones
\begin{equation}
    \tilde\beta = \tilde\beta^E\,,\quad \bar\beta=\bar\beta^O+\bar\beta^E\,, \quad D^\tau \bar\beta^E=0\,.
\end{equation}

In conclusion, upon imposing the parity condition and restricting the FHT solution space by setting $\bar\lambda_A=0$, we recover the same residual symmetries. As we will discuss in the next section, even though the boundary conditions employed are different, the resulting charge algebra is nevertheless isomorphic.

\subsection{Charges}
We could now compare the conserved charge of FHT to ours. Their charge, with the above parameters, can readily be used, as our transformation is a field-independent combination of theirs so that no issues with integrability arise. However, even upon using our boundary conditions, we actually find that the charge apparently disagrees with ours. The same applies to the precise expression of the central terms. This is not an issue, per se. 
First, note we have not made a precise comparison between our symplectic structures. This discrepancy therefore simply notifies us of a corner piece in the structure that distinguishes their phase space from ours. This is a common phenomenon in the presence of boundary Lagrangians that contain derivatives of the fields, and as such is not surprising given the GHY term \cite{Freidel:2020xyx} and finite corner ambiguities that we use.

These details matter in the precise comparison of algebras, see for instance~\cite{Grumiller:2019fmp}, and therefore may need dedicated attention to be resolved properly. We leave this issue for future work.
Nevertheless, we stress the following: as seen already in the main text, we recover (under parity conditions) a symmetry algebra isomorphic to a copy of Poincar\'{e} and higher mode Heisenberg algebras, which matches the result of FHT. Therefore, while the explicit expressions do not match, we can see the resulting algebra does, and this gives us confidence that a more thorough matching of the two phase spaces can be achieved in principle. We leave this more detailed analysis for future work.

\section{Redefinition of Lorentz charge}
\label{App:Redef}

The Poisson bracket of the Lorentz charges with the other charges is 
\begin{equation}
    \{Q_{\mathcal{Y}}, Q_{(\omega,H)} \} = Q_{(\mathcal L_{\mathcal Y}\omega, \mathcal L_{\mathcal Y}H)}\,.
\end{equation}
As already argued and demonstrated in FHT, one can re-organize this algebra in a productive way by providing an alternative Lorentz charge $\tilde{Q}_{\mathcal{Y}}$ which, together with the odd translations $T^O$ (corresponding to the parameters $\omega^O_{\ell < 2}$) forms a Lie algebra ideal $\mathfrak{iso}(1,3)$. The algebra is then of the form
\begin{equation}
\mathfrak{iso}(1,3)_{(\mathcal Y,\omega_{\text{reg}})}\oplus
{\mathfrak{h}_3}_{(\omega_{\text{sing}},H_{\text{reg}})}\oplus {\mathfrak{h}_3}_{(\omega_{\text{super}}^{\text{even}},H_{\text{super}}^{\text{odd}})} \oplus {\mathfrak{h}_3}_{(\omega_{\text{super}}^{\text{odd}},H_{\text{super}}^{\text{even}})}
\end{equation}
which is a direct sum of Heisenberg algebras between supertranslations and log-supertranslations,
regular log-translations and singular translations and a 
Poincar\'{e} algebra.

We note a simple interpretation of these alternative Lorentz charges: they represent a center-of-mass angular momentum which is unchanged by all transformations but the global Poincar\'{e} translations. In essence, the redefinition removes all contributions from degrees of freedom in the orbits of (log-)supertranslations. 

The improved Lorentz charge has the general form (expressed through the field modes),
\begin{equation}
    \tilde{Q}_{\mathcal{Y}}
    =
    Q_{\mathcal{Y}}
    +
    \sum_{\ell\ge 2,\ell'\ge 2,m,m'} \hat\sigma^E_{\ell,m}g^{\mathcal{Y}}_{\ell,m;\ell',m'}\hat\beta^O_{\ell',m'}
    +\sum_{\ell,\ell',m,m'} \hat\sigma^O_{\ell,m}h^{\mathcal{Y}}_{\ell,m;\ell',m'}\hat\beta^E_{\ell',m'}\,.
\end{equation}
We exclude the term corresponding to $\hat\sigma^E_{\ell<2,m}$ because we cannot pair it with $\hat\beta^O_{\ell<2,m}$ (see the brackets~\eqref{centralchargeinmodecomposition}). 
The coefficient matrices $g,h$ are determined from transformation matrices of the Lorentz group, in the specific representations that even/odd solutions of $(D^2+3)F=0$ transform in. 
To demonstrate the logic, we proceed schematically. We reduce the indices $(\ell,m)=a$ to block indices, i.e.
\begin{equation}
    \tilde{Q}_{\mathcal{Y}}
    =
    Q_{\mathcal{Y}}
    +
    \sum_{a,b} \pqty{\hat\sigma^E_{a}g^{\mathcal{Y}}_{a;b}\hat\beta^O_{b}+\hat\sigma^O_{a}h^{\mathcal{Y}}_{a;b}\hat\beta^E_{b}}\,\,.
\end{equation}
Then we write the field mode transformation under the unmodified Lorentz charge as
\begin{equation}
\begin{split}
    &\{ Q_{\mathcal{Y}}, \hat\sigma^E_{a}\} = \sum_b f^E_{ab} \,\hat\sigma^E_{b},
    \qquad
    \{ Q_{\mathcal{Y}}, \hat\beta^O_{a}\} = \sum_{b} f^O_{ab} \,\hat\beta^O_{b}\,,\\
    &\{ Q_{\mathcal{Y}}, \hat\sigma^O_{a}\} = \sum_b f^O_{ab} \,\hat\sigma^O_{b},
    \qquad
    \{ Q_{\mathcal{Y}}, \hat\beta^E_{a}\} = \sum_{b} f^E_{ab} \,\hat\beta^E_{b}\,
\end{split}
\end{equation}
with some representation matrices $f^{E|O}$ (which depend on $\mathcal Y$) whose specific form we do not need. However, consistency of the charge algebra
\begin{equation}
\begin{split}
\{\hat\sigma^O_{\ell,m},\hat\beta^E_{\ell',m'}\}=-\frac{\kappa^2}{2 \mathcal C_\ell}\delta_{\ell,\ell'}\delta_{m,-m'} &\Longleftrightarrow \{\hat\sigma^O_{a},\hat\beta^E_{b}\}\equiv -K_{ab} \,,\\
\{\hat\sigma^E_{\ell\geq2,m},\hat\beta^O_{\ell'\geq2,m'}\}=\frac{\kappa^2}{2 \mathcal C_\ell}\delta_{\ell,\ell'}\delta_{m,-m'}&\Longleftrightarrow \{\hat\sigma^E_{a},\hat\beta^O_{b}\}\equiv K_{ab}\,,
\end{split}
\end{equation}
informs us that these representation matrices must be related to each other by
\begin{equation}\label{consistencyJacobi}
    \sum_c\pqty{f^E_{ac}K_{cb}+f^O_{bc}K_{ac}}=0\Longleftrightarrow\mathcal C_{\ell'}f^{E}_{\ell', -m';\ell,m}+\mathcal C_{\ell}f^{O}_{\ell, -m;\ell',m'}=0\,.
\end{equation}

Now, we try to  schematically determine the form of the $g$ matrix by requiring that the action of the modified Lorentz charges leave $\hat\sigma^E_a$ and $\hat\beta^O_a$ invariant. The explicit calculation gives, respectively,
\begin{equation}
    f^E_{ab}-\sum_c g^{\mathcal Y}_{bc}K_{ac}=0= f^O_{ab}+\sum_c K_{ca}g^{\mathcal Y}_{cb}.
\end{equation}
These equations have a unique consistent solution, given schematically by
\begin{equation}
    g^{\mathcal Y} =- K^{-1} f^O=(f^{E})^T K^{-1}\,,
\end{equation}
or, more concretely, by
\begin{equation}
    g^\mathcal{Y}_{\ell, m, \ell', m'}
    =-\frac{2}{\kappa^2}\mathcal C_\ell f^{O}_{\ell,-m;\ell',m'}=\frac{2}{\kappa^2} \mathcal C_{\ell'} f^{E}_{\ell',-m';\ell,m}\,,
\end{equation}
which is clearly possible thanks to the consistency condition~\eqref{consistencyJacobi}.
The intermediate result is that the charge
\begin{equation}
    Q_{\mathcal{Y}}
    -
    \sum_{a,b} \hat\sigma^E_{a}K^{-1}_{ab} \{ Q_{\mathcal{Y}}, \hat\beta^O_{b}\}
\end{equation}
commutes with $\hat\sigma^E,\hat\beta^O$. Repeating the same procedure to determine $h^{\mathcal Y}$ gives
\begin{equation}
    h^{\mathcal Y}=-K^{-1}f^E,
\end{equation}
which gets us to the final schematic version of the modified Lorentz charge
\begin{equation}
    \tilde{Q}_{\mathcal{Y}}
    =Q_{\mathcal{Y}}
    -\sum_{a,b}(\hat \sigma^E_a K^{-1}_{ab}\{ Q_{\mathcal{Y}}, \hat\beta^O_{b}\}+\hat \sigma^O_a K^{-1}_{ab}\{ Q_{\mathcal{Y}}, \hat\beta^E_{b}\})\,.
\end{equation}
Expanding the indices and recalling that we exclude $\hat\sigma^E_{\ell<2,m}$, we can write
\begin{equation}
\begin{split}
        \tilde{Q}_{\mathcal{Y}}
    &=
    Q_{\mathcal{Y}}\,+\\
    &\quad-
    \frac{2}{\kappa^2}\sum_{\ell\ge 2, \ell'\geq 2, m, m'} \pqty{
    \hat\sigma^E_{\ell,m}
    \mathcal{C}_{ \ell} f^{O}_{\ell,-m; \ell', m'}
    \hat\beta^O_{\ell',m'}+
    \hat\sigma^O_{\ell,m}
    \mathcal{C}_{ \ell} f^{E}_{\ell,-m; \ell', m'}
    \hat\beta^E_{\ell',m'}}+\\
    &\quad-
   \frac{2}{\kappa^2} \sum_{\ell<2,\ell'<2, m,m'} 
    \hat\sigma^O_{\ell,m}
    \mathcal{C}_{\ell} f^{E}_{\ell,-m;\ell', m'} \hat\beta^E_{\ell',m'}+\\
    &\quad-
   \frac{2}{\kappa^2} \sum_{\ell\ge  2,\ell'< 2, m,m'} 
    \hat\sigma^O_{\ell,m}
    \mathcal{C}_{ \ell} f^{E}_{\ell,-m; \ell', m'}
    \hat\beta^E_{\ell',m'}\,.
\end{split}
\end{equation}

The only mixing between low ($\ell<2$) and high ($\ell\geq 2$) modes in the modified Lorentz charge is between low $\hat\sigma^O$ and high $\hat\beta^E$ in the third line, due to the non-vanishing of $f^E_{\ell<2,m;\ell'\ge 2,m'}$. When typical parity conditions are imposed, $\hat\sigma^O$ and $\hat\beta^E$ are removed from the phase space, removing this type of mixing as well. In our setup though, it must be included to properly decouple the Lorentz sector from (log)supertranslation.

The resulting algebra is then
\begin{equation}
    \{\tilde{Q}_{\mathcal{Y}}, \hat\sigma^{E|O}_{\ell\geq 2,m}  \}=\{\tilde{Q}_{\mathcal{Y}}, \hat\beta^{E|O}_{\ell\geq 2,m}  \}= \{\tilde{Q}_{\mathcal{Y}}, \hat\sigma^{O}_{\ell<2 ,m}  \}=\{\tilde{Q}_{\mathcal{Y}}, \hat\beta^{E}_{\ell<2 ,m}  \}=0\,,
\end{equation}
while the action on the remaining fields is left intact
\begin{equation}
    \{\tilde{Q}_{\mathcal{Y}},\tilde{Q}_{\mathcal{Y}'}\}= 
    \tilde{Q}_{[\mathcal{Y},\mathcal{Y}']}\,
    \qquad 
    \{\tilde{Q}_{\mathcal{Y}}, \hat\sigma^{E}_{\ell<2 ,m}  \}=\sum_{\ell' <2,m'} f^{E}_{\ell, m; \ell', m'}\hat\sigma^{E}_{\ell',m'}\,.
\end{equation}
The last two relations tell us that the $\tilde{Q}_{\mathcal Y}$ still form a representation of the Lorentz group with the semidirect action on odd (regular) translations which is expected to form the Poincar\'{e} algebra.

In this construction, we did not need to know the precise expression of the matrices $f^E,f^O$, even if it can be worked out from the relations~\eqref{oddLorentzalgebra} and~\eqref{evenLorentzalgebra}. What we do need, however, is the fact that the mode operators $\hat{\sigma},\hat{\beta}$ transform (parity) oppositely from the parameters $\omega,H$ of the charges - this is due to e.g. the pairing of even and odd modes of $\omega$ and $\bar\sigma$ in the charge. The only property that we used in the end is the specific, parity-dependent, mixing pattern of low and high modes
\begin{equation}
    f^O_{\ell\geq 2,m;\ell'<2,m'}=0=f^E_{\ell<2,m;\ell'\geq 2,m'}\,,
\end{equation}
which states that boosting a high odd mode $\hat{\sigma},\hat{\beta}$ does not produce low odd modes, and boosting a low even mode does not produce high even modes. This property is a consequence of the particular Lorentz representation that the charges $Q_{\omega,H}$ and their parameters $\omega,H$ transform under, see \eqref{oddLorentzalgebra},\eqref{evenLorentzalgebra}.

\section{Identities on the three dimensional hyperboloid}\label{app:idhyp}

We report here, for convenience of the reader, a list of useful identities that hold on the unit hyperboloid $\mathcal H$. We use the convention $[D_a,D_b]V^c=R_{bad}{}^c V^d$, where the Riemann tensor is
\begin{equation}
    R_{abcd}=\hyp_{ac}\hyp_{bd}-\hyp_{ad}\hyp_{bc}.
\end{equation}
Therefore, the following identities hold for generic scalar $S$, vector $V$ and tensors $A$ on $\mathcal H$
\begin{gather}
    [D_a,D_b]V^c=\delta^c_a V_b-\delta^c_b V_a\,,\\
    D^a \tldd{a}{b}S =\frac{2}{3}D_b(D^2+3)S\,,\\
    \tluu{a}{b} \tldd{a}{b}S =\frac{2}{3}D^2(D^2+3)S\,,\\
    \tldd{a}{b}(D^2+3)S=(D^2-3)\tldd{a}{b}S\,,\\
    \curl \tldd{a}{b} S=\frac{1}{3}\epsilon_{abc}D^c(D^2+3)S\,, \\
    D^a (\curl A)_{ab} =-\epsilon_{b}{}^{ij}A_{ij}\,,\\
    D^b (\curl A)_{ab} = -\epsilon_{abi}D^iD_j A^{bj}-\epsilon_{a}{}^{ij}A_{ij}\,,\\
    (\curl\curl A)_{ab}=(D^2-2)A_{ab}-A_{ba}+A \hyp_{ab}-D_a D_i A^i{}_b\,,
\end{gather}
where $(\curl\cdot)_{ab}\equiv \epsilon_a{}^{cd}D_c\cdot{}_{db}$ is the curl on the hyperboloid and $A=h_{(0)}^{ab}A_{ab}$. 
For two scalars $f,g$ we also have 
\begin{equation}
    D^a D_{\langle c}D_{d\rangle}f\, D^{\langle c}D^{d\rangle}g 
    =
    D^a f\, \frac{2}{3}D^2(D^2+3)g
    - \frac{2}{3}f D^a (D^2+3)g
    + D_c W^{ca}\,,
\end{equation}
\begin{equation}
    W^{ca} := D_d D^a f D^{\langle c}D^{d\rangle}g 
        -  D^a f D_d D^{\langle c}D^{d\rangle}g
        + f D^{\langle c}D^{a \rangle} g  \,.
\end{equation}

\bibliography{references.bib}

@article{ChruscielMacCallumSingleton1994,
  author  = {Chru{\'s}ciel, Piotr T. and MacCallum, Malcolm A. H. and Singleton, David B.},
  title   = {Existence of polyhomogeneous expansions at null infinity},
  journal = {Classical and Quantum Gravity},
  volume  = {11},
  number  = {7},
  pages   = {1601--1620},
  year    = {1994}
}

@article{ChruscielDelay2000,
  author  = {Chru{\'s}ciel, Piotr T. and Delay, Erwann},
  title   = {Existence of non-trivial, vacuum, asymptotically simple spacetimes},
  journal = {Journal of Mathematical Physics},
  volume  = {41},
  number  = {10},
  pages   = {6716--6732},
  year    = {2000}
}

@article{Chrusciel1992Asymptotic,
  author  = {Chru{\'s}ciel, Piotr T.},
  title   = {On the asymptotic structure of gravitational fields},
  journal = {Proceedings of the Royal Society of London. Series A},
  volume  = {436},
  number  = {1896},
  pages   = {299--316},
  year    = {1992}
}

@article{Nguyen:2021ydb,
    author = "Nguyen, Kevin and Salzer, Jakob",
    title = "{Celestial IR divergences and the effective action of supertranslation modes}",
    eprint = "2105.10526",
    archivePrefix = "arXiv",
    primaryClass = "hep-th",
    doi = "10.1007/JHEP09(2021)144",
    journal = "JHEP",
    volume = "09",
    pages = "144",
    year = "2021"
}

@article{Compere:2011db,
    author = "Compere, Geoffrey and Dehouck, Francois and Virmani, Amitabh",
    title = "{On Asymptotic Flatness and Lorentz Charges}",
    eprint = "1103.4078",
    archivePrefix = "arXiv",
    primaryClass = "gr-qc",
    reportNumber = "ULB-TH-11-08",
    doi = "10.1088/0264-9381/28/14/145007",
    journal = "Class. Quant. Grav.",
    volume = "28",
    pages = "145007",
    year = "2011"
}

@article{Virmani:2011gh,
    author = "Virmani, Amitabh",
    title = "{Asymptotic Flatness, Taub-NUT, and Variational Principle}",
    eprint = "1106.4372",
    archivePrefix = "arXiv",
    primaryClass = "hep-th",
    reportNumber = "ULB-TH-11-15",
    doi = "10.1103/PhysRevD.84.064034",
    journal = "Phys. Rev. D",
    volume = "84",
    pages = "064034",
    year = "2011"
}

@article{Javadinezhad:2022hhl,
    author = "Javadinezhad, Reza and Kol, Uri and Porrati, Massimo",
    title = "{Supertranslation-invariant dressed Lorentz charges}",
    eprint = "2202.03442",
    archivePrefix = "arXiv",
    primaryClass = "hep-th",
    doi = "10.1007/JHEP04(2022)069",
    journal = "JHEP",
    volume = "04",
    pages = "069",
    year = "2022"
}

@article{Fuentealba:2023hzq,
    author = "Fuentealba, Oscar and Henneaux, Marc",
    title = "{Simplifying (super-)BMS algebras}",
    eprint = "2309.07600",
    archivePrefix = "arXiv",
    primaryClass = "hep-th",
    doi = "10.1007/JHEP11(2023)108",
    journal = "JHEP",
    volume = "11",
    pages = "108",
    year = "2023"
}

@article{Henneaux:2018hdj,
    author = "Henneaux, Marc and Troessaert, C{\'e}dric",
    title = "{Hamiltonian structure and asymptotic symmetries of the Einstein-Maxwell system at spatial infinity}",
    eprint = "1805.11288",
    archivePrefix = "arXiv",
    primaryClass = "gr-qc",
    doi = "10.1007/JHEP07(2018)171",
    journal = "JHEP",
    volume = "07",
    pages = "171",
    year = "2018"
}

@article{Compere:2019gft,
    author = "Comp{\`e}re, Geoffrey and Oliveri, Roberto and Seraj, Ali",
    title = "{The Poincar{\'e} and BMS flux-balance laws with application to binary systems}",
    eprint = "1912.03164",
    archivePrefix = "arXiv",
    primaryClass = "gr-qc",
    doi = "10.1007/JHEP10(2020)116",
    journal = "JHEP",
    volume = "10",
    pages = "116",
    year = "2020",
    note = "[Erratum: JHEP 06, 045 (2024)]"
}

@article{Raclariu:2021zjz,
    author = "Raclariu, Ana-Maria",
    title = "{Lectures on Celestial Holography}",
    eprint = "2107.02075",
    archivePrefix = "arXiv",
    primaryClass = "hep-th",
    month = "7",
    year = "2021"
}

@article{Ruzziconi:2026bix,
    author = "Ruzziconi, Romain",
    title = "{Carrollian Physics and Holography}",
    eprint = "2602.02644",
    archivePrefix = "arXiv",
    primaryClass = "hep-th",
    month = "2",
    year = "2026"
}

@article{Fuentealba:2025ekj,
    author = "Fuentealba, Oscar and Henneaux, Marc",
    title = "{Logarithmic angle-dependent gauge transformations at null infinity}",
    eprint = "2504.05385",
    archivePrefix = "arXiv",
    primaryClass = "hep-th",
    doi = "10.1007/JHEP07(2025)112",
    journal = "JHEP",
    volume = "07",
    pages = "112",
    year = "2025"
}

@article{Javadinezhad:2023mtp,
    author = "Javadinezhad, Reza and Porrati, Massimo",
    title = "{Three Puzzles with Covariance and Supertranslation Invariance of Angular Momentum Flux and Their Solutions}",
    eprint = "2312.02458",
    archivePrefix = "arXiv",
    primaryClass = "hep-th",
    doi = "10.1103/PhysRevLett.132.151604",
    journal = "Phys. Rev. Lett.",
    volume = "132",
    number = "15",
    pages = "151604",
    year = "2024"
}

@article{McNees:2024iyu,
    author = "McNees, Robert and Zwikel, C\'eline",
    title = "{The symplectic potential for leaky boundaries}",
    eprint = "2408.13203",
    archivePrefix = "arXiv",
    primaryClass = "hep-th",
    doi = "10.1007/JHEP01(2025)049",
    journal = "JHEP",
    volume = "01",
    pages = "049",
    year = "2025"
}

@article{Fuentealba:2024lll,
    author = "Fuentealba, Oscar and Henneaux, Marc",
    title = "{Logarithmic matching between past infinity and future infinity: The massless scalar field in Minkowski space}",
    eprint = "2412.05088",
    archivePrefix = "arXiv",
    primaryClass = "gr-qc",
    doi = "10.1007/JHEP03(2025)081",
    journal = "JHEP",
    volume = "03",
    pages = "081",
    year = "2025"
}

@article{Capone:2022gme,
    author = "Capone, Federico and Nguyen, Kevin and Parisini, Enrico",
    title = "{Charge and antipodal matching across spatial infinity}",
    eprint = "2204.06571",
    archivePrefix = "arXiv",
    primaryClass = "hep-th",
    doi = "10.21468/SciPostPhys.14.2.014",
    journal = "SciPost Phys.",
    volume = "14",
    number = "2",
    pages = "014",
    year = "2023"
}

@article{Prabhu:2019fsp,
    author = "Prabhu, Kartik",
    title = "{Conservation of asymptotic charges from past to future null infinity: Supermomentum in general relativity}",
    eprint = "1902.08200",
    archivePrefix = "arXiv",
    primaryClass = "gr-qc",
    doi = "10.1007/JHEP03(2019)148",
    journal = "JHEP",
    volume = "03",
    pages = "148",
    year = "2019"
}

@article{danieliADMFormalismHamiltoniana,
  title = {{{ADM}} Formalism: A {{Hamiltonian}} Approach to {{General Relativity}}},
  author = {Danieli, Alessandro and Molinari, Luca Guido},
  langid = {english}
}

@article{Henneaux:2018cst,
    author = "Henneaux, Marc and Troessaert, C\'edric",
    title = "{BMS Group at Spatial Infinity: the Hamiltonian (ADM) approach}",
    eprint = "1801.03718",
    archivePrefix = "arXiv",
    primaryClass = "gr-qc",
    doi = "10.1007/JHEP03(2018)147",
    journal = "JHEP",
    volume = "03",
    pages = "147",
    year = "2018"
}

@article{Compere:2023qoa,
    author = "Comp\`ere, Geoffrey and Gralla, Samuel E. and Wei, Hongji",
    title = "{An asymptotic framework for gravitational scattering}",
    eprint = "2303.17124",
    archivePrefix = "arXiv",
    primaryClass = "gr-qc",
    doi = "10.1088/1361-6382/acf5c1",
    journal = "Class. Quant. Grav.",
    volume = "40",
    number = "20",
    pages = "205018",
    year = "2023"
}

@article{Compere:2017knf,
    author = "Comp\`ere, Geoffrey and Fiorucci, Adrien",
    title = "{Asymptotically flat spacetimes with BMS$_3$ symmetry}",
    eprint = "1705.06217",
    archivePrefix = "arXiv",
    primaryClass = "hep-th",
    doi = "10.1088/1361-6382/aa8aad",
    journal = "Class. Quant. Grav.",
    volume = "34",
    number = "20",
    pages = "204002",
    year = "2017"
}

@article{McNees:2025acf,
    author = "McNees, Robert and Zwikel, C{\'e}line",
    title = "{(Anti)-de Sitter with leaky boundaries and corners}",
    eprint = "2512.03170",
    archivePrefix = "arXiv",
    primaryClass = "hep-th",
    month = "12",
    year = "2025"
}

@article{Beig:1982ifu,
    author = "Beig, R. and Schmidt, B. G.",
    title = "{Einstein's equations near spatial infinity}",
    doi = "10.1007/BF01211056",
    journal = "Commun. Math. Phys.",
    volume = "87",
    number = "1",
    pages = "65--80",
    year = "1982"
}

@article{Geiller:2025dqe,
    author = "Geiller, Marc and Mao, Pujian and Vincenti, Antoine",
    title = "{Twisting asymptotically-flat spacetimes}",
    eprint = "2511.13814",
    archivePrefix = "arXiv",
    primaryClass = "gr-qc",
    month = "11",
    year = "2025"
}

@article{Mishra:2025nmd,
    author = "Mishra, Sharad and Banerjee, Kinjal and Bhattacharyya, Jishnu",
    title = "{Asymptotic symmetries at spatial infinity}",
    eprint = "2504.19910",
    archivePrefix = "arXiv",
    primaryClass = "gr-qc",
    doi = "10.1088/1361-6382/ae1096",
    journal = "Class. Quant. Grav.",
    volume = "42",
    number = "21",
    pages = "215009",
    year = "2025"
}

@article{Andrade:2006pg,
    author = "Andrade, Tomas and Banados, Maximo and Rojas, Francisco",
    title = "{Variational Methods in AdS/CFT}",
    eprint = "hep-th/0612150",
    archivePrefix = "arXiv",
    doi = "10.1103/PhysRevD.75.065013",
    journal = "Phys. Rev. D",
    volume = "75",
    pages = "065013",
    year = "2007"
}

@article{Kehrberger:2024aak,
    author = "Kehrberger, Leonhard and Masaood, Hamed",
    title = "{The Case Against Smooth Null Infinity V: Early-Time Asymptotics of Linearised Gravity Around Schwarzschild for Fixed Spherical Harmonic Modes}",
    eprint = "2401.04179",
    archivePrefix = "arXiv",
    primaryClass = "gr-qc",
    month = "1",
    year = "2024"
}

@article{Geiller:2024ryw,
    author = "Geiller, Marc and Laddha, Alok and Zwikel, C{\'e}line",
    title = "{Symmetries of the gravitational scattering in the absence of peeling}",
    eprint = "2407.07978",
    archivePrefix = "arXiv",
    primaryClass = "gr-qc",
    doi = "10.1007/JHEP12(2024)081",
    journal = "JHEP",
    volume = "12",
    pages = "081",
    year = "2024"
}

@article{Geiller:2024amx,
    author = "Geiller, Marc and Zwikel, C\'eline",
    title = "{The partial Bondi gauge: Gauge fixings and asymptotic charges}",
    eprint = "2401.09540",
    archivePrefix = "arXiv",
    primaryClass = "hep-th",
    doi = "10.21468/SciPostPhys.16.3.076",
    journal = "SciPost Phys.",
    volume = "16",
    pages = "076",
    year = "2024"
}

@article{Freidel:2021fxf,
	archiveprefix = {arXiv},
	author = {Freidel, Laurent and Oliveri, Roberto and Pranzetti, Daniele and Speziale, Simone},
	doi = {10.1007/JHEP07(2021)170},
	eprint = {2104.05793},
	journal = {JHEP},
	pages = {170},
	primaryclass = {hep-th},
	title = {{The Weyl BMS group and Einstein\textquoteright{}s equations}},
	volume = {07},
	year = {2021}}

@article{Fiorucci:2024ndw,
    author = "Fiorucci, Adrien and Matulich, Javier and Ruzziconi, Romain",
    title = "{Superrotations at spacelike infinity}",
    eprint = "2404.02197",
    archivePrefix = "arXiv",
    primaryClass = "hep-th",
    doi = "10.1103/PhysRevD.110.L061502",
    journal = "Phys. Rev. D",
    volume = "110",
    number = "6",
    pages = "L061502",
    year = "2024"
}

@article{McNees:2023tus,
	archiveprefix = {arXiv},
	author = {McNees, Robert and Zwikel, C\'eline},
	doi = {10.1007/JHEP08(2023)154},
	eprint = {2306.16451},
	journal = {JHEP},
	pages = {154},
	primaryclass = {hep-th},
	title = {{Finite charges from the bulk action}},
	volume = {08},
	year = {2023}}

@article{Barnich:2001jy,
	archiveprefix = {arXiv},
	author = {Barnich, Glenn and Brandt, Friedemann},
	doi = {10.1016/S0550-3213(02)00251-1},
	eprint = {hep-th/0111246},
	journal = {Nucl. Phys. B},
	pages = {3--82},
	reportnumber = {ULB-TH-01-19, MPI-MIS-94-2001},
	title = {{Covariant theory of asymptotic symmetries, conservation laws and central charges}},
	volume = {633},
	year = {2002}}

@article{Fuentealba:2022xsz,
	archiveprefix = {arXiv},
	author = {Fuentealba, Oscar and Henneaux, Marc and Troessaert, C\'edric},
	doi = {10.1007/JHEP02(2023)248},
	eprint = {2211.10941},
	journal = {JHEP},
	pages = {248},
	primaryclass = {hep-th},
	title = {{Logarithmic supertranslations and supertranslation-invariant Lorentz charges}},
	volume = {02},
	year = {2023}}

@article{Geiller:2022vto,
	archiveprefix = {arXiv},
	author = {Geiller, Marc and Zwikel, C\'eline},
	doi = {10.21468/SciPostPhys.13.5.108},
	eprint = {2205.11401},
	journal = {SciPost Phys.},
	pages = {108},
	primaryclass = {hep-th},
	title = {{The partial Bondi gauge: Further enlarging the asymptotic structure of gravity}},
	volume = {13},
	year = {2022}}

@article{Kehrberger:2021vhp,
	archiveprefix = {arXiv},
	author = {Kehrberger, Leonhard M. A.},
	eprint = {2105.08084},
	month = {5},
	primaryclass = {gr-qc},
	title = {{The Case Against Smooth Null Infinity II: A Logarithmically Modified Price's Law}},
	year = {2021}}

@article{Kehrberger:2021azo,
	archiveprefix = {arXiv},
	author = {Kehrberger, Leonhard M. A.},
	doi = {10.1007/s40818-022-00129-2},
	eprint = {2106.00035},
	journal = {Ann. PDE},
	number = {2},
	pages = {12},
	primaryclass = {gr-qc},
	title = {{The Case Against Smooth Null Infinity III: Early-Time Asymptotics for Higher $\ell $-Modes of Linear Waves on a Schwarzschild Background}},
	volume = {8},
	year = {2022}}

@article{Gajic:2022pst,
	archiveprefix = {arXiv},
	author = {Gajic, Dejan and Kehrberger, Leonhard M. A.},
	doi = {10.1088/1361-6382/ac8863},
	eprint = {2202.04093},
	journal = {Class. Quant. Grav.},
	number = {19},
	pages = {195006},
	primaryclass = {gr-qc},
	title = {{On the relation between asymptotic charges, the failure of peeling and late-time tails}},
	volume = {39},
	year = {2022}}

@article{Ashtekar:1985aa,
	author = {Ashtekar, Abhay},
	da = {1985/04/01},
	doi = {10.1007/BF01889278},
	id = {Ashtekar1985},
	isbn = {1572-9516},
	journal = {Foundations of Physics},
	number = {4},
	pages = {419--431},
	title = {Logarithmic ambiguities in the description of spatial infinity},
	ty = {JOUR},
	url = {https://doi.org/10.1007/BF01889278},
	volume = {15},
	year = {1985}}

@article{Friedrich1998,
  author       = {Friedrich, Helmut},
  title        = {Gravitational Fields Near Space-Like and Null Infinity},
  journal      = {Journal of Geometry and Physics},
  volume       = {24},
  number       = {2},
  pages        = {83--163},
  year         = {1998},
  doi          = {10.1016/S0393-0440(97)00032-1}
}

@article{Compere:2011ve,
	archiveprefix = {arXiv},
	author = {Compere, Geoffrey and Dehouck, Fran},
	doi = {10.1088/0264-9381/28/24/245016},
	eprint = {1106.4045},
	journal = {Class. Quant. Grav.},
	note = {[Erratum: Class.Quant.Grav. 30, 039501 (2013)]},
	pages = {245016},
	primaryclass = {hep-th},
	title = {{Relaxing the Parity Conditions of Asymptotically Flat Gravity}},
	volume = {28},
	year = {2011}}

@article{Prabhu:2021cgk,
    author = "Prabhu, Kartik and Shehzad, Ibrahim",
    title = "{Conservation of asymptotic charges from past to future null infinity: Lorentz charges in general relativity}",
    eprint = "2110.04900",
    archivePrefix = "arXiv",
    primaryClass = "gr-qc",
    doi = "10.1007/JHEP08(2022)029",
    journal = "JHEP",
    volume = "08",
    pages = "029",
    year = "2022"
}

@article{Troessaert:2017jcm,
	archiveprefix = {arXiv},
	author = {Troessaert, C\'edric},
	doi = {10.1088/1361-6382/aaae22},
	eprint = {1704.06223},
	journal = {Class. Quant. Grav.},
	number = {7},
	pages = {074003},
	primaryclass = {hep-th},
	title = {{The BMS4 algebra at spatial infinity}},
	volume = {35},
	year = {2018}}

@article{Andersson:1993we,
	archiveprefix = {arXiv},
	author = {Andersson, Lars and Chrusciel, Piotr T.},
	doi = {10.1103/PhysRevLett.70.2829},
	eprint = {gr-qc/9304019},
	journal = {Phys. Rev. Lett.},
	pages = {2829--2832},
	reportnumber = {TRITA-MAT-92-0038},
	title = {{On 'hyperboloidal' Cauchy data for vacuum Einstein equations and obstructions to smoothness of 'null infinity'}},
	volume = {70},
	year = {1993}}

@article{Valiente-Kroon:2002xys,
	archiveprefix = {arXiv},
	author = {Valiente-Kroon, Juan Antonio},
	doi = {10.1007/s00220-003-0967-5},
	eprint = {gr-qc/0211024},
	journal = {Commun. Math. Phys.},
	pages = {133--156},
	title = {{A New class of obstructions to the smoothness of null infinity}},
	volume = {244},
	year = {2004}}

@article{Campiglia:2020qvc,
	archiveprefix = {arXiv},
	author = {Campiglia, Miguel and Peraza, Javier},
	doi = {10.1103/PhysRevD.101.104039},
	eprint = {2002.06691},
	journal = {Phys. Rev. D},
	number = {10},
	pages = {104039},
	primaryclass = {gr-qc},
	title = {{Generalized BMS charge algebra}},
	volume = {101},
	year = {2020}}

@article{Campiglia:2014yka,
	archiveprefix = {arXiv},
	author = {Campiglia, Miguel and Laddha, Alok},
	doi = {10.1103/PhysRevD.90.124028},
	eprint = {1408.2228},
	journal = {Phys. Rev. D},
	number = {12},
	pages = {124028},
	primaryclass = {hep-th},
	title = {{Asymptotic symmetries and subleading soft graviton theorem}},
	volume = {90},
	year = {2014}}

@article{Campiglia:2015yka,
	archiveprefix = {arXiv},
	author = {Campiglia, Miguel and Laddha, Alok},
	doi = {10.1007/JHEP04(2015)076},
	eprint = {1502.02318},
	journal = {JHEP},
	pages = {076},
	primaryclass = {hep-th},
	title = {{New symmetries for the Gravitational S-matrix}},
	volume = {04},
	year = {2015}}

@article{Barnich:2009se,
	archiveprefix = {arXiv},
	author = {Barnich, Glenn and Troessaert, Cedric},
	doi = {10.1103/PhysRevLett.105.111103},
	eprint = {0909.2617},
	journal = {Phys. Rev. Lett.},
	pages = {111103},
	primaryclass = {gr-qc},
	reportnumber = {ULB-TH-09-24},
	title = {{Symmetries of asymptotically flat 4 dimensional spacetimes at null infinity revisited}},
	volume = {105},
	year = {2010}}

@article{Sachs:1961zz,
	author = {Sachs, R. K.},
	doi = {10.1098/rspa.1961.0202},
	journal = {Proc. Roy. Soc. Lond. A},
	pages = {309--338},
	title = {{Gravitational waves in general relativity. 6. The outgoing radiation condition}},
	volume = {264},
	year = {1961}}

@article{Kehrberger:2021uvf,
	archiveprefix = {arXiv},
	author = {Kehrberger, Leonhard M. A.},
	doi = {10.1007/s00023-021-01108-2},
	eprint = {2105.08079},
	journal = {Annales Henri Poincare},
	number = {3},
	pages = {829--921},
	primaryclass = {gr-qc},
	title = {{The Case Against Smooth Null Infinity I: Heuristics and Counter-Examples}},
	volume = {23},
	year = {2022}}

@book{Erdelyi1956,
  author    = {Arthur Erdélyi},
  title     = {Asymptotic Expansions},
  publisher = {Dover},
  year      = {1956},
  address   = {New York}
}

@book{Hinch1991,
  author    = {E. J. Hinch},
  title     = {Perturbation Methods},
  publisher = {Cambridge University Press},
  year      = {1991},
  address   = {Cambridge}
}

@book{Melrose1995,
  author    = {Richard B. Melrose},
  title     = {Geometric Scattering Theory},
  publisher = {Cambridge University Press},
  year      = {1995},
  address   = {Cambridge}
}

@article{Flanagan:2015pxa,
	archiveprefix = {arXiv},
	author = {Flanagan, \'Eanna \'E. and Nichols, David A.},
	doi = {10.1103/PhysRevD.95.044002},
	eprint = {1510.03386},
	journal = {Phys. Rev. D},
	number = {4},
	pages = {044002},
	primaryclass = {hep-th},
	title = {{Conserved charges of the extended Bondi-Metzner-Sachs algebra}},
	volume = {95},
	year = {2017}}

@article{Compere:2018ylh,
	archiveprefix = {arXiv},
	author = {Comp\`ere, Geoffrey and Fiorucci, Adrien and Ruzziconi, Romain},
	doi = {10.1007/JHEP11(2018)200},
	eprint = {1810.00377},
	journal = {JHEP},
	note = {[Erratum: JHEP 04, 172 (2020)]},
	pages = {200},
	primaryclass = {hep-th},
	title = {{Superboost transitions, refraction memory and super-Lorentz charge algebra}},
	volume = {11},
	year = {2018}}

@article{Barnich:2016lyg,
	archiveprefix = {arXiv},
	author = {Barnich, Glenn and Troessaert, C\'edric},
	doi = {10.1007/JHEP03(2016)167},
	eprint = {1601.04090},
	journal = {JHEP},
	pages = {167},
	primaryclass = {gr-qc},
	title = {{Finite BMS transformations}},
	volume = {03},
	year = {2016}}

@article{Barnich:2011mi,
	archiveprefix = {arXiv},
	author = {Barnich, Glenn and Troessaert, Cedric},
	doi = {10.1007/JHEP12(2011)105},
	eprint = {1106.0213},
	journal = {JHEP},
	pages = {105},
	primaryclass = {hep-th},
	reportnumber = {ULB-TH-11-10},
	title = {{BMS charge algebra}},
	volume = {12},
	year = {2011}}

@article{Compere:2008us,
	archiveprefix = {arXiv},
	author = {Compere, Geoffrey and Marolf, Donald},
	doi = {10.1088/0264-9381/25/19/195014},
	eprint = {0805.1902},
	journal = {Class. Quant. Grav.},
	pages = {195014},
	primaryclass = {hep-th},
	title = {{Setting the boundary free in AdS/CFT}},
	volume = {25},
	year = {2008}}

@article{Freidel:2020xyx,
	archiveprefix = {arXiv},
	author = {Freidel, Laurent and Geiller, Marc and Pranzetti, Daniele},
	doi = {10.1007/JHEP11(2020)026},
	eprint = {2006.12527},
	journal = {JHEP},
	pages = {026},
	primaryclass = {hep-th},
	title = {{Edge modes of gravity. Part I. Corner potentials and charges}},
	volume = {11},
	year = {2020}}

@article{Barnich:2010eb,
	archiveprefix = {arXiv},
	author = {Barnich, Glenn and Troessaert, Cedric},
	doi = {10.1007/JHEP05(2010)062},
	eprint = {1001.1541},
	journal = {JHEP},
	pages = {062},
	primaryclass = {hep-th},
	reportnumber = {ULB-TH-09-28},
	title = {{Aspects of the BMS/CFT correspondence}},
	volume = {05},
	year = {2010}}

@article{Grumiller:2019fmp,
	archiveprefix = {arXiv},
	author = {Grumiller, Daniel and P\'erez, Alfredo and Sheikh-Jabbari, M.M. and Troncoso, Ricardo and Zwikel, C\'eline},
	doi = {10.1103/PhysRevLett.124.041601},
	eprint = {1908.09833},
	journal = {Phys. Rev. Lett.},
	number = {4},
	pages = {041601},
	primaryclass = {hep-th},
	reportnumber = {Preprint: CECS-PHY-18/01, IPM/P-2019/009, TUW-18-03},
	title = {{Spacetime structure near generic horizons and soft hair}},
	volume = {124},
	year = {2020}}

@article{Strominger:2013jfa,
	archiveprefix = {arXiv},
	author = {Strominger, Andrew},
	doi = {10.1007/JHEP07(2014)152},
	eprint = {1312.2229},
	journal = {JHEP},
	pages = {152},
	primaryclass = {hep-th},
	title = {{On BMS Invariance of Gravitational Scattering}},
	volume = {07},
	year = {2014}}

@article{Barnich:1991tc,
    author = "Barnich, Glenn and Henneaux, Marc and Schomblond, Christiane",
    title = "{On the covariant description of the canonical formalism}",
    reportNumber = "ULB-TG2-91-01",
    doi = "10.1103/PhysRevD.44.R939",
    journal = "Phys. Rev. D",
    volume = "44",
    pages = "R939--R941",
    year = "1991"
}

@article{Friedrich:1983vx,
	author = {Friedrich, Helmut},
	date = {1983/12/01},
	doi = {10.1007/BF01206015},
	id = {Friedrich1983},
	isbn = {1432-0916},
	journal = {Communications in Mathematical Physics},
	number = {4},
	pages = {445--472},
	title = {Cauchy problems for the conformal vacuum field equations in general relativity},
	url = {https://doi.org/10.1007/BF01206015},
	volume = {91},
	year = {1983}}

@article{PhysRevD.19.3483,
	author = {Walker, Martin and Will, Clifford M.},
	doi = {10.1103/PhysRevD.19.3483},
	issue = {12},
	journal = {Phys. Rev. D},
	month = {Jun},
	numpages = {0},
	pages = {3483--3494},
	publisher = {American Physical Society},
	title = {Relativistic Kepler problem. I. Behavior in the distant past of orbits with gravitational radiation damping},
	url = {https://link.aps.org/doi/10.1103/PhysRevD.19.3483},
	volume = {19},
	year = {1979}}

@article{PhysRevD.19.3495,
	author = {Walker, Martin and Will, Clifford M.},
	doi = {10.1103/PhysRevD.19.3495},
	issue = {12},
	journal = {Phys. Rev. D},
	month = {Jun},
	numpages = {0},
	pages = {3495--3508},
	publisher = {American Physical Society},
	title = {Relativistic Kepler problem. II. Asymptotic behavior of the field in the infinite past},
	url = {https://link.aps.org/doi/10.1103/PhysRevD.19.3495},
	volume = {19},
	year = {1979}}

@article{Kehrberger:2024clh,
	archiveprefix = {arXiv},
	author = {Kehrberger, Leonhard},
	doi = {10.1098/rsta.2023.0039},
	eprint = {2401.04170},
	journal = {Phil. Trans. Roy. Soc. Lond. A},
	number = {2267},
	pages = {20230039},
	primaryclass = {gr-qc},
	title = {{The case against smooth null infinity IV: Linearized gravity around Schwarzschild -- an overview}},
	volume = {382},
	year = {2024}}

@phdthesis{Kehrberger:2023btg,
	author = {Kehrberger, Leonhard},
	doi = {10.17863/CAM.99689},
	school = {Department of Applied Mathematics And Theoretical Physics, Cambridge U.},
	title = {{Mathematical Studies on the Asymptotic Behaviour of Gravitational Radiation in General Relativity}},
	year = {2023}}

@article{doi:10.1098/rspa.1981.0101,
	author = {Porrill, J. and Stewart, J. M. and Penrose, Roger},
	doi = {10.1098/rspa.1981.0101},
	eprint = {https://royalsocietypublishing.org/doi/pdf/10.1098/rspa.1981.0101},
	journal = {Proceedings of the Royal Society of London. A. Mathematical and Physical Sciences},
	number = {1766},
	pages = {451-463},
	title = {Electromagnetic and gravitational fields in a Schwarzschild space-time},
	url = {https://royalsocietypublishing.org/doi/abs/10.1098/rspa.1981.0101},
	volume = {376},
	year = {1981}}

@article{Kroon:2004me,
	archiveprefix = {arXiv},
	author = {Kroon, Juan Antonio Valiente},
	doi = {10.1088/0264-9381/22/9/015},
	eprint = {gr-qc/0412045},
	journal = {Class. Quant. Grav.},
	pages = {1683--1707},
	title = {{Time asymmetric spacetimes near null and spatial infinity. II. Expansions of developments of initial data sets with non-smooth conformal metrics}},
	volume = {22},
	year = {2005}}

@article{PhysRev.124.274,
  title = {"Gauge-Invariant" Variables in General Relativity},
  author = {Bergmann, Peter G.},
  journal = {Phys. Rev.},
  volume = {124},
  issue = {1},
  pages = {274--278},
  numpages = {0},
  year = {1961},
  month = {Oct},
  publisher = {American Physical Society},
  doi = {10.1103/PhysRev.124.274},
  url = {https://link.aps.org/doi/10.1103/PhysRev.124.274}
}

@article{Beig1984Integration,
  author       = {Beig, R.},
  title        = {Integration of Einstein's Equations Near Spatial Infinity},
  journal      = {Proceedings of the Royal Society A: Mathematical, Physical and Engineering Sciences},
  volume       = {391},
  number       = {1801},
  pages        = {295--308},
  year         = {1984},
  doi          = {10.1098/rspa.1984.0015}
}

@article{Bieri:2023cyn,
    author = "Bieri, Lydia",
    title = "{Radiation and Asymptotics for Spacetimes with Non-Isotropic Mass}",
    eprint = "2304.00611",
    archivePrefix = "arXiv",
    primaryClass = "gr-qc",
    doi = "10.4310/pamq.2024.v20.n4.a4",
    journal = "Pure Appl. Math. Quart.",
    volume = "20",
    number = "4",
    pages = "1601--1634",
    year = "2024"
}

@article{Brown:1986ed,
    author = "Brown, J. David and Henneaux, M.",
    title = "{On the Poisson Brackets of Differentiable Generators in Classical Field Theory}",
    doi = "10.1063/1.527249",
    journal = "J. Math. Phys.",
    volume = "27",
    pages = "489--491",
    year = "1986"
}

@article{Boschetti:2025tru,
    author = "Boschetti, Gianni and Campiglia, Miguel",
    title = "{Log translation invariance of log soft gravitational radiation}",
    eprint = "2508.17919",
    archivePrefix = "arXiv",
    primaryClass = "gr-qc",
    doi = "10.1007/JHEP10(2025)105",
    journal = "JHEP",
    volume = "10",
    pages = "105",
    year = "2025"
}

@article{Compere:2025bnf,
    author = "Comp{\`e}re, Geoffrey and Robert, S{\'e}bastien",
    title = "{Scalar, vector and tensor fields on $dS_3$ with arbitrary sources: harmonic analysis and antipodal maps}",
    eprint = "2512.15578",
    archivePrefix = "arXiv",
    primaryClass = "hep-th",
    month = "12",
    year = "2025"
}

@article{Ashtekar:1978zz,
	author = {Ashtekar, A. and Hansen, R. O.},
	doi = {10.1063/1.523863},
	journal = {J. Math. Phys.},
	pages = {1542-1566},
	slaccitation = {%%CITATION = JMAPA,19,1542;%%},
	title = {{A unified treatment of null and spatial infinity in general relativity. I - Universal structure, asymptotic symmetries, and conserved quantities at spatial infinity}},
	volume = {19},
	year = {1978}}

@article{Sachs:1962zza,
	author = {Sachs, R.},
	doi = {10.1103/PhysRev.128.2851},
	journal = {Phys. Rev.},
	pages = {2851-2864},
	slaccitation = {%%CITATION = PHRVA,128,2851;%%},
	title = {{Asymptotic symmetries in gravitational theory}},
	volume = {128},
	year = {1962}}

@article{Bondi:1962px,
	author = {Bondi, H. and van der Burg, M. G. J. and Metzner, A. W. K.},
	doi = {10.1098/rspa.1962.0161},
	journal = {Proc. Roy. Soc. Lond.},
	pages = {21-52},
	slaccitation = {%%CITATION = PRSLA,A269,21;%%},
	title = {{Gravitational waves in general relativity. 7. Waves from axisymmetric isolated systems}},
	volume = {A269},
	year = {1962}}

@article{Regge:1974zd,
	author = {Regge, Tullio and Teitelboim, Claudio},
	doi = {10.1016/0003-4916(74)90404-7},
	journal = {Annals Phys.},
	pages = {286},
	reportnumber = {Print-74-0988 (IAS,PRINCETON)},
	slaccitation = {%%CITATION = APNYA,88,286;%%},
	title = {{Role of Surface Integrals in the Hamiltonian Formulation of General Relativity}},
	volume = {88},
	year = {1974}}

@article{Iyer:1994ys,
	archiveprefix = {arXiv},
	author = {Iyer, Vivek and Wald, Robert M.},
	doi = {10.1103/PhysRevD.50.846},
	eprint = {gr-qc/9403028},
	journal = {Phys. Rev.},
	pages = {846-864},
	primaryclass = {gr-qc},
	slaccitation = {%%CITATION = GR-QC/9403028;%%},
	title = {{Some properties of Noether charge and a proposal for dynamical black hole entropy}},
	volume = {D50},
	year = {1994}}

@article{Barnich:2007bf,
	archiveprefix = {arXiv},
	author = {Barnich, Glenn and Compere, Geoffrey},
	doi = {10.1063/1.2889721},
	eprint = {0708.2378},
	journal = {J. Math. Phys.},
	pages = {042901},
	primaryclass = {gr-qc},
	reportnumber = {ULB-TH-06-30},
	slaccitation = {%%CITATION = ARXIV:0708.2378;%%},
	title = {{Surface charge algebra in gauge theories and thermodynamic integrability}},
	volume = {49},
	year = {2008}}

@article{Brown:1986nw,
	author = {Brown, J. David and Henneaux, M.},
	doi = {10.1007/BF01211590},
	journal = {Commun. Math. Phys.},
	pages = {207-226},
	slaccitation = {%%CITATION = CMPHA,104,207;%%},
	title = {{Central Charges in the Canonical Realization of Asymptotic Symmetries: An Example from Three-Dimensional Gravity}},
	volume = {104},
	year = {1986}}

@misc{McNees-Useful,   
    title = {Conventions, Definitions, Identities, and Formulas},  
    howpublished = "\url{http://jacobi.luc.edu/Useful.html}",   
    author = {McNees, Robert},   
    year = {},   
    note = {} 
}

@incollection{Ashtekar:1990gc,
title = {The covariant phase space of asymptotically flat gravitational fields},
editor = {Mauro Francaviglia},
booktitle = {Mechanics, Analysis and Geometry: 200 Years After Lagrange},
publisher = {Elsevier},
address = {Amsterdam},
pages = {417-450},
year = {1991},
series = {North-Holland Delta Series},
issn = {09275029},
doi = {https://doi.org/10.1016/B978-0-444-88958-4.50021-5},
url = {https://www.sciencedirect.com/science/article/pii/B9780444889584500215},
author = {Abhay Ashtekar and Luca Bombelli and Oscar Reula}
}
\bibliographystyle{Biblio}
\end{document}